\documentclass[aps,twocolumn,showpacs,preprintnumbers,amsmath,amssymb,nofootinbib,superscriptaddress,showkeys]{revtex4}

\usepackage{epsfig}
\usepackage{graphicx}

\begin{document}

\title{Renormalization of NN Interaction with Chiral Two Pion Exchange
Potential. Central Phases and the Deuteron.}  \author{M. Pav\'on
Valderrama}\email{mpavon@ugr.es} \affiliation{Departamento de
F\'{\i}sica At\'omica, Molecular y Nuclear, Universidad de Granada,
E-18071 Granada, Spain.}  \author{E. Ruiz
Arriola}\email{earriola@ugr.es} \affiliation{Departamento de
F\'{\i}sica At\'omica, Molecular y Nuclear, Universidad de Granada,
E-18071 Granada, Spain.}

\date{\today}

\begin{abstract} 
\rule{0ex}{3ex} We analyze the renormalization of the NN interaction
at low energies and the deuteron bound state through the Chiral Two
Pion Exchange Potential assumed to be valid from zero to infinity. The
short distance Van der Waals singularity structure of the potential as
well as the requirement of orthogonality conditions on the wave
functions determines that, after renormalization, in the $^1S_0$
singlet channel and in the $^3S_1-^3D_1 $ triplet channel one can use
the deuteron binding energy, the asymptotic $D/S$ ratio and the
$S$-wave scattering lengths as well as the chiral potential parameters
as independent variables. We use then the asymptotic wave function
normalization $A_S$ of the deuteron and the singlet and triplet
effective ranges to determine the chiral constants yielding $c_1 $,
$c_3$ and $c_4$. The role of finite cut-off corrections, the loss of
predictive power due to uncertainties in the input data and the
connection to OPE distorted waves perturbative approaches is also
discussed.
\end{abstract}


\maketitle



\section{Introduction} 

The possibility suggested by Weinberg~\cite{Weinberg:1990rz} and
pioneered by Ray, Ordo\~nez and Van
Kolck~\cite{Ordonez:1992xp,VanKolck:1993ee,Ordonez:1995rz} of making
model independent predictions for NN scattering using Effective Field
Theory (EFT) methods and, more specifically, Chiral Perturbation
Theory (ChPT) has triggered a lot of activity in recent years (for a
review see e.g. Ref.~\cite{Bedaque:2002mn}). In addition to the
previous works, most subsequent calculations dealing with the specific
consequences of ChPT have been focused in making predictions for NN
scattering phase-shifts and deuteron properties based in the genuine
Two Pion Exchange (TPE) chiral
potentials~\cite{Rijken:1995pu,Kaiser:1997mw,Kaiser:1998wa,Epelbaum:1998ka,Epelbaum:1999dj,Rentmeester:1999vw,Friar:1999sj,Richardson:1999hj,Kaiser:1999ff,Kaiser:1999jg,
Kaiser:2001at,Kaiser:2001pc,Kaiser:2001dm,Entem:2001cg,
Entem:2002sf,Rentmeester:2003mf,Epelbaum:2003gr,Epelbaum:2003xx,Entem:2003cs,
Higa:2003jk,Higa:2003sz, Higa:2004cr} although some incipient work has
also recently been started implementing Three Pion Exchange
effects~\cite{Kaiser:1999ff,Kaiser:1999jg,Kaiser:2001dm,
Entem:2003ft,Epelbaum:2004fk}. In a given partial wave (coupled)
channel with good total angular momentum the reduced NN potential
($U(r)=M V(r) $) in configuration space can schematically be written
in local and energy independent
form~\cite{Kaiser:1997mw,Friar:1999sj,Rentmeester:1999vw} (see
Refs.~\cite{Ordonez:1992xp,VanKolck:1993ee,Ordonez:1995rz} for an
energy dependent representation) for any distances larger than a
finite short distance radial regulator, $r_c$,
\begin{eqnarray}
U (r) &=& \frac{M m^3}{f^2} W_{\rm LO} (mr, g) + \frac{M m^5}{f^4}
W_{\rm NLO} (mr, g, \bar d) \nonumber \\ &+& \frac{m^6}{f^4} W_{\rm
NNLO} (mr, g, \bar c_1 , \bar c_3 , \bar c_4 ) + \dots \, 
\label{eq:pot_chpt}
\end{eqnarray}
where $W(x)$ are known dimensionless functions which are everywhere
finite except for the origin where they exhibit power law
divergencies, which demand the use of some regularization. In writting
the previous expression we have disregarded distributional contact
terms (deltas and derivatives of deltas) which strengths are scheme
dependent but do not contribute for $ r \ge r_c > 0 $. The potential
is completely specified by the pion mass, $m$, the pion weak decay
constant, $f$, the nucleon mass $M$, the axial coupling constant $g$,
the Goldberger-Treimann discrepancy $\bar d_{18}$ and three additional
low energy constants $ \bar c_1 = c_1 M $, $ \bar c_3 = c_3 M $ and $
\bar c_4 = c_4 M$ which can directly be deduced from the analysis of
low energy $\pi N $ scattering within
ChPT~\cite{Fettes:1998ud,Buettiker:1999ap,GomezNicola:2000wk,Nicola:2003zi}. Given
this information one can then solve the single or coupled channel
Schr\"odinger equation imposing a regularity condition of the wave
function at the origin for each separate channel. In this paper we
will work under the assumption that the long range pieces of the
potential should be iterated to all orders, but some perturbative
analysis will also be done. Our motivation is to describe long range
correlations between observables in the NN problem in a model
independent way. As we will see below there is still some additional
physical information required in the form of either counter-terms or
short distance boundary conditions to make the problem well posed if
one indeed wants to go to remove the regularization.  They depend
exclusively on the short distance behaviour of the potential through
phases of the wave function. The number of phases depends crucially on
the repulsive or attractive character of the potential. In this sense
the power counting for the short distance interactions cannot be
regarded as independent on the power counting of the singular chiral
potentials. Non-perturbatively this materializes, after
renormalization, in non-integer power counting for physical
observables. This imposes severe limitations on the admissible
structure of counterterms and the corresponding renormalization
conditions of the quantum mechanical problem. Right away we hast to
emphasize that {\it our approach is not the conventional EFT one} of
allowing all possible short distance counterterms allowed by the
symmetry, and to a certain extent our viewpoints are admittedly
heterodox within the conventional EFT framework. However, based on the
physical requirement of having small wave functions in the short range
unknown region the basic and orthodox quantum mechanical requirements
of completeness and orthoghonality of states are deduced, providing a
justification for the additional restrictions. We remind here the
series of works by Phillips and
Cohen~\cite{Phillips:1996ae,Phillips:1999am} (see also
Re.~\cite{Scaldeferri:1996nx}), where restrictions on zero range
interactions were deduced for non-singular potentials based on the
Wigner causality conditions. Here we extend their results also to the
singular NN interactions of Eq.~(\ref{eq:pot_chpt}).

The theorem underlying the EFT developments is that if chiral symmetry
is spontaneously broken down in QCD, the true NN potential at long
distances is embedded in the parameter envelope of the general chiral
NN potential, Eq.~(\ref{eq:pot_chpt}), and the chiral expansion
provides a reliable hierarchy at those long distances. The hope is
that compatible and perhaps accurate determinations of both $\pi N $
and $NN$ low energy data, bound states and resonances can be achieved
with the same sets of parameters. The problem is that, in order to
make truly model independent predictions, short distance ambiguities
should be under control and their size smaller than the experimental
data uncertainties used as input of the calculation.  Only then can
the renormalization program be carried out satisfactorily as it was
done in the OPE
case~\cite{Frederico:1999ps,Beane:2001bc,PavonValderrama:2003np,PavonValderrama:2004nb,PavonValderrama:2005gu,Nogga:2005hy},
although, as recognized by Nogga, Timmermans and Van Kolck, this may be
done at the expense of modifying the power
counting~\cite{Nogga:2005hy} of the counterterms in favor of
renormalizability (see also Ref.~\cite{PavonValderrama:2005gu} for a
complementary formulation in terms of boundary conditions). The
present work analyses this problem extending our previous OPE
renormalized calculations to NNLO TPE and its implication in the
values of the chiral constants.

The determination of the chiral constants $c_1$, $c_3$ and $c_4$ (in
units of ${\rm GeV}^{-1}$ from now on) from $\pi N$ scattering has
been undertaken in several works and shows significant systematic
discrepancies depending on the details of the analysis. In Heavy
Baryon ChPT for low energy $\pi N $ scattering~\cite{Fettes:1998ud}
the values $c_1=-1.23 \pm 0.16 $, $ c_3=-5.94 \pm 0.09 $ and $c_4 =
3.47 \pm 0.05$ were deduced with a sigma term of $\sigma(0)=70 {\rm
MeV}$.  In Ref.~\cite{Buettiker:1999ap} the analysis of low energy
$\pi N $ scattering inside the Mandelstam triangle yields $c_1 = -0.81
\pm 0.15 $, $c_3 = -4.69 \pm 1.34 $ and $c_4=3.40 \pm 0.04 $ with,
however, a bit too low sigma term $\sigma (0) = 40 {\rm MeV}$ as
compared to ChPT. Unitarization methods reproducing the
phase-shifts~\cite{GomezNicola:2000wk,Nicola:2003zi} from threshold to
the $\Delta$ resonance region conclude $c_1=-0.43 \pm 0.04 $,
$c_3=-3.10 \pm 0.05 $ and $c_4=1.51 \pm 0.04 $.

The values of the chiral constants $c_1$, $c_3$ and $c_4$ also depend
on regularization details of the $NN$ chiral interaction. The $\pi N$
values from Ref.~\cite{Buettiker:1999ap} were taken in the
nucleon-nucleon NNLO calculation of Ref.~\cite{Epelbaum:1999dj} with
sharp and gaussian cut-offs $\Lambda=0.6-0.8 {\rm GeV} $ in momentum
space, and momentum dependent counter-terms were supplemented and
determined from a fit to the NN data base based on the Partial Wave
Analysis (PWA) of Ref.~\cite{Stoks:1993tb,Stoks:1994wp}. Likewise,
Ref.~\cite{Entem:2001cg} constructs a NNLO chiral potential where
channel dependent gaussian momentum space cut-offs in the range
$\Lambda=0.4-0.5 {\rm GeV}$ were used to fit the NN
database~\cite{Arndt:1994br}. The N$^3$LO extension of this
work~\cite{Entem:2003ft} uses only one common cut-off and fixing
$c_1=-0.81$ produces $c_3=-3.20$ and $c_4=5.40$.  In
Ref.~\cite{Rentmeester:1999vw} the NNLO calculation was done in
configuration space with a short distance cut-off at $r=1.4 {\rm fm}$
where an energy and channel dependent boundary condition was imposed
and the fixed value $c_1 = -0.76 \pm 0.07 $ was used to make a PWA to
pp data yielding $c_3 = -5.08 \pm 0.24 $ and $c_4 =4.70 \pm 0.70 $. An
update of this calculation also including np
data~\cite{Rentmeester:2003mf} generates $c_3 = -4.78 \pm 0.10 $ and
$c_4 = 3.96 \pm 0.22 $.  The calculations of
Ref.~\cite{Epelbaum:2003gr,Epelbaum:2003xx} improve the cut-off
dependence of the potential in momentum space by using spectral
regularization and take again gaussian cut-offs and fix $c_1=-0.81 $
yielding, after fitting the counter-terms to the NN
PWA~\cite{Stoks:1993tb,Stoks:1994wp}, the values $c_3=-3.40 $ and
$c_4=3.40$.  The extension of this work to N$^3$LO has been done in
Ref.~\cite{Epelbaum:2004fk} keeping the same values for $c_3$ and
$c_4$ and readjusting the counter-terms.

In a renormalized theory results should be insensitive to the
auxiliary regularization method if the regulator is removed at the
end. If a fit to the database proves successful, then the resulting
parameters should be cut-off independent or at least the systematic
uncertainty induced by the regularization should be smaller than the
statistical errors induced by experimental data. Otherwise, the
cut-off becomes a physical parameter.  The first indication that
finite cut-offs effects are sizeable in present calculations has to do
with the variety of values that have been used in the literature for
the low energy constants $c_1$, $c_3$ and $c_4$ to adjust NN partial
waves and deuteron
properties~\cite{Rentmeester:1999vw,Epelbaum:1999dj,Epelbaum:2003gr,Epelbaum:2003xx,Entem:2003ft,Rentmeester:2003mf}
(see also the comment in Ref.\cite{Entem:2003cs}). Obviously, we do
not expect the values of the c's to agree exactly, but the
discrepancies should be at the level of the difference in the
approximation~\footnote{For instance, pion loops at NLO modify the
contribution to $c_3$ by $\sim 3g_A^2 m^2/(64\pi f^2) \sim 0.4/{\rm
GeV}$. This contribution must be taken into account when comparing
numbers between~\cite{Rentmeester:1999vw,Rentmeester:2003mf},
\cite{Buettiker:1999ap} and the present approach. Only the extractions
using the N3LO $NN$ potential in Ref.~\cite{Entem:2003ft} are made at
the same order as those from \cite{Buettiker:1999ap}. }.  Since the
data base is the same but the regularization schemes are different,
one unavoidably suspects that these determinations of the LECS may
perhaps be regularization and hence cut-off dependent.

To get a proper perspective on the issue of renormalization let us
consider the size of the contributions of the chiral potential in
configuration space at different distances. For instance, at $r=1.4
{\rm fm }$ in the $^1S_0$ channel each order in the expansion is about
an order of magnitude smaller than the preceding one. At short
distances, however, the situation is exactly the opposite, higher
orders dominate over the lower orders. In the previous example of the
$^1S_0$ channel, LO and NLO become comparable at $r \sim 0.9 {\rm fm
}$, and NLO and NNLO become comparable at distances which value $r
\sim 0.1-0.4 {\rm fm }$ depends strongly on the particular choice of
low energy constants $c_3 $ and $c_4$. Actually, a general feature of
the chiral NN potentials at NNLO has to do with their short distance
behavior; they develop an attractive Van der Waals singularity $U \sim
-M C_6 /r^6 $ similar to the one found for neutral atomic systems. In
such a situation, the standard regularity condition at the origin only
specifies the wave function uniquely if the potential is repulsive,
but some additional information is required if the potential is
attractive~\cite{Case:1950} (for a comprehensive review in the one
channel case see e.g.  Ref.~\cite{Frank:1971}.). Within the EFT
framework the problem has been revisited in Ref.~\cite{Beane:2000wh}.
The net result is that the regularity condition at the origin tames
the singularity~\cite{PavonValderrama:2005gu} and, in fact, more
singular potentials become less important at low energies.

In this work we reanalyze the NN chiral potential including TPE
potential at NNLO. We carry out the analysis entirely in coordinate
space following the ideas developed in our previous
work~\cite{PavonValderrama:2005gu} for the OPE potential.  In
configuration space the (renormalized) potential is finite except at
the origin, a point which should carefully be handled, requiring a
delicate numerical limiting procedure. For a singular potential
in coordinate space, the corresponding potential in momentum space is
not finite unless a short distance cut-off or a subtraction procedure
at the level of the potential is implemented, hence modifying the
potential everywhere and not just at high energies. This results
generally in {\it two} cut-offs: one for the irreducible two point
function and another for the Lippmann-Schwinger
iteration~\cite{Epelbaum:1999dj,Entem:2001cg,Entem:2002sf,Epelbaum:2003gr,Epelbaum:2003xx,Entem:2003ft,Epelbaum:2004fk}. The
short distance coordinate space cut-off is unique and common both to
the potential and the scattering solution.

Unlike previous works on the TPE potential we try to remove the
cut-off completely taking the consequences seriously.  This does not
mean that finite cut-off calculations are necessarily incorrect or not
entitled to describe all or some part of the data, but there are also
good reasons for removing the cut-off and looking at the physical
consequences. In the first place the limit exists in strict
mathematical sense under well defined conditions, as the analysis
below shows. This is a non-trivial fact, because calculations done in
momentum space can only address this question numerically by adding
counterterms suggested by an {\it a priori} power counting on the
short distance potential. As shown in Ref.~\cite{Nogga:2005hy} this
does not always work, and calculations may require some trial and
error. Secondly, this is the only way we know how to get rid of short
distance ambiguities, and thus to make truly model independent
calculations. Third, the study of peripheral waves has proven to be
successful by using perturbative renormalized amplitudes corresponding
to irreducible TPE and iterated OPE where the cut-off has been removed
in the intermediate
state~\cite{Kaiser:1997mw,Entem:2002sf}. Peripheral waves mainly probe
large distances in the Born approximation but they also see some of
the short distance interaction due to re-scattering effects. Fourth,
the advantage of renormalization is that one should obtain the same
results provided one uses as input the same physical information,
regardless whether the calculation is done in coordinate or momentum
space, and also regardless on the particular regularization. Finally,
a reliable estimate on the errors and convergence rate of the chiral
expansion can be done, without any spurious cut-off contamination. In
principle, the higher order in the chiral expansion the better,
provided there is perfect errorless data to fit the increasing number
of low energy constants appearing at any order. However, the chiral
expansion may reach a limited predictive power because of finite
experimental accuracy in the low energy constants used as input. The
output inherits a propagated error which may eventually become larger
than the experimental uncertainty~\footnote{This issue has been
illustrated in
Refs.~\cite{Nieves:1999zb,Colangelo:2000jc,Colangelo:2001df} for the
case of $\pi\pi$ scattering at two loops, and will become clear in NN
scattering below.}. Finite cut-off uncertainties are not a substitute
for propagating input experimental errors to the predictions of the
theory, and can be regarded at best as a lower bound on systematic
errors. In this paper we regard this possible cut-off dependence as
purely numerical inaccuracies of the calculation, and not as a measure
of the uncertainty in the predictions of the theory, so we make any
effort to minimize these cut-off induced systematic errors.

In the process of eliminating the cut-off we find some surprises, and
effects not explored up to now become manifest. Even for low energy
scattering parameters and deuteron properties, where the description
should be more reliable and robust we find systematic discrepancies in
our calculation with values quoted in the literature and which we
conclusively identify as finite cut-off effects. This might provide a
natural explanation why calculations with different cut-off methods
fitting the NN phase
shifts~\cite{Stoks:1993tb,Stoks:1994wp,Arndt:1994br} obtain different
results for the chiral constants $c_1$, $c_3 $ and $c_4 $ or why
different values of the constants yield good fits to the data.
According to our study, for the lowest phases the reason can partly be
related to the dominance of short distance Van der Waals singularities
for a system with unnaturally large scattering lengths or a weakly
bound state as it is the case for the $^1S_0$ and $^3S_1-^3D_1$
channels. In some cases, they are as large as a $30\%$ effect like in
the effective range of the triplet $^3S_1$ channel. The size of the
effect depends on the value of the low energy $\pi N $ constants
$c_1$, $c_3$ and $c_4$. Given the significant sensitivity of low
energy NN properties and deuteron properties on these low energy $\pi
N $ constants we try to make a fit to some {\it low energy properties
which uncertainties are reliably known and where we expect the chiral
theory to be most reliable}.  At this point we depart from the
standard large scale fits to all phase shifts or partial wave analysis
where the low energy threshold parameters are determined {\it a
posteriori}. The assignment of statistical errors on the fitting
parameters $c_1 $, $c_3$ and $c_4$ is often not addressed (see however
Refs.~\cite{Rentmeester:1999vw,Rentmeester:2003mf}) because the NN
data bases used to fit the phase
shifts~\cite{Stoks:1993tb,Stoks:1994wp,Arndt:1994br} are treated as errorless. 
We also try to improve on
this point within our framework.

The paper is organized as follows. In Sect.~\ref{sec:small_wf} we
discuss the basic assumption of the smallness of the wave function in
the short range unknown region and its consequences. We also analyze
the constraints based on causality and analyticity of the $S-$matrix.
In Sect.~\ref{sec:short} we introduce the classification of boundary
conditions which will be used along the paper to effectively
renormalize the amplitudes both in the one-channel as well as the
coupled channel case. We will also review the orthogonality
constraints for singular potentials already used in our previous
work~\cite{PavonValderrama:2005gu} for the OPE
potential. Sect.~\ref{sec:singlet} deals with the description of the
singlet $^1S_0$ channel. From the superposition principle of boundary
conditions we show how a universal form of a low energy theorem for
the threshold parameters as well as for the phase-shift arises. In
Sect.~\ref{sec:triplet} we discuss the interesting triplet
$^3S_1-^3D_1$ channel both for the deuteron bound state as well as the
corresponding scattering states, where full use of the orthogonality
constraints as well as the superposition principle of boundary
conditions generates interesting analytical relations connecting
deuteron and scattering properties. In Sect.~\ref{sec:error} a careful
discussion of errors for our cut-off independent results is carried
out. Also, a determination of the chiral constants based on low energy
data and deuteron properties is made.  In Sect.~\ref{app:a} we present
a simplified study on the significance of the chiral Van der Waals
forces and the striking similarities with the full calculations for
the s-waves. In Sect.~ \ref{app:b} we show some puzzling results for
the NLO calculation in the deuteron channel. We also comment the
relation to finite cut-off calculations and the conflict between
Weinberg counting and non-perturbative renormalization at NLO. We also
outline possible solutions to this problem. In Sect.~\ref{sec:pert} we
analyze our results on the light of long distance perturbation theory,
reinforcing the usefulness of non-perturbative renormalization due to
an undesirable proliferation of counterterms. Finally, in
Sect.~\ref{sec:conc} we summarize our conclusions.

For numerical calculations we take $f_\pi=92.4 {\rm MeV}$, $m=138.03
{\rm MeV}$, $M = M_p M_n /(M_p+M_n) = 938.918 {\rm MeV}$, $ g_{\pi NN}
=13.083 $ in the OPE piece to account for the Goldberger-Treimann
discrepancy according to the Nijmegen phase shift analysis NN
scattering~\cite{deSwart:1997ep} and $g_A=1.26 $ in the TPE piece of
the potential.  The values of the coefficients $c_1$, $c_3$ and $c_4$
used along this paper are listed in Table \ref{tab:table_vdw} for
completeness. The potentials in configuration space used in this paper
are exactly those provided in
Ref.~\cite{Kaiser:1997mw,Friar:1999sj,Rentmeester:1999vw} but disregarding
relativistic corrections, $M/E \to 1$.

\section{Short distance insensitivity conditions and Renormalization} 
\label{sec:small_wf}

In this section we ellaborate on the essential role played by standard
quantum mechanical orthogonality and completeness properties of the
wave functions in the rest of this paper. As we have already
mentioned, our approach is unconventional from an EFT perspective, and
at present it is unclear whether such properties have an EFT
justification. At the same time one should say that many
self-denominated EFT calculations do indeed normalize deuteron wave
functions to unity and use energy independent regulators from which
orthogonality relations follow automatically. 

\subsection{The inner and the outer regions}

Similar to EFT, our basic assumption is that low energy physics should
not depend on short distance fine details. This rather general
principle can be made into a precise quantitative statement {\it in
practice} for a quantum mechanical system. For the sake of clarity let
us consider the singlet $^1S_0$ channel for positive energies. If we
assume a short distance regulator $r_c$, above which our long distance
(local) potential acts, the reduced Schr\"odinger equation in the
outer region reads
\begin{eqnarray}
-u_{k, {\rm L}}'' (r) + U_{\rm L} (r) u_{k,{\rm L}} (r) = k^2 u_{k,{\rm L}} (r) \, , \qquad r > r_c\end{eqnarray} 
where the label L stands for long. Asymptotically behaves as 
\begin{eqnarray}
u_{k,{\rm L}} (r) \to A \sin ( kr + \delta ( k, r_c ) )
\end{eqnarray} 
where $A$ is an arbitrary normalization constant and the dependence on
the short distance regulator $r_c$ has been explicitly highlighted. In
the inner region, the dynamics is {\it unknown} but we also expect it
to be {\it irrelevant} provided $k r_c \ll 1 $, i.e. if we assume the
corresponding wavelength to be larger than the short distance
scale. The potential can be deduced from perturbation theory in the
full amplitude,
\begin{eqnarray}
U (\vec x) = C_0 \delta (\vec x) + C_2 \left\{ \nabla^2 , \delta (\vec
x) \right\} +  \dots + U_{\rm L} (x)
\end{eqnarray} 
where $U_L(x)$ corresponds to the expansion in
Eq.~(\ref{eq:pot_chpt})~\cite{Kaiser:1997mw,Friar:1999sj,Rentmeester:1999vw}.
The distributional contact terms are regularization scheme dependent
and correspond to polynomial terms in momentum space. Obviously, they
do not contribute to the region $r> r_c$ for $r_c > 0$. The very
nature of such a calculation already implies that $C_0$, $C_2$,
etc. are perturbative corrections to the short range physics, but do
not include possible non-perturbative effects. As will become clear
below (Sect.~\ref{sec:short}), finiteness of the physical phase shitf
in the limit $r_c \to 0$, implies a highly non-perturbative
reinterpretation of the short range terms, even if the long range
pieces are computed perturbatively.

Following~\cite{Phillips:1996ae,Scaldeferri:1996nx} it is useful to
use a nonlocal and energy independent potential to describe the short
distance dynamics
\begin{eqnarray}
-u_{k,{\rm S}} '' (r) + \int_0^{r_c} U_{\rm S} (r,r') u_{k,{\rm
 S}} (r')&=& k^2 u_{k,{\rm S}} (r) \, ,  \nonumber \\ &&  r < r_c
\label{eq:short-nl}
\end{eqnarray} 
where the label S stands for short. This holds provided we are below
any inelastic channel such as $\pi NN$. Above such threshold, any open
channel should be included explicitly.  The nonlocal short distance
potential $U_{\rm S} (r,r') $ encodes, in particular, contact terms
(deltas and derivatives of deltas) which appear when the long distance
potential $U_{\rm L} (r)$ is computed in perturbation theory. These
terms are in fact ambiguous (and hence unphysical) and depend on the
regularization scheme used in the perturbative calculation, but are
not essential since they do not contribute to the absortive part and
hence to the corresponding spectral function~\cite{Kaiser:1997mw}. The
important point is that these ambiguous distributional terms never
contribute to the long range part provided $r_c > 0$ since in this
case a compact support for distributional terms is guaranteed. This is
a clear advantage of the radial cut-off we are using. In fact, our
main motivation for carrying out the analysis in coordinate space is
this clean separation between short and long distances. Many
regulators, mainly those in momentum space, do not fulfill this
condition, since the regulator effectively smears these short distance
terms and intrudes somewhat into the long distance
region~\footnote{For instance, the widely used gaussian regulator in
momentum space~\cite{Entem:2003ft,Epelbaum:2004fk,Nogga:2005hy} of a
local potential in cordinate space corresponds indeed to a convolution
of the original potential smeared over the region of size
$1/\Lambda$.}. 

Finally, we need a matching condition connecting the inner and outer
regions,
\begin{eqnarray}
\frac{u_{k,{\rm L}}'(r_c)}{u_{k,{\rm L}}(r_c)}
=\frac{u_{k,{\rm S}}'(r_c)}{u_{k,{\rm S}}(r_c)}
\label{eq:matching_ls}
\end{eqnarray} 
Viewed from the outer region this relation corresponds to an energy
dependent boundary condition at a given short distance cut-off radius,
$r_c$. Because of elastic unitarity we expect the state to be
normalized, so that if we use a box of size $a$ as an infrared
regulator, we have 
\begin{eqnarray}
\int_0^{r_c} u_{k,{\rm S}} (r)^2 dr + \int_{r_c}^a u_{k,{\rm L}} 
(r)^2 dr = 1
\end{eqnarray}
with $a$ much larger than the range of the potential.  This equation
gives a quantitative separation between long distance and known
physics and short distance and unknown physics. Obviously, any
effective description based on the long range part should fulfill
\begin{eqnarray}
\int_0^{r_c} u_{k,{\rm S}} (r)^2 dr \ll  1 
\label{eq:small_wf}
\end{eqnarray} 
which corresponds to the requirement of a {\it small wave function in
the inner unknown region}. This is our basic condition from which most
of our results follow. It should be realized that here long and short
distances are intertwined through the matching condition,
Eq.~(\ref{eq:matching_ls}). In particular, an arbitrarily growing
function at the origin cannot fulfill this condition even if value of
the short distance cut-off is taken.

\begin{figure}[ttt]
\begin{center}
\epsfig{figure=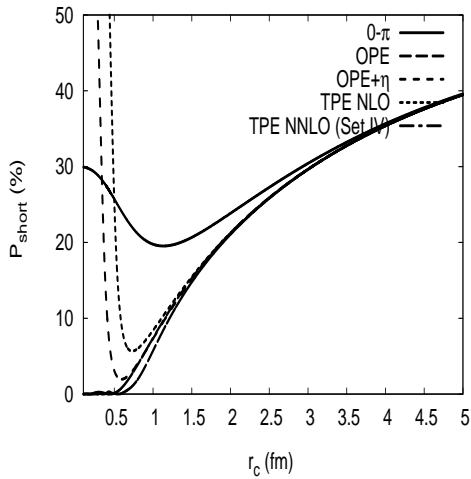,height=6.5cm,width=6.5cm}
\end{center}
\caption{Probability inside the short distance region as a function of
the short distance cut-off radius $r_c$ for the deuteron state for
different approximations to the potential. In all cases the deuteron
binding energy is fixed to the experimental value.  $0-\pi $ means a
pionless theory where in addition $\eta=0.0256$ is also fixed to the
experimental number. OPE means our previous
work~\cite{PavonValderrama:2005gu} where $\eta=0.02633$ is deduced
from the regularity condition of the wave function at the
origin. $OPE+\eta$ coresponds to the OPE case when $\eta=0.0256$ (the
experimental value). TPE NLO and TPE NNLO (Set IV) correspond to the
higher terms in Eq.~(\ref{eq:pot_chpt}) where also the experimental
$\eta$ value is taken.}
\label{fig:prob-short}
\end{figure}

This kind of pathological situation actually occurrs when dealing with
the deuteron channel in the theory with no explicit
pions~\cite{Phillips:1999am}, with OPE
potential~\cite{PavonValderrama:2005gu} and TPE potential at NLO in
the Weinberg counting (see Sect.~\ref{app:b} and \ref{sec:finite}
below). As an instructive and enlightening example, we illustrate the
situation in Fig.~\ref{fig:prob-short} in the deuteron state for the
quantity
\begin{eqnarray}
P (r_c) &=& \int_0^{r_c} (u (r) ^2+ w(r) ^2 ) dr
\end{eqnarray} 
(for notation see Sect.~\ref{sec:triplet}) using the LO, NLO and NNLO
potentials, Eq.~(\ref{eq:pot_chpt}), in the outer region, $r > r_c$,
and matching to a free particle in the inner region $r < r_c$ (the
precise form of the wave function inside turns out not to be
essential)~\footnote{In a momentum space formulation this is somewhat
equivalent to cut-off the Lippmann-Schwinger equation above a given
value $\Lambda$.}. As one sees, there are cases such as a pure short
distance theory where the asymptotic $D/S$ ratio, $\eta$, and the
deuteron binding energy are fixed to their experimental value. For
$r_c < 1.4 {\rm fm}$ one has a miminum probability of about $20\%$ in
the inner region, and then $P(r_c) $ starts to grow. In this case the
description cannot be considered effective below that critical
value. A similar situation occurs in the NLO TPE case where one has a
decreasing inside probability until one reaches $r_c \sim 0.8 {\rm
fm}$ where it takes its minimum value, about $7 \%$ and then starts
increasing. In contrast, for the LO (OPE) case where the deuteron
binding energy is fixed to the experimental one and $\eta$ is
predicted~\cite{PavonValderrama:2005gu} to have the value
$\eta=0.02633$ from the regularity condition of the wave function at
the origin and also the NNLO (TPE) case where both numbers are fixed
to experiment, the probability in the interior region is controlled
and generally decreasing. This is a first and transparent illustration
that the description based on any preconceived power counting is not
necessarily consistent with the fact that short distance ambiguities
are under control, since the wave function in the inner and unknown
region does not become arbitrarily small as the cut-off is removed.
Another possibility is to keep a finite cut-off, and we will analyze
this point in more detail in Sect.~\ref{sec:finite}.

As we will discuss in the next Sect.~\ref{sec:short}, tight
constraints on the structure of short distance counterterms must be
fulfilled if the requirement $P(r_c) \to 0 $ as $r_c \to 0^+$ is
imposed.

\medskip

\subsection{Wigner bounds on the short distance contributions}

To proceed further, we derive with respect to the energy
Eq.~(\ref{eq:short-nl}) and one immediately gets after some algebra,
\begin{eqnarray}
\frac{d}{d k^2} \left[\frac{u_{k,{\rm S}}'(r_c)}{u_{k,{\rm
S}}(r_c)} \right] = - \frac{
\int_0^{r_c} u_{k,{\rm S}} (r)^2 dr}{u_{k,{\rm S}} (r_c)^2} \le 0
\label{eq:wigner}
\end{eqnarray} 
whenever $r_c$ is not a zero of the wave function.

\begin{figure}[ttt]
\begin{center}
\epsfig{figure=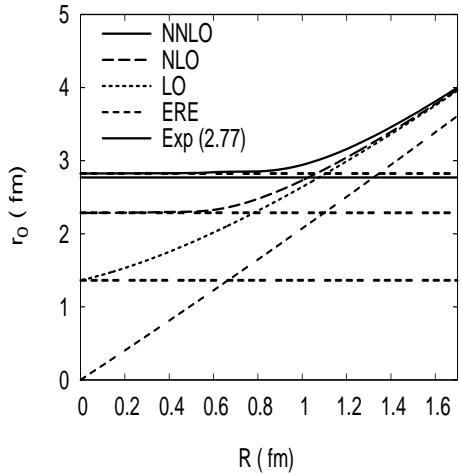,height=6.5cm,width=6.5cm}
\end{center}
\caption{Wigner bound on the effective range of the short distance
interaction, (r.h.s. of Eq.~(\ref{eq:wigner_r0})) as a function of the
short distance cut-off radius for the contact interaction (ERE), the
OPE interaction (LO), the TPE (NLO) and NNLO. We use the conditions
$\alpha_0 = -23.74 {\rm fm}$ and $r_0 = 2.77 {\rm fm}$ when $r_c \gg
1/ m_\pi $. Causality is violated when the curve crosses the
experimental value. Straight horizontal lines correspond to the
values obtained for the effective range when the short distance
initial condition $r_{0,S} (0^+)=0$ is assumed.}
\label{fig:wigner-violation}
\end{figure}

The short range theory can be characterized by an {\it accumulated}
phase shift $\delta_S (k, r_c)$, given by the solution of the
truncated short range problem,
\begin{eqnarray} 
u_{k,{\rm S}} (r) = \sin \left( k r + \delta_S (k,r_c) \right) \, ,
\quad r > r_c
\end{eqnarray} 
which fulfills, from Eq.~(\ref{eq:wigner}),
\begin{eqnarray}
\frac{d}{d k^2} \left[ k \cot \left( k r_c + \delta_S (k,r_c) \right)
\right] \le 0
\end{eqnarray} 
a condition equivalent to Wigner's causality
condition~\cite{Phillips:1996ae}. Using an effective range expansion
for the short distance phase shift
\begin{eqnarray}
k \cot \delta_S = - \frac1{\alpha_{0,S}}+ \frac12 r_{0,S} k^2 + \dots
\end{eqnarray} 
we get the Wigner bound for the effective range 
\begin{eqnarray}
r_{0,S} \le 2 r_c \left[ 1 - \frac{r_c}{\alpha_{0,S}}+ \frac{r_c^2}{3
\alpha_{0,S}} \right] 
\label{eq:wigner_r0}
\end{eqnarray} 
where $\alpha_{0,S} $ and $r_{0,S}$ represent the scattering length
and effective range when the short distance potential $U_S (r,r')$ is
switched on from the origin up to the scale $r_c$ or equivalently when
the long distance potential is switched off from infinity down to
$r_c$ (see Refs.~\cite{PavonValderrama:2003np,PavonValderrama:2004nb}
for more details).  In a theory where the long distant potential is
absent $U_{\rm L}(r)=0$, i.e., a pure short distance description,
we have the obvious result that the short distance threshold
parameters coincide with the physical parameters
$\alpha_{0,S}=\alpha_0 $ and $r_{0,S}= r_0$. Thus,
\begin{eqnarray}
r_{0} \le 2 r_c \left[ 1 - \frac{r_c}{\alpha_{0}}+ \frac{r_c^2}{3
\alpha_{0}} \right] \, , \quad U_{\rm L} (r)=0  
\end{eqnarray} 
which implies that $r_0 \le 0 $ for $r_c \to 0 $. With the
experimental values one gets the lowest short distance cut-off
compatible with causality to be $r_c=1.4{\rm fm}$. For a given long
distance potential we just solve the equations from infinity inwards
and look for the point where the Wigner condition is first violated.
In Fig.~\ref{fig:wigner-violation} we plot the evolution of the Wigner
bound on the effective range for the different approximations to the
potential according to the expansion (\ref{eq:pot_chpt}) as a function
of the short distance cut-off radius. As we see, the lower bound on
the radius is pushed towards the origin, and in fact {\it for the NNLO
approximation there is no lower bound at all}. Thus, only for the NNLO TPE 
potential can one build the full strength of the experimental
effective range without violation of the Wigner condition. We will see
more on this in Sect.~\ref{sec:singlet}.

So far the discussion has been carried out for a fixed value of $r_c$.
If we change the short distance radius, $r_c \to r_c + \Delta r_c$, we
can use the matching condition, Eq.~(\ref{eq:matching_ls}), to
evaluate the change seen from the outer region. This results into a
variable phase equation which has been analyzed extensively in our
previous works~\cite{PavonValderrama:2003np,PavonValderrama:2004nb}.
If we take the limit $r_c \to 0 $, we get that the outer wave
functions fulfills
\begin{eqnarray}
\frac{d}{d k^2} \left[\frac{u_k '(0^+)}{u_k( 0^+)} \right] =0  
\end{eqnarray} 
where the label L has been supressed.  Thus, the boundary condition
becomes {\it energy independent} when the limit $r_c \to 0^+$ is taken
if the inner wave function becomes arbitrarily small. Note that the
limit is taken {\it from above} such that $r_c > 0 $~\footnote{To
illustrate this point, let us note that for potentials which do not
diverge too strongly at the origin $r^2 U(r) \to 0$ such as OPE in the
$^1 S_0$ channel, there is some irreversibility in the process of
integrating from {\it exactly} $r_c=0$ out (which requires the regular
solution, $u(0)=0$) or integrating in towards the origin from above $r
\to 0^+$ (which generally involves the irregular solution, $u(0) \neq
0 $). See the discussion in Ref.~\cite{PavonValderrama:2003np} and in
Sect.~\ref{sec:power_counting}}. As we have said already, this
justifies not considering contact terms in the potential. A direct
consequence is that we get also that with the exception of the short
distance scattering length $\alpha_{S,0} (0^+)$, which can be fixed by
some renormalization condition (see Sect.~\ref{sec:short}), the
remaining short distance threshold parameters are also zero in this
limit,
\begin{eqnarray}
r_{0,S} (0^+)=0  \, \qquad v_{2,S}(0^+) =0 \, , \, \dots 
\end{eqnarray} 
This energy independence of the boundary condition at the origin 
insures the orthogonality conditions between different energy states,  
\begin{eqnarray}
\int_{0^+}^\infty u_{k} (r) u_{k'} (r) dr = \delta (k-k') \, .
\end{eqnarray}

\medskip

\section{Short distance behavior of CHIRAL POTENTIALS, orthogonality constraints and the number of independent constants} 
\label{sec:short} 

As we have said, chiral NN potentials, Eq.~(\ref{eq:pot_chpt}), 
although decay exponentially at large distances, become singular at
short distances, where one has
\begin{eqnarray} 
U(r) = \frac{M C_n }{r^n} \left( 1 + a_1 r + a_2 r^2 + \dots \right) 
\end{eqnarray} 
To avoid any misconception let us emphasize that the short distance
behaviour of a long distance potential should be regarded as a long
distance feature, i.e. a long wavelength property, since different
long distance potentials yield different short distance
behaviours. The short distance properties of chiral potentials have
nothing to do with short distance properties of the ``true''
potential, but renormalization and finiteness requires a very precise
behaviour of the wave function when approaching the origin from long
distances. In this section we classify the undetermined constants
depending on the attractive or repulsive nature of the corresponding
potentials in the single channel case and the eigenvalues of the
potential matrix in the coupled channel case. In coordinate space and
disregarding relativistic corrections the potentials in
Eq.~(\ref{eq:pot_chpt}) are local and energy independent~\footnote{In
momentum space and up to NNLO the long distance part of the potential
depends on the momentum transfer $q$ only and not on the total
momentum $k$. Essential non-localities, i.e. contributions of the form
$ V(q,k) = L(q) k^2 $ with $L(q)$ a non-polynomial function, depend
weakly on the total momentum and appear first at
N$^3$LO~\cite{Kaiser:1999ff,Kaiser:1999jg,Kaiser:2001at,Kaiser:2001pc,Kaiser:2001dm,Entem:2002sf}
due to relativistic $1/M^2$ one loop contributions. In coordinate
space this weak non-locality corresponds to a modification of the
kinetic energy term in the form of a general self-adjoint
Sturm-Liouville operator, $-u''(r) \to - (p(r) u'(r))' $, with a
singular $p(r)$ function at the origin and exponentially decaying at
long distances. The present formalism can in principle be extended to
include these features, and will be discussed elsewhere. Nevertheless,
according to the results of Sect.~\ref{sec:error} (see
Table~\ref{tab:table_errors} ) on the loss of predictive power already
at NNLO there is a lack of phenomenological motivation.}. An important
condition on the short distance behavior of the wave functions are the
orthogonality constraints between states of different energy. For a
regular energy independent potential these constraints are
automatically satisfied, but for singular potentials they generate new
relations relevant to the NN interaction. Our approach is not the
conventional one of adding short distance counterterms following an
{\it a priori} power counting on the long distance potential. Rather
it is the power counting in the potential what uniquely determines the
admissible form of the short distance physics if we want to reach a
finite limit when the regulator is removed. This can only be achieved
by choosing the {\it regular solution} at the origin,
i.e. $u(0)=0$. In Sect.~\ref{sec:finite} we will show that irregular
solutions generate divergent results after renormalization.  Although
this may look as a potential drawback of removing the cut-off it
provides valuable insight on the form of the potential and the
validity of the expansion (see Sect.~\ref{sec:delta}).

\subsection{One channel case} 

Let us first review the single channel case in a way that results for
the coupled channel situation can be easily stated. The reduced
Schr\"odinger equation for angular momentum $l$ is
\begin{eqnarray}
-u'' + U(r) u + \frac{l(l+1)}{r^2} = k^2 u \, . 
\end{eqnarray} 
 For a power law singular potential at the origin of the form $U (r) =
M C_n /r^n = \pm (R/r)^n /R^2 $ with $n > 2 $ and $R$ the length scale
dimension, the de Broglie wavelength is given by $ 1/k(r) =1 /
\sqrt{|U(r)|} $ and the applicability condition for the WKB
approximation reads $(1/k(r))' \ll 1$, so that for distances $ r \ll R
(n/2)^{2/(2+n)} $ one has a semiclassical wave
function~\cite{Case:1950,Frank:1971,Beane:2000wh}. Keeping the leading
short distance behavior one gets for attractive and repulsive singular
potentials and any angular momentum the following behaviour for the
{\it regular} solutions
\begin{eqnarray}
u_A (r) &\to & C_A \left(\frac{r}{R}\right)^{n/4} \sin\left[
\frac{2}{n-2} \left(\frac{R}{r}\right)^{\frac{n}2-1} + \varphi \right] \, ,\nonumber \\  \label{eq:uA} \\
{\rm for} \qquad U_A &\to& -\frac{1}{R^2} \left(\frac{R}{r}\right)^n  \, \\   
u_R (r) & \to & C_R \left(\frac{r}{R}\right)^{n/4} \exp \left[- \frac{2}{n-2}
\left(\frac{R}{r}\right)^{\frac{n}2-1}  \right] \, ,\label{eq:uR} \\ {\rm for} \qquad 
U_R &\to& +\frac{1}{R^2} \left(\frac{R}{r}\right)^n \, ,   
\end{eqnarray} 
respectively. Here $C_A$ and $C_R$ are normalization constants and
$\varphi$ an arbitrary short distance phase. In the repulsive case we
have discarded the irregular solution (a similar exponential with a
positive sign) which would not allow to normalize states. For an
attractive singular potential there is a short distance unknown
parameter. This phase could, in principle, be energy dependent. Chiral
potentials are, however, local and energy independent at NNLO at all
distances, and become genuinely energy dependent at N$^3$LO, due to
relativistic $1/M^2$ corrections~\cite{Kaiser:1999ff,Kaiser:1999jg,
Kaiser:2001at,Kaiser:2001pc,Kaiser:2001dm}~\footnote{The subthreshold
energy dependence from the virtual pion production channel $NN \to NN
\pi $ which is in principle N$^3$LO, disappears since in the heavy
baryon limit the threshold $s_{\pi NN} = (2M+m)^2 = 4(M^2+k^2)$
translates into a CM momentum $ k = \sqrt{m (M+m/4)} \to \infty
$.}. Thus, if we require orthogonality of states with different energy
(positive or negative) we get
\begin{eqnarray} 
0 &=& u_k' u_p - u_p u_k' \Big|_0 \nonumber \\
&=& \frac1{R} \sin \left( \varphi(k) - \varphi(p) \right) \, . 
\end{eqnarray} 
Hence, the phase $\varphi $ is energy independent and could be
fixed by matching the solution to the asymptotic large distance region
(we assume a short range potential), e.g., by requiring a given value
of the scattering length, $\alpha_l$, at zero energy. In this way, a
new and physical scale appears into the problem which is not specified
by the potential. This is equivalent to the well known phenomenon of 
dimensional transmutation. Another possibility is to fix $\varphi $
from a given bound state energy, $E=-B$. The new scale entering the
problem is the corresponding wave number, $\gamma= \sqrt{M B}$.  Note
that although neither $\alpha_l$ nor $\gamma$ can be predicted from a
singular potential, the orthogonality constraint does predict a
correlation between them through the potential. Likewise, the phase
shifts $\delta_l$ can be deduced from either $\alpha_l$ or $\gamma$ by
taking the same short distance phase $\varphi$. In the repulsive case
there is no dimensional transmutation since the orthogonality
condition follows from regularity at the origin, and the potential
fully specifies the wave function. In this case, the scattering length
and the spectrum are completely determined from the potential as for
standard regular potentials.

\subsection{Coupled channel case} 

We turn now to the two coupled channel case where the wave functions
are denoted by a column vector $(u,w) $ (for some particular cases see
e.g. Refs.~\cite{Martorell94,Beane:2000wh,PavonValderrama:2005gu}).  If we
assume that at short distances the reduced potential behaves as
\begin{eqnarray}
U \to M \frac{\bf C_n}{r^n} \,
\end{eqnarray} 
where ${\bf C}_n$ is a symmetric matrix of Van der Waals
coefficients. Diagonalizing the matrix ${\bf C_n} $ we get
\begin{eqnarray} 
{\bf C_n}
&=&  
\begin{pmatrix}
\cos\theta & \sin\theta  \\ -\sin\theta &  \cos \theta   
\end{pmatrix}  
\begin{pmatrix}
C_{n,+}  & 0  \\ 0 &  C_{n,-}     
\end{pmatrix}  
\begin{pmatrix}
\cos\theta & -\sin\theta  \\ \sin\theta &  \cos \theta   
\end{pmatrix} \nonumber \\ \, 
\end{eqnarray}  
where $C_{n,\pm} $ are the corresponding eigenvalues and $\theta$ the
mixing angle. Thus, at short distances we can decouple the equations
to get 
\begin{eqnarray} 
\begin{pmatrix}
u \\ w  
\end{pmatrix}  
&\to &
\begin{pmatrix}
\cos\theta & \sin\theta  \\ -\sin\theta &  \cos \theta   
\end{pmatrix}  
\begin{pmatrix}
u_+   \\ u_-     
\end{pmatrix}  \, 
\end{eqnarray}  
where $( u_+ , u_- ) $ are regular solutions as in the single channel
case. So, in the two channel situation we have three possible cases
depending upon the sign of the eigenvalues.
\begin{enumerate} 
\item Both eigenvalues are negative, i.e., both eigenpotentials are
attractive and $M C_{n,+} = -R_+^{n-2} $ and $M C_{n,-} = -R_-^{n-2} $
with $R_\pm $ the corresponding scale dimension. In this case the
short distance eigensolutions are oscillatory and there are two
undetermined short distance phases, $\varphi_+ $ and
$\varphi_-$. Moreover for two states, $(u_k,w_k) $ and $(u_p,w_p) $,
with different energies we get the orthogonality constraint
\begin{eqnarray} 
0 &=& u_k' u_p - u_p u_k' + w_k' w_p - w_p w_k' \Big|_0 \nonumber \\  &=& \frac1{R_+}
 \sin \left( \varphi_+ (k) - \varphi_+ (p) \right) + \frac1{R_-} \sin
\left( \varphi_- (k) - \varphi_- (p) \right) \, . \nonumber \\ 
\label{eq:orth}
\end{eqnarray} 
\item One eigenvalue is negative and the other is positive, $M C_{n,+}
= R_+^{n-2} $ and $M C_{n,-} = -R_-^{n-2} $.  One short distance
eigensolution is a decreasing exponential and the other is
oscillatory, so we have one short distance phase $\varphi $.
In this case for two states $(u_k,w_k) $
and $(u_p,w_p) $ with different energies we get the orthogonality
constraint
\begin{eqnarray} 
0 &=& u_k' u_p - u_p u_k' + w_k' w_p - w_p w_k' \Big|_0 \nonumber \\
 &=& \frac1{R_+} \sin \left( \varphi_+ (k) - \varphi_+ (p) \right)\, .
\end{eqnarray} 
\item Both eigenvalues are positive, $M C_{n,+} = R_+^{n-2} $ and $M
C_{n,-} = R_-^{n-2} $. Then, both short distance eigensolutions are
decreasing exponentials. There are no short distance phases. In this
case the orthogonality relations are automatically satisfied. 
\end{enumerate}  

This simple argument can be easily generalized to any number of
coupled channels. The number of undetermined short distance phases
corresponds to the number of attractive eigenpotentials at short
distances. Orthogonality of the wave functions requires that all these
short distance phases fulfill a generalized condition of the form of
Eq.~(\ref{eq:orth}).

The orthogonality conditions require the determination of the short
distance phases, as we did in Ref.~~\cite{PavonValderrama:2005gu} for
the OPE case. This requires in general an improvement on the short
distance behaviour to high orders. An alternative method is to 
impose the orthogonality constraints either in the single or coupled
channel case by integrating in from infinity for a fixed energy,
either positive or negative, and then impose the condition at a
sufficiently short distance cut-off radius $r=r_c$. In the single
channel case one would get the condition, 
\begin{eqnarray}
\frac{u'_k (r_c)}{u_k(r_c)} = \frac{u'_0 (r_c)}{u_0(r_c)} \, , 
\end{eqnarray} 
if the zero energy state is taken as the reference state. An analogous
relation holds for the coupled channel situation, namely 
\begin{eqnarray}
0 &=& u_k(r_c) u_0 (r_c)' - u_k(r_c) ' u_0 (r_c) \nonumber \\ &+& w_k
(r_c) w_0 (r_c)' - w_k (r_c) ' w_0 (r_c) \, .
\end{eqnarray} 
Obviously, in this procedure cut-off independence must be checked. For
the TPE chiral potentials analyzed in this paper we find that
$r_c=0.1-0.2 {\rm fm} $ proves a sufficiently small value of the short
distance cut-off.

\subsection{Power Counting, counterterms and short distance parameters} 
\label{sec:power_counting} 

As we see, the number of independent parameters is determined from the
potential, although their value can be fixed arbitrarily, by some
renormalization condition like, e.g., fixing scattering lengths to
their physical value. This removes the cut-off in a way that short
distances become less and less important.  Now, if the potential is
regular, i.e., $r^2 |U(r)| \to 0 $, one may {\it choose} between the
regular and irregular solution~\footnote{Both cases comply to the
normalizability condition at the origin, Eq.~(\ref{eq:small_wf})}. In
the first case the scattering length is predicted while in the second
case the scattering length becomes an input of the calculation. In
either case the wave function is still normalizable at the origin.
Singular potentials at the origin, i.e. fullfiling, $r^2 |U(r)| \to
\infty $, do not allow this choice if one insists on normalizability
of the wave function at the origin.  If the potential is repulsive,
the scattering length is fixed while for an attractive potential the
scattering length {\it must} be an input parameter. Furthermore,
orthogonality of different energy solutions requires an energy
independence of the boundary condition, so that in {\it all cases} the
effective range, and higher order threshold parameters cannot be taken
as independent parameters, in addition to the scattering lengths.

This can be translated into the language of counterterms quite
straightforwardly. In momentum space, fixing $\alpha_0$ arbitrarily
corresponds to take a constant $C_0$ cut-off dependent and energy
independent contribution to the potential $V_0(k',k)$ in the
Lippmann-Schwinger equation. Likewise, fixing $r_0$ can be mapped as
adding a term $C_2 ( k^2 + {k'}^2 )$ to the potential. For higher
coupled channel partial waves one fixes the scattering length
$\alpha_{l,l'}$ one has instead terms of the form $ C_{l',l} k'^{l'} k^l
$ in the potential $V_{l',l} (k',k) $. 

The OPE potential in the singlet $^1S_0$ is regular at the origin and
hence one can take $\alpha_0$ as an independent parameter or not (see
Refs.~\cite{PavonValderrama:2003np,PavonValderrama:2004nb}.) Actually,
the smallness of the scattering length for the regular solution,
suggests using the irregular solution. In the Weinberg's counting of
the potential, at NLO one has TPE contributions in the potential. At
short distances they behave as an attractive $1/r^5$ potential (see
Sect.~~\ref{app:b}), and then $\alpha_0$ {\it must} be an
independent parameter.  At NNLO one has, again, a singular attractive
$1/r^6$ potential (see Sect.~\ref{sec:singlet}), and thus an
adjustable scattering length. This looks quite natural because
increasing the order in the potential has a meaning and we can always
compare the effect in the phase shifts of having a higher order
potential with the same scattering length (See
Sect.~~\ref{app:b}). In this construction, if the next term in the
expansion turned out to be more singular and repulsive the scattering
length would be fully predicted from the potential.

The OPE potential in the triplet $^3S_1-^3D_1 $ coupled channel
corresponds to case 2) and hence one has {\it one} free parameter in
addition to the OPE potential parameters. One may choose this
parameter to be the deuteron binding energy (or alternatively the
triplet S-wave scattering length). Any other bound state or scattering
observables are predicted. This case was treated in great detail in
our previous work~\cite{PavonValderrama:2005gu}~\footnote{Relevant
previous work on this channel was also presented in
Ref.~\cite{Beane:2001bc} and \cite{PavonValderrama:2004nb} where the
orthogonality conditions where not considered. See the discussion at
the end of Sect. V in  Ref.~\cite{PavonValderrama:2005gu} }. In the
NLO TPE potential we have case 3) because both eigen potentials
present a repulsive $1/r^5$ singularity (see Sect.~~~\ref{app:b}) and
one would predict all observables from the potential
parameters. Finally, in the NNLO TPE potential we have case 1)
corresponding to an attractive-attractive (see
Sect.~\ref{sec:triplet}) and two additional parameters need to be
specified for a state with a given energy. The orthogonality condition
imposes a relation between two states of different energy, so that for
all energies in the triplet channel one has {\it three} independent
parameters. We will take these three parameters to be the deuteron
binding energy, the asymptotic D/S ratio and the $S-$wave scattering
length. The trend one observes when going from LO to NNLO is quite
natural; as usual in ChPT one has more parameters at any order of the
approximation. The NLO approximation poses, however, a problem since
one seems to have more predictive power than at LO (See
Sect.~~\ref{app:b} for more details on the consequences of using our
renormalization ideas literally for the conventional NLO potential).

\begin{table*}[ttt] 
\caption{\label{tab:table_vdw} Short distance Van der Waals coefficients
for the NNLO chiral potential in singlet $^1S_0 $ and the $^3S_1-^3D_1
$ triplet channels.}
\begin{ruledtabular}
\begin{tabular}{|c|c|c|c|c|c|c|c|c|}
\hline  Set & Source & $c_1 ({\rm GeV}^{-1}) $ & $c_3 ({\rm GeV}^{-1}) $ & $c_4
({\rm GeV}^{-1}) $&  $ M C_{6} ({\rm fm}^4) $ & $ M C_{6,+} ({\rm fm}^4) $ &$ M C_{6,-} ({\rm fm}^4) $
& $\theta $ (degrees) \\ \hline
Set I & (BM) $\pi N$~\cite{Buettiker:1999ap}  & -0.81$\pm$ 0.15  & -4.69$\pm$ 1.34   & 3.40 $\pm$ 0.04 & -8.74 & -16.96 & -5.63  & 140.8 \\ 
Set II & (RTdS) $NN $~\cite{Rentmeester:1999vw} 
& -0.76  & -5.08   & 4.70  & -10.19 & -21.45 & -5.58  & 170.0 \\ 
Set III & (EMa) $NN $~\cite{Entem:2002sf}  
& -0.81  & -3.40    & 3.40   & -6.45 & -14.68 & -3.35  &  140.7 \\ 
Set IV & (EMb) $NN$~\cite{Entem:2003ft} & -0.81  & -3.20   & 5.40  &  -7.28 & -20.18 & -1.86   & 182.6  \\ 
\end{tabular}
\end{ruledtabular}
\end{table*}

For the NNLO TPE triplet $^3S_1-^3D_1$ channel we fix the deuteron
binding energy, or equivalently $\gamma$, and the asymptotic D/S ratio
$\eta$ by their experimental values. This fixes the short distance
phases $\varphi_+ (\gamma) $ and $ \varphi_-( \gamma) $. Next, if we
use an $\alpha $ or $\beta $ (see below for a definition) zero energy
scattering state we have in principle two short distance phases
$(\varphi_{\alpha,+}(0),\varphi_{\alpha,-}(0))$ and
$(\varphi_{\beta,+}(0)$, $\varphi_{\beta,-}(0))$ which can be related
to the $(\alpha_0 , \alpha_{02} ) $ and $(\alpha_{02} , \alpha_{2} ) $
scattering lengths respectively. Using the orthogonality constraints
to the deuteron bound state one can then eliminate $\alpha_{02}$ and
$\alpha_2$ and treat $\alpha_0$ as a free parameter.  Thus, in the
triplet $^3S_1-^3D_1$ channel we can treat $\gamma$, $\eta$ and $
\alpha_0$ as independent parameters. Once these parameters have been
fixed we can actually predict the corresponding phase shifts since any
positive energy state must be orthogonal both to the deuteron bound
state and the zero energy scattering states. This result is a direct
consequence of the singular Van der Waals attractive behavior of the
TPE potential at the origin. It is remarkable that this same set of
independent parameters was also adopted in Ref.~\cite{deSwart:1995ui}
within the realistic potential model treatment.

Conflicts between naive dimensional power counting and renormalization
have been reported recently already at the LO (OPE)
level~\cite{Nogga:2005hy} where it is shown that even the $^3P_0$
partial wave depends strongly on the cut-off in momentum space (a
gaussian regulator is used) if according to the standard Weinberg
counting no counterterm is added. The requirement of renormalizability
makes the promotion of one counter term unavoidable for channels which
present an attractive singularity. This promotion is the minimal
possible one compatible with finiteness, because in a coupled channel
problem one could think in general of three counterterms. From this
viewpoint the choice of just one counterterm in triplet channels is a
bit mysterious. Our discussion in coordinate space agrees with these
authors in the OPE potential, and actually allows to identify {\it a
priori} the necessarily promotable counterterms as non trivial
boundary conditions at the origin for singular attractive
potentials. Moreover, we also see that the promotion of {\it only one
counterterm} in the triplet channels with an attractive-repulsive
singularity invoked in Ref.~\cite{Nogga:2005hy} is also maximal, since
any additional counterterm would also produce divergent results (see
Sect.~\ref{sec:finite}). Thus, we see that although power counting
determines the long distance potential, the short distance singular
character of the potential does not allow to fix the counterterms
arbitrarily.

To conclude this discussion, let us mention that the short distance
phases, whenever they become relevant play the role of some
dimensionless constants which depend exclusively on the form of the
potential, but not on the potential
parameters~\cite{PavonValderrama:2005gu}. For the same reason they can
be taken to be zeroth order in the power counting used to generate the
chiral potential in Eq.~(\ref{eq:pot_chpt}), although they are
subjected in general to higher order corrections. In this sense, the
form of the short distance interaction is dictated by the potential
only, and cannot be considered as independent information (See also
Ref.~\cite{Griesshammer:2005ga} for a similar view on the three body
problem in the absence of long distance potentials).

\section{The Singlet $^1S_0$-Channel}
\label{sec:singlet} 

For the singlet $^1S_0$ channel one has to solve 
\begin{eqnarray}
-u '' (r) + U_{^1S_0} (r) u (r)  &=& k^2 u
 (r) 
\label{eq:sch_singlet} 
\end{eqnarray}
At short distances the NNLO NN chiral potential behaves
as~\cite{Kaiser:1997mw,Friar:1999sj,Rentmeester:1999vw}
\begin{eqnarray} 
U_{^1 S_0} (r) & \to & \frac{3 g^2}{128 f^4 \pi^2 r^6} (-4 + 15 g^2 +
24 \bar c_3 - 8 \bar c_4 )  \nonumber \\  &=& - \frac{R^4}{r^6}
\label{eq:pot_sing_short}
\end{eqnarray} 
which is a Van der Waals type interaction with typical length scale
$R=(-M C_6)^{1/4}$. Here, $\bar c_i = M c_i$. The value of the
coefficient is negative for the four parameter sets of
Table~\ref{tab:table_vdw}, so the solution at short distances is of
oscillatory type, Eq.~(\ref{eq:uA}) with $n=6$, and
\begin{eqnarray}
u (r) &\to & A \left(\frac{r}{R}\right)^{3/2} \sin\left[
- \frac{1}{2} \left(\frac{R}{r}\right)^{2} + \varphi  \right]
\end{eqnarray} 
where there is a undetermined energy independent phase, $\varphi$, and
$A $ is a normalization constant. Note that the corresponding Van der
Waals radius is quite sensitive to the  choice of chiral
parameters.

\subsection{Low energy parameters} 

For the zero energy state we use the asymptotic normalization at large
distances
\begin{eqnarray} 
u_0 (r) &\to & 1- \frac{r}{\alpha_0}  \, . 
\end{eqnarray} 
Then, the effective range is given by 
\begin{eqnarray} 
r_0 &=& 2 \int_0^\infty dr \left[ \left(1-\frac{r}{\alpha_0} \right)^2-
u_0 (r)^2 \right] 
\label{eq:r0_singlet}
\end{eqnarray} 
We can use the superposition principle for boundary conditions 
\begin{eqnarray} 
u_0 (r) &= & u_{0,c} (r) - \frac1{\alpha_0} u_{0,s} (r)  
\end{eqnarray} 
where $u_{0,c} (r) \to 1 $ and $ u_{0,s} (r) \to r $ correspond
to cases where the scattering length is either infinity or zero
respectively. Using this decomposition one gets
\begin{eqnarray} 
r_0  &=&  A + \frac{B}{\alpha_0}+ \frac{C}{\alpha_0^2}  \, ,    
\label{eq:r0_univ} 
\end{eqnarray} 
where 
\begin{eqnarray}
A &=& 2 \int_0^\infty dr ( 1 - u_{0,c}^2 ) \, , \\    
B &=& -4 \int_0^\infty dr ( r - u_{0,c} u_{0,s} ) \, , \\    
C &=& 2 \int_0^\infty dr ( r^2 - u_{0,s}^2 )    \, ,  
\end{eqnarray} 
depend on the potential parameters only. The interesting thing is that
all explicit dependence on the scattering length $\alpha_0$ is
displayed by Eq.~(\ref{eq:r0_univ} ). In a sense this is the
non-perturbative universal form of a low energy theorem, which applies
to {\it any} potential regular or singular at the origin which decays
faster than a certain power of $r$ at large distances (for an
analytical example with the pure Van der Waals potential $U=- R^4 /r^6
$ see Sect.~ \ref{app:a}). Since the potential is known accurately at
long distances we can visualize Eq.~(\ref{eq:r0_univ}) as a long
distance correlation between $r_0$ and $\alpha_0$. Naturally, if there
is scale separation between the different contributions in the
potential, Eq.~(\ref{eq:pot_chpt}), we expect the coefficients A,B and
C to display a converging pattern. This is exactly what happens (see
Eq.~(\ref{eq:r0_todos}) and Eq.~(\ref{eq:r0_NLO}) below) although not
compatible with a naive and perturbative power counting (see
Sect.~\ref{sec:pert}).

In the $^1S_0$ the TPE potential becomes singular and attractive at
short distances. Nevertheless, already at this point one can see the
dramatic difference between attractive and repulsive singular
potentials. In the attractive case the short distance phase allows to
choose the scattering length {\it independently} on the potential,
hence the coefficients $A$,$B$ and $C$ are uncorrelated with
$\alpha_0$. For a singular repulsive potential, however, $\alpha_0$ as
well as $A$,$B$ and $C$ are determined by the potential. If one
assumes $A$,$B$ and $C$ to be independent on $\alpha_0$ in the
repulsive case, this can only be possible due to an admixture of both
the regular and irregular solutions, the latter will dominate at short
distances and the effective range will diverge $r_0 \to -\infty$. This
fact will become relevant in Sect.~\ref{sec:finite}.

Obviously, for the chiral TPE potential, Eq.~(\ref{eq:pot_chpt}), the
coefficients have to be evaluated by numerical means and they are
finite. We expect that these coefficients scale with the relevant
scale of the potential. If long distances dominate $A \sim 1/m $,
$B\sim 1/m^2 $ and $C \sim 1/m^3 $ but then $r_0 \sim 1/m $. On the
contrary if short distances dominate $A \sim R $, $B\sim R^2 $ and $C
\sim R^3 $ and $r_0 \sim R $. The real situation is somewhat in
between, but it is clear that $A$ is far more sensitive to short
distances than $C$.  Actually, for a large scattering length, as it is
the case in the $^1S_0$ channel, the coefficient $A$ dominates. Note
that unlike the standard approaches, where a short distance
contribution to the effective range is allowed (in the form of a
momentum dependent counterterm $C_2 (k^2 + k'^2 ) $ ), we build $r_0$
{\it solely} from the potential and the scattering length
$\alpha_0$. This is a direct consequence of the orthogonality
relations, which preclude energy dependent boundary conditions for the
local and energy independent chiral TPE potential. 

\begin{table}[ttt] 
\caption{\label{tab:table_singlet} Threshold parameters in the singlet
$^1S_0 $ channel for the different sets of parameters $c_1$, $c_3$ and
$c_4$ given in Table~\ref{tab:table_vdw}. We compare our renormalized
results given by the cut-off independent universal formula
(\ref{eq:r0_univ}) for $r_0$ and its extension for $v_2$ to finite
cut-off NN calculations using their scattering length as an input.
The difference is attributed to finite cut-off effects.}
\begin{ruledtabular}
\begin{tabular}{|c|c|c|c|c|c|}
\hline Set & Calculation & $\Lambda=1/r_c$ & $\alpha_0 ({\rm
fm}) $ & $r_0 ({\rm fm}) $ & $v_2 ({\rm fm}^3 ) $\\ \hline
Set I & NNLO \cite{Epelbaum:1999dj} & 0.6-1 GeV & -23.72 & 2.68 & -0.61\\ 
Set I & NNLO \cite{Entem:2001cg} & 0.5 GeV & -23.75  & 2.70 & - \\ 
Set I  & This work & $\infty $& Input  & 2.92 & -0.30 \\ \hline 
Set II & NNLO \cite{Rentmeester:1999vw} & 1/1.4 fm & - & - & - \\
Set II & This work & $\infty $ & Input & 2.97 & -0.23 \\  \hline 
%
Set III & NNLO \cite{Epelbaum:2003xx} & 0.65 GeV & -23.4  & 2.67 & -0.50\\ 
Set III & ${\rm N}^3$LO \cite{Epelbaum:2004fk}  & 0.7 GeV & -23.6  & 2.66  & -0.50 \\ 
Set III & This work & $\infty $& Input  & 2.83 & -0.43 \\  \hline  
Set IV   & ${\rm N}^3$LO  \cite{Entem:2003ft} & 0.5 GeV & -23.73  & 2.73 & -  
\\ 
Set IV & This work & $\infty $& Input   & 2.87 & -0.38   \\ \hline 
Nijm II  & \cite{Stoks:1993tb,Stoks:1994wp}   & - & -23.73   & 2.67 & -0.48   
\\ 
Reid 93  & \cite{Stoks:1993tb,Stoks:1994wp}   & - & -23.74   & 2.75 & -0.49   \\ \hline 
Exp. & -- & --  & -23.74(2) & 2.77(5) & - 
\end{tabular}
\end{ruledtabular}
\end{table}

\begin{figure}[ttt]
\begin{center}
\epsfig{figure=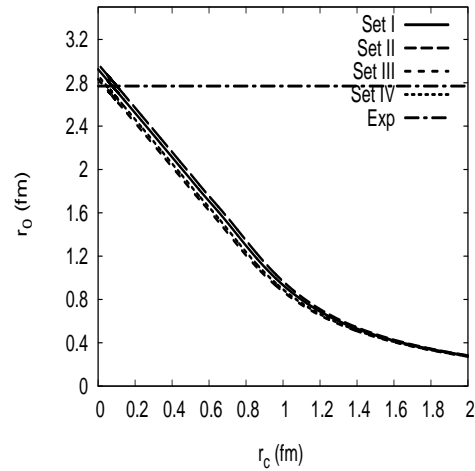,height=6.5cm,width=6.5cm}
\end{center}
\caption{Effective range for the TPE potential in the singlet $^1S_0$
channel when the lower limit of integration in
Eq.~(\ref{eq:r0_singlet}) is taken to be a short distance cut-off and
the experimental value of the scattering length $\alpha_0 = -23.74
{\rm fm}$ is taken. We show Sets I,II,III and IV (see main text).}
\label{fig:r0-1S0_Rc.2fm.eps}
\end{figure}

In Fig.~(\ref{fig:r0-1S0_Rc.2fm.eps}) we show the dependence of the
effective range as a function of the short distance cut-off radius
$r_c$, i.e.  replacing the lower limit of integration in
Eq.~(\ref{eq:r0_singlet}), for values between $2 {\rm }$ and $0.1 {\rm
fm} $ and taking the experimental value of the scattering length
$\alpha_0 = -23.74 {\rm fm}$. As we see, the short distance behaviour
is well under control and nicely convergent towards the experimental
value. This dependence also illustrates that an error estimate based
on varying the cut-off between certain range is only a measure on the
size of finite cut-off effects, rather than a measure on the
error. The linear behaviour observed at small $r_c$ is a consequence
of the dominance of the first term in Eq.~(\ref{eq:r0_singlet}) as
compared to the second term where the wave function contribution
vanishes as $ \sim r_c^4$. Let us remind that in the conventional
treatments a counterterm $C_2$ is added to provide a short distance
contribution to the effective range parameter and the result is fitted
to experiment so that $r_0$ becomes an input of the
calculation. Obviously one expects $C_2$ to depend on the
regularization scale. As shown in Fig.~(\ref{fig:r0-1S0_Rc.2fm.eps})
the size of the counterterm $C_2$ at $r_c \to 0 $ must be numerically
small, since the TPE potential provides the bulk of the contribution.
This agrees with the orthogonality constraint which requires $C_2 \to
0 $ when $r_c \to 0$.

\medskip

Numerically we find the following renormalized relations in the
singlet channel for the OPE and NNLO TPE,
\begin{eqnarray} 
r_0 &=& 1.308062 -
\frac{4.547741}{\alpha_0}+\frac{5.192606}{\alpha_0^2} \, \qquad ({\rm
OPE} ) \, , \nonumber \\ 
r_0 &=& 2.670963 -
\frac{5.755234}{\alpha_0}+\frac{6.031119}{\alpha_0^2} \, \qquad ({\rm
Set \, I} ) \, , \nonumber \\ 
r_0 &=& 2.715075 -
\frac{5.847358}{\alpha_0}+\frac{6.093430}{\alpha_0^2} \, \qquad ({\rm
Set \, II} ) \, , \nonumber \\ 
r_0 &=& 2.586862 -
\frac{5.584383}{\alpha_0}+\frac{5.916900}{\alpha_0^2} \, \qquad ({\rm
Set \, III } ) \, , \nonumber \\ 
r_0 &=& 2.616830 -
\frac{5.640921}{\alpha_0}+\frac{5.952694}{\alpha_0^2} \, \qquad ({\rm
Set \, IV} ) \nonumber \\ 
\label{eq:r0_todos}
\end{eqnarray} 
As we see, the coefficient dependent of $\alpha_0^{-2}$ is not very
sensitive to the choice of the coefficients $c_1$, $c_3$, $c_4$ and
the OPE potential already provides the bulk of the contribution. On
the other hand, the coefficient independent on $\alpha_0$ changes
dramatically when going from OPE to TPE, suggesting that the effect is
clearly non-perturbative. A direct inspection of the integrands for
the A,B and C coefficients shows that $A$ picks its main contribution
from the short distance region around 1 fm, whereas for $B$ and $C$
the most important contribution is located around 3 fm. One expects
that different choices of coefficients $c_3$ and $c_4$ influence
mostly the A coefficient. We confirm this expectation analytically by
only keeping the Van der Waals contribution to the full potential in
Sect.~ \ref{app:a}. We emphasize, again, that $A$, $B$ and $C$ are
intrinsic information of the potential; these values for the effective
range stem solely from the NNLO chiral potential and the scattering
length $\alpha_0$, without any additional short distance
contribution. The closeness of these numbers to the experimental value
suggests that there is perhaps no need to make the boundary condition
energy dependent if the cut-off is indeed removed, and that the
missing $0.1 {\rm fm} $ contribution can be clearly attributed to
N$^3$LO contributions in the potential.

The results are summarized in Table \ref{tab:table_singlet}. For $pn$
we have the experimental values $ \alpha_0 = -23.74(2) $ and $r_0 =
2.77(5) $. The previous formula, Eq.~(\ref{eq:r0_univ}) yields $r_0 =
2.92, 2.97, 2.83, 2.87 $ for Sets I,II, III and IV respectively, which
show a systematic discrepancy with the published values in several
works (see References at the Table \ref{tab:table_singlet}) and also a
systematic trend to discrepancy with respect to the experimental
value.  Our renormalized values are always larger than the finite
cut-off results. This seems natural since finite cut-off corrections
diminish the integration region.  Note also that the size of the
discrepancy is {\it larger} than the experimental uncertainties and
hence is statistically significant.

The value of the effective range was not given in the coordinate space
calculation of Ref.~\cite{Rentmeester:1999vw} but the quality of the
fit suggests that they get a value very close to the experimental one,
$r_0 = 2.75$. The contribution to the effective range from the origin
to $0.1 {\rm fm}$ is about ${\rm 0.2}$. In
Ref.~\cite{Rentmeester:1999vw} the cut-off is in coordinate space and 
an energy dependent boundary condition is considered. This means in
practice cutting-off the lower integration in
Eq.~(\ref{eq:r0_singlet}) at $a=1.4 {\rm fm} $ and adding a short
distance contribution $r_S $ as to reproduce the experimental
value. This introduces a new potential independent parameter.  As we
have argued, in the limit $a\to 0$, the short distance contribution of
the effective range should go to zero, as implied by the orthogonality
constraints. For finite $a$, the orthogonality constraint does not
imply a vanishing short distance contribution to the effective range.

For Set IV one could reach the upper experimental value by flipping
the sign of $c_1$ and keeping $c_3$ and $c_4$ unchanged. For $ c_1 =
2.43{\rm GeV}^{-1}$ one gets $r_0=2.78 {\rm fm} $. The full
experimental range would be covered by letting $0.81 {\rm GeV}^{-1} <
c_1 < 4.90{\rm GeV}^{-1}$. This is in total contradiction to the
expectations of $\pi N $ scattering
studies~\cite{Buettiker:1999ap}. The insensitivity of our results with
respect to the $c_1$ coefficient has to do with the fact that $c_1$
only enters in the potential at short distances at order $ 1/r^4$
which is sub-leading as compared to the leading Van der Waals
singularity. This is another confirmation on the short distance
dominance in the effective range parameter $r_0$.

Thus, according to our analysis, for the accepted values of chiral
constants of Sets I,II, III and IV used in previous works, the
difference in the value of the $^1S_0 $ effective range could only be
attributed to three pion exchange, relativistic effects and
electromagnetic corrections. Another possibility, of course, is to
refit the chiral constants to our renormalized, cut-off free
results. This will be discussed in Sect.~\ref{sec:error}.



\medskip
\begin{figure}[]
\begin{center}
\epsfig{figure=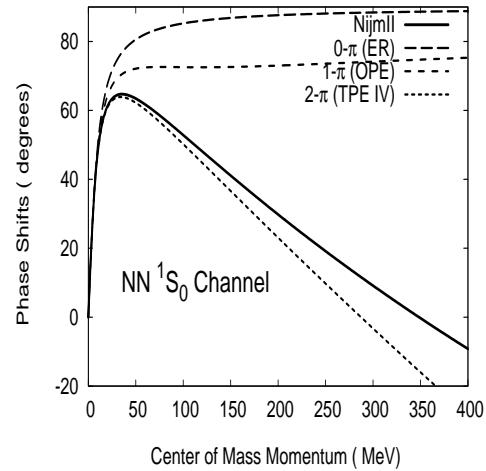,height=6.5cm,width=6.5cm}
\end{center}
\caption{Renormalized phase shifts for the OPE and TPE potentials as a
function of the CM np momentum $k$ in the singlet $^1S_0$ channel
compared to the Nijmegen results~\cite{Stoks:1993tb} for different
parameter sets. The regular scattering wave functions for finite $k$
are orthogonal to the zero energy  wave functions. For the TPE
potential we have taken the chiral couplings of set IV.}
\label{fig:phase-shift_1S0}
\end{figure}

\subsection{Phase Shift}

For a finite energy scattering state we solve for the chiral TPE
potential with the normalization
\begin{eqnarray}
u_k (r) \to \frac{\sin(kr + \delta_0)}{\sin \delta_0}   
\label{eq:norm}
\end{eqnarray} 
Again, if we use the superposition principle 
\begin{eqnarray}
u_k (r) = u_{k,c} (r)  +  k \cot \delta_0 \,  u_{k,s} (r)    
\end{eqnarray} 
with $ u_{k,c} \to \cos (k r) $ and $ u_{k,s} \to \sin (k r) /k $ and
impose the orthogonality constraint with the zero energy state to get
\begin{eqnarray}
\frac{u'_{k,c} (a) + k \cot \delta_0 u_{k,s}' (a) }
{u_{k,c} (a) +  k \cot \delta_0 u_{k,s} (a) } = 
\frac{- \alpha_0 u'_{0,c} (a) + u_{0,s}' (a) }
{- \alpha_0 u_{0,c} (a) + u_{0,s} (a) }
\end{eqnarray} 
Note that the dependence of the phase-shift on the scattering length
is {\it explicit}; $\cot \delta_0 $ is a bilinear rational mapping of
$\alpha_0$. Taking the limit $a \to 0 $ we get
\begin{eqnarray}
k \cot \delta_0 = \frac{ \alpha_0 {\cal A} ( k) - {\cal B} (k)}{ \alpha_0 {\cal C} ( k) - {\cal D} (k)}
\label{eq:phase_singlet}
\end{eqnarray} 
whereas the functions ${\cal A}$, ${\cal B}$, ${\cal C}$ and ${\cal
D}$ are even functions of $k$ which depend only  on the potential and
are given by
\begin{eqnarray}
{\cal A}(k) &=& \lim_{a \to 0} \left( u_{0,c} (a)  u'_{k,c} (a) - 
u_{0,c}' (a)  u_{k,c} (a) \right)  \nonumber \\  
{\cal B}(k) &=& \lim_{a \to 0} \left( u_{k,c} (a)  u'_{0,s} (a) - 
u_{0,s} (a)  u_{k,c}' (a) \right)  \nonumber \\  
{\cal C}(k) &=& \lim_{a \to 0} \left( u_{0,c}' (a)  u_{k,s} (a) - 
u_{0,c} (a)  u'_{k,s} (a) \right)  \nonumber \\  
{\cal D}(k) &=& \lim_{a \to 0} \left( u_{0,s} (a)  u'_{k,s} (a) - 
u_{0,s}' (a)  u_{k,s} (a) \right)  \nonumber \\ 
\end{eqnarray}  
The obvious conditions ${\cal A}(0)={\cal D}(0)=0$  and  
${\cal B}(0)={\cal C}(0)=1$  are satisfied. Expanding the expression 
for small $k$ one gets the well known effective range expansion 
\begin{eqnarray}
k \cot \delta = - \frac1{\alpha_0} + \frac12 r_0 k^2 + v_2 k^2 + \dots  
\end{eqnarray} 
where $v_k$ is a polynomial in $1/\alpha_0 $ of degree $k+1$.

The renormalized phase shift is presented in
Fig.~\ref{fig:phase-shift_1S0} for Set IV. As we see the trend in the
effective range $r_0$ and the $v_2$ parameter is reflected in the
behavior of the phase shift~\footnote{Let us mention that the momentum
space calculation of Ref.~\cite{Nogga:2005hy} {\it does not} reproduce
the physical and well measured scattering length. We have checked by
an explicit solution of the Lipmann-Schwinger equation that the
problem is more related to an insufficient number of Gauss points
rather than to the value of the cut-off.}

\section{The Triplet $^3S_1-^3D_1$ Channel} 
\label{sec:triplet}

The coupled channel $^3S_1 - ^3D_1 $ set of equations read
\begin{eqnarray}
-u '' (r) + U_{^3S_1} (r) u (r) + U_{E_1} (r) w (r) &=& k^2 u
 (r) \, ,\nonumber  \\ -w '' (r) + U_{E_1} (r) u (r) + \left[U_{^3D_1} (r) +
 \frac{6}{r^2} \right] w (r) &=& k^2 w (r) \, , \nonumber \\
\label{eq:sch_coupled} 
\end{eqnarray}
At short distances the NN chiral NNLO potential behaves as~\cite{Kaiser:1997mw,Friar:1999sj,Rentmeester:1999vw} 
\begin{eqnarray} 
U_{^3S_1}(r) &\to& \frac{M C_{6,^3S_1}}{ r^6 } \nonumber \\ 
U_{E_1}(r) &\to& \frac{M C_{6,E_1}}{ r^6 } \nonumber \\ 
U_{^3D_1}(r) &\to& \frac{M C_{6,^3D_1}}{ r^6 } \nonumber \\ 
\end{eqnarray} 
which is a coupled channels Van der Waals type interaction where the coefficients are given by 
\begin{eqnarray} 
M C_{^3S_1} &=& \frac{3 g^2}{128 f^4 \pi^2 } ( 4 - 3 g^2 + 24
\bar c_3 - 8 \bar c_4 ) \nonumber \\ M C_{E_1} &=& -
\frac{3 \sqrt{2} g^2}{128 f^4 \pi^2 } (-4 + 3 g^2 - 16 \bar c_4 ) \nonumber \\ 
M C_{^3D_1}&=&   \frac{9 g^2}{32 f^4 \pi^2  } (-1+2 g^2 + 2 \bar
c_3 - 2 \bar c_4 ) \nonumber \\ 
\label{eq:vdw_triplet}
\end{eqnarray} 
If we diagonalize the corresponding matrix we get 
\begin{eqnarray} 
\begin{pmatrix}
C_{6,^3S_1} & C_{6,E_1} \\ C_{6,E_1}  & C_{6,^3D_1}  
\end{pmatrix}  
&=&  
\begin{pmatrix}
\cos\theta & \sin\theta  \\ -\sin\theta &  \cos \theta   
\end{pmatrix}  
\begin{pmatrix}
C_{6,+}  & 0  \\ 0 &  C_{6,-}     
\end{pmatrix} \nonumber \\ &\times&   
\begin{pmatrix}
\cos\theta & -\sin\theta  \\ \sin\theta &  \cos \theta   
\end{pmatrix} \nonumber \\ 
\end{eqnarray}  
where $C_{6,\pm} $ are the corresponding eigenvalues and $\theta$ the
mixing angle. They are listed in Table~\ref{tab:table_vdw} for
different parameters choices of the chiral couplings $c_1$, $c_3$ and
$c_4$. We see that in all cases both eigenpotentials are attractive at
short distances and hence the short distance behavior of the wave
functions is of oscillatory type with $n=6$. Defining the Van der
Waals scales 
\begin{eqnarray} 
R_{\pm} = ( -MC_{6, \pm})^{1/4}   
\end{eqnarray} 
the short distance solutions read 
\begin{eqnarray} 
\begin{pmatrix}
u \\ w 
\end{pmatrix}  
&      \to & 
\begin{pmatrix}
\cos\theta & \sin\theta  \\ -\sin\theta &  \cos \theta   
\end{pmatrix}  
\begin{pmatrix}
\left(\frac{r}{R_+}\right)^{\frac32} \sin\Big[
\frac{1}{2} \left(\frac{R_+}{r}\right)^{2} + \varphi_{+}  \Big]  \\ 
\left(\frac{r}{R_-}\right)^{\frac32} \sin\Big[
\frac{1}{2} \left(\frac{R_-}{r}\right)^{2} + \varphi_{-}  \Big] \nonumber  \\   \end{pmatrix}  
\end{eqnarray}  
Thus, we have two arbitrary short distance phases $\varphi_\pm $ for a
given fixed energy which cannot be deduced from the potential and
hence have to be treated as independent parameters. We will fix them
to some physical observables by integrating
Eqs.~(\ref{eq:sch_coupled}) from infinity down to the origin.

\subsection{The deuteron} 

In the deuteron $k^2 = - \gamma^2 $ and we solve
Eq.~(\ref{eq:sch_coupled}) together with the asymptotic condition at
infinity
\begin{eqnarray}
u (r) &\to & A_S e^{-\gamma r} \, , \nonumber \\ w (r) & \to & A_D
e^{-\gamma r} \left( 1 + \frac{3}{\gamma r} + \frac{3}{(\gamma r)^2}
\right) \, ,
\label{eq:bcinfty_coupled} 
\end{eqnarray}
where $ \gamma = \sqrt{M B} $ is the deuteron wave number, $A_S$ is
the normalization factor and the asymptotic D/S ratio parameter is
defined by $\eta=A_D/A_S$. In what follows we use $\gamma$ and $\eta$
as {\it input} parameters thus fixing the short distance phases
$\varphi_\pm$ automatically. 

\begin{figure}[]
\begin{center}
\epsfig{figure=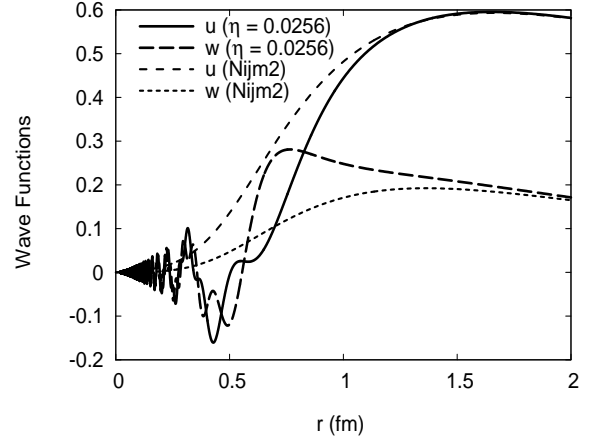,height=6cm,width=8cm}
\end{center}
\caption{The TPE deuteron wave functions, u and w, as a function of
the distance (in {\rm fm} ) compared to the Nijmegen II wave
functions~\cite{Stoks:1993tb,Stoks:1994wp}. The asymptotic normalization 
$u \to e^{-\gamma r}$ has been adopted and the asymptotic D/S ratio is
taken $\eta = 0.0256 (4)$. We use the set IV of chiral couplings.}
\label{fig:u+w_TPE}
\end{figure}

\begin{figure*}[]
\begin{center}
\epsfig{figure=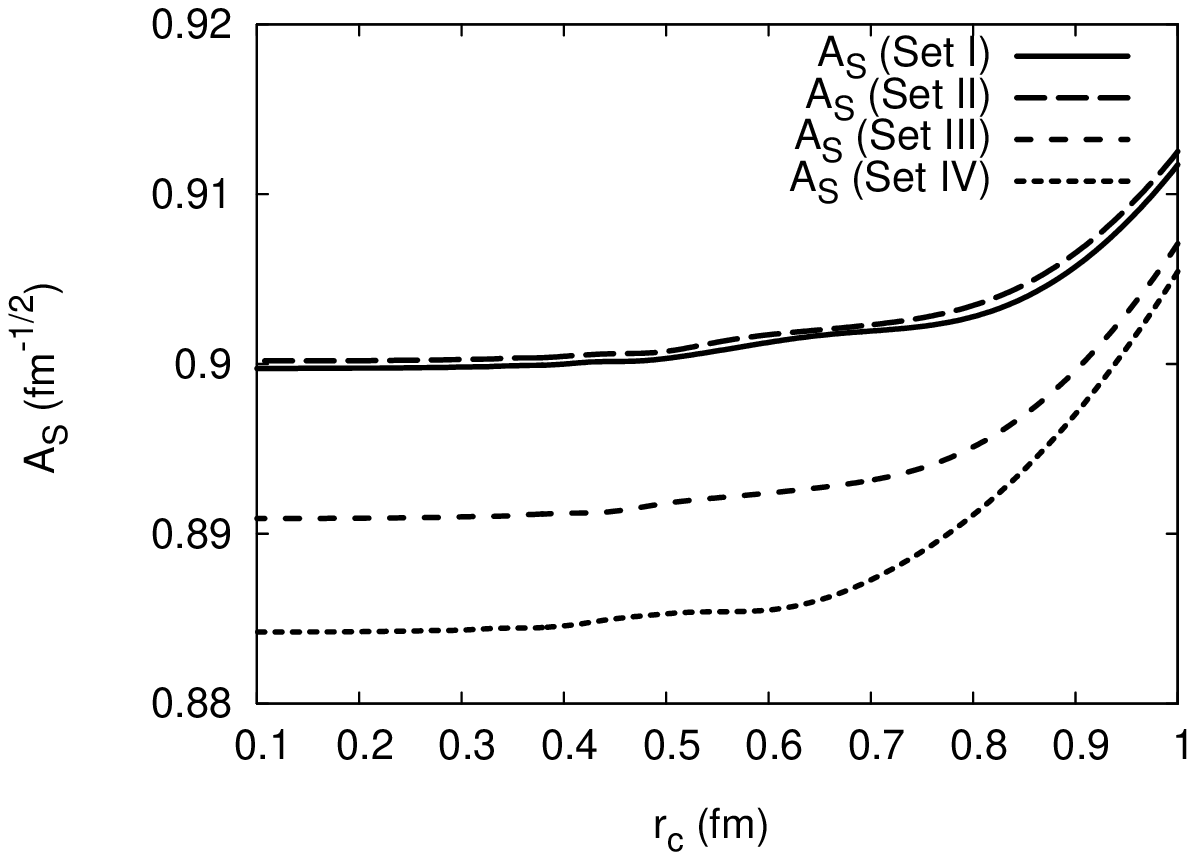,height=5.5cm,width=5.5cm}
\epsfig{figure=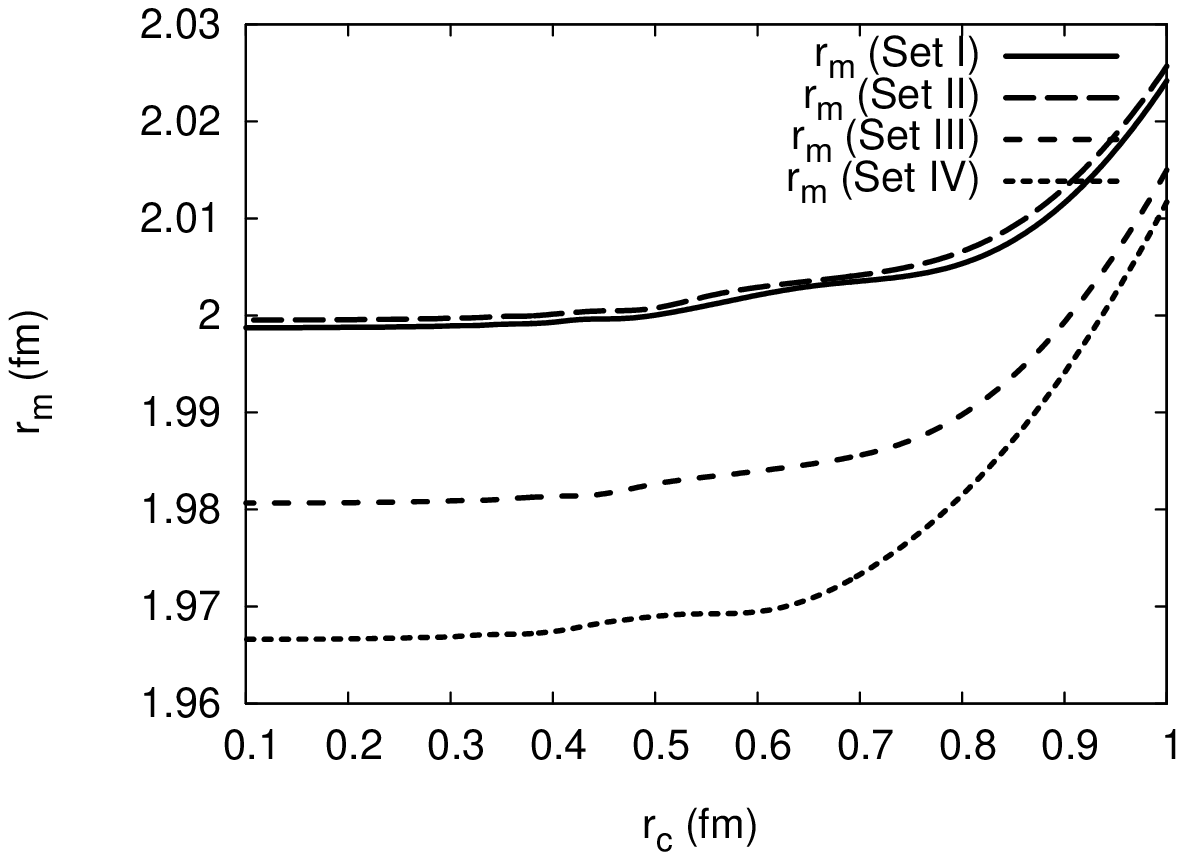,height=5.5cm,width=5.5cm}\\  
\epsfig{figure=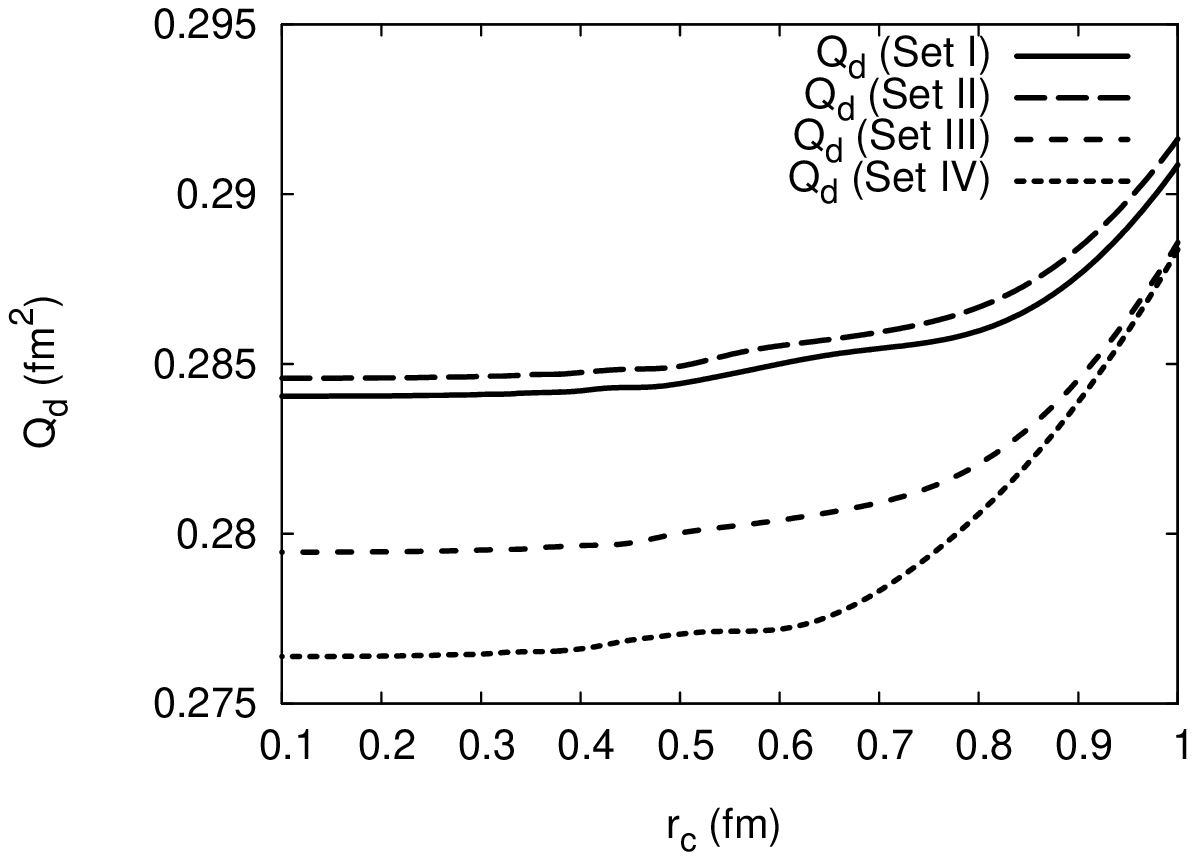,height=5.5cm,width=5.5cm}
\epsfig{figure=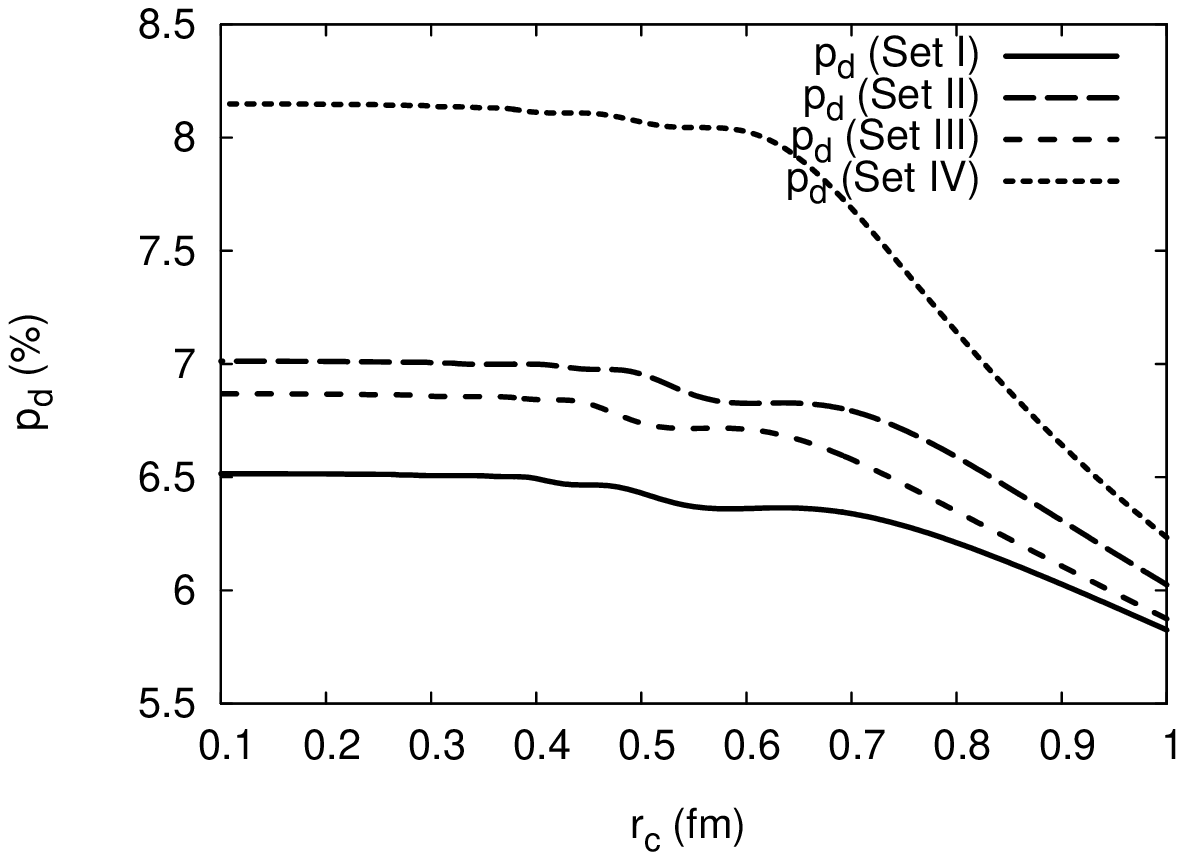,height=5.5cm,width=5.5cm}
\epsfig{figure=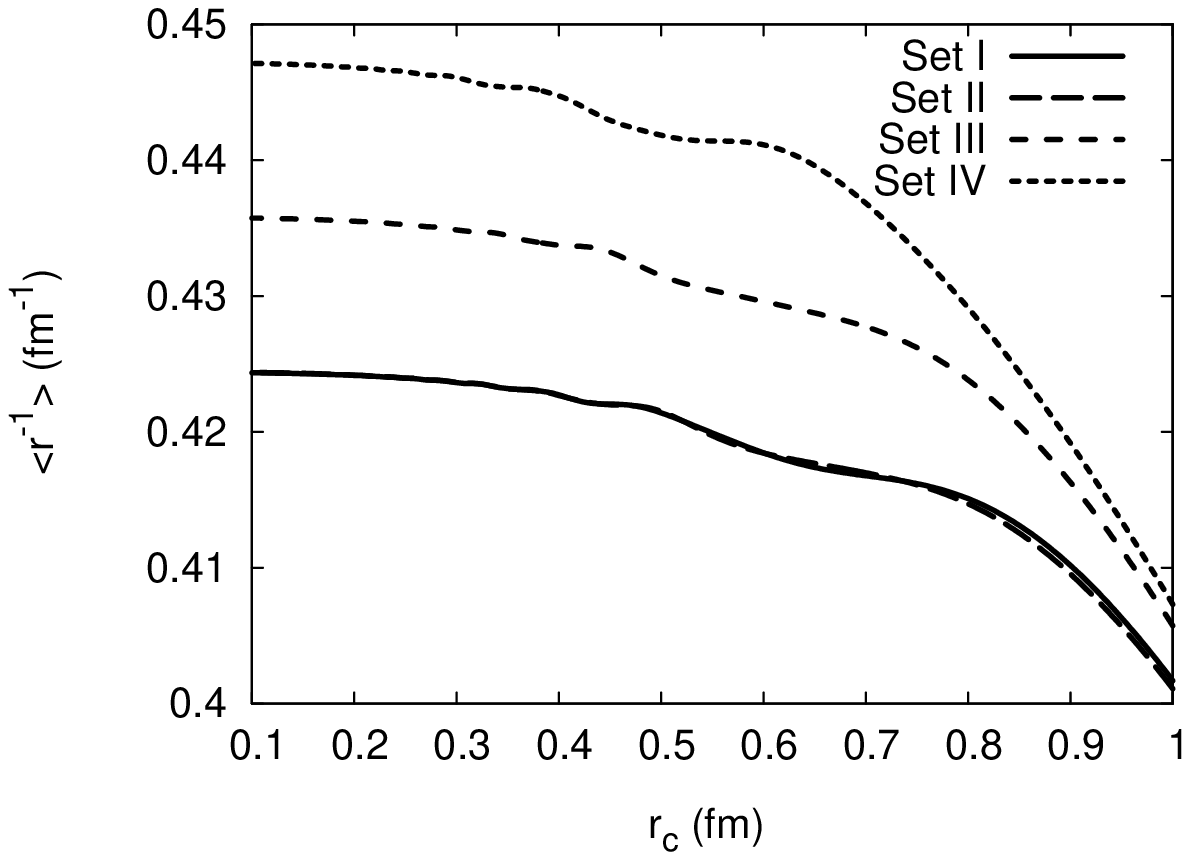,height=5.5cm,width=5.5cm}
\end{center}
\caption{The dependence of the S-wave normalization $A_S$ (in ${\rm
fm}^{-1/2} $, upper left panel) , the matter radius $r_m$ (in ${\rm
fm} $,upper right panel), the quadrupole moment $ Q_d $ (in ${\rm
fm}^2 $, lower left panel), the $D$-state probability (lower mid
panel) and the inverse radius $\langle r^{-1} \rangle $ (in ${\rm
fm}^{-1}$ lower right panel) for the TPE potential on the short
distance cut-off, $r_c$ (in ${\rm fm} $) for Sets I,II,III and IV of
low energy chiral constants.}
\label{fig:finite_cutoff}
\end{figure*}

In this paper we compute the matter radius, which reads,
\begin{eqnarray}
r_m^2 = \frac{\langle r^2 \rangle}{4} = \frac14 \int_0^\infty r^2 ( u(r)^2 +
w(r)^2 ) dr \, , 
\end{eqnarray} 
the quadrupole moment (without meson
exchange currents) 
\begin{eqnarray}
Q_d  = \frac1{20} \int_0^\infty r^2 w(r) ( 2\sqrt{2} u(r)-w(r) ) dr  \, , 
\end{eqnarray} 
the deuteron inverse radius
\begin{eqnarray}
\langle r^{-1} \rangle = \int_0^\infty dr \frac{u(r)^2 + w(r)^2}{r} \, ,  
\end{eqnarray} 
which appears in low energy pion-deuteron scattering, 
and the $D$-state probability   
\begin{eqnarray}
P_D = \int_0^\infty w(r)^2  dr \, . 
\end{eqnarray} 
Following Ref.~\cite{PavonValderrama:2005gu} we use the superposition
principle of boundary conditions and write
\begin{eqnarray}
u (r) &=& u_S (r) + \eta \, u_D (r) \, , \nonumber \\ w (r) &=& w_S
(r) + \eta \, w_D (r) \, ,
\label{eq:sup_bound} 
\end{eqnarray}
where $(u_S,w_S)$ and $(u_D,w_D)$ correspond to the boundary
conditions at infinity, Eq.~(\ref{eq:bcinfty_coupled}), with $A_S=1$
and $A_D=0$ and with $A_S=0$ and $A_D=1$ respectively. Obviously,
$u_S,u_D,w_S $ and $w_D $ depend on the potential and the deuteron
binding energy only, so that the dependence on the asymptotic D/S
ratio $\eta $ can de determined analytically. The value is taken as a
free parameter. The resulting deuteron wave functions for Set IV are
displayed in Fig.~\ref{fig:u+w_TPE} and compared to the Nijmegen II 
results~\cite{Stoks:1993tb,Stoks:1994wp}. One clearly sees the
incommensurable ever increasing oscillations already below $r=0.6 {\rm fm}.$  

The short distance cut-off dependence of these deuteron properties
using the experimental values for the deuteron binding energies and
the asymptotic D/S ratio, $\eta=0.0256$, can be looked up in
Fig.~\ref{fig:finite_cutoff}. As one sees the cut-off dependence is
well under control, so the infinite cut-off limit can be extracted
without difficulty.

Using the superposition principle of boundary conditions,
Eq.~(\ref{eq:sup_bound} ) the asymptotic S-wave ratio depends 
quadratically on $\eta$ as follows
\begin{eqnarray}
\frac{1}{A_S^2} &=& \int_0^\infty dr ( u_S^2 + w_S^2 ) + 2 \eta 
\int_0^\infty dr ( u_S u_D  + w_S w_D  ) \nonumber \\ &+& \eta^2 
\int_0^\infty dr ( u_D^2 + w_D^2 )  \, .
\end{eqnarray}
The coefficients of this second order polynomial depends on the
potential and the deuteron binding energy. Similar relations hold for
other observables. Evaluating the integrals numerically we get the
following analytic correlations

\underline{Set I}
\begin{eqnarray} 
1 / A_S^2 &=& 3.78888 - 214.675\,\eta + 4489.43\,{\eta}^2 \nonumber \\
r_m^2 / A_S^2 & =& 5.47297 - 54.1956\,\eta + 1295.89\,{\eta}^2
\nonumber\\ Q_d / A_S^2 &=&-0.342883 + 36.6449\,\eta -
372.841\,{\eta}^2 \nonumber \\ P_D / A_S^2 &=& 2.10904 - 184.824\,\eta
+ 4124.37\,{\eta}^2 \nonumber \\ \langle r^{-1} \rangle / A_S^2 &=&
3.58173 - 252.20 \,\eta + 5186.53\,{\eta}^2
\end{eqnarray} 

\underline{Set II}
\begin{eqnarray} 
1 / A_S^2 &=& 3.01271 - 155.591\,\eta + 3363.94\,{\eta}^2 \nonumber \\
r_m^2 / A_S^2 & =& 5.34737 - 44.8896\,\eta + 1122.59\,{\eta}^2 \nonumber\\ 
Q_d / A_S^2 &=& -0.296852 + 33.2406\,\eta - 309.624\,{\eta}^2 \nonumber \\ 
P_D / A_S^2 &=& 1.44293 - 132.314\,\eta + 3098.89\,{\eta}^2
\nonumber \\ \langle r^{-1} \rangle / A_S^2 &=&
2.52815 - 171.36 \,\eta + 3635.18\,{\eta}^2
\end{eqnarray} 

\underline{Set III}
\begin{eqnarray} 
1 / A_S^2 &=& 4.65049 - 283.545\,\eta + 5902.53\,{\eta}^2 \nonumber \\
r_m^2 / A_S^2 & =& 5.58929 - 63.1854\,\eta + 1481.50\,{\eta}^2 \nonumber \\ 
Q_d / A_S^2 &=& -0.377779 + 39.2691\,\eta - 420.250\,{\eta}^2 \nonumber \\ 
P_D / A_S^2 &=& 2.77521 - 241.491\,\eta + 5330.67\,{\eta}^2
\nonumber \\ \langle r^{-1} \rangle / A_S^2 &=&
4.87639 - 356.21 \,\eta + 7311.77 \,{\eta}^2
\end{eqnarray} 

\begin{figure*}[]
\begin{center}
\epsfig{figure=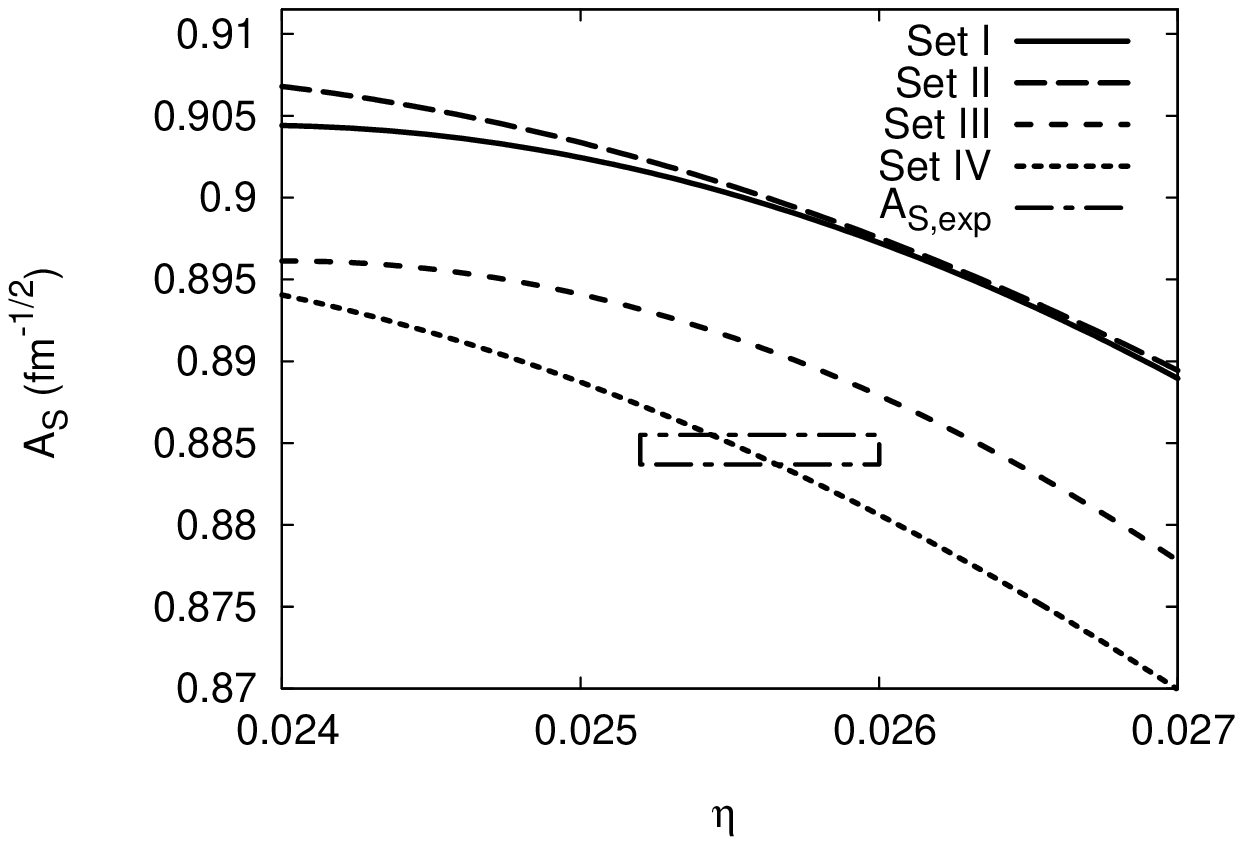,height=5.5cm,width=5.5cm}
\epsfig{figure=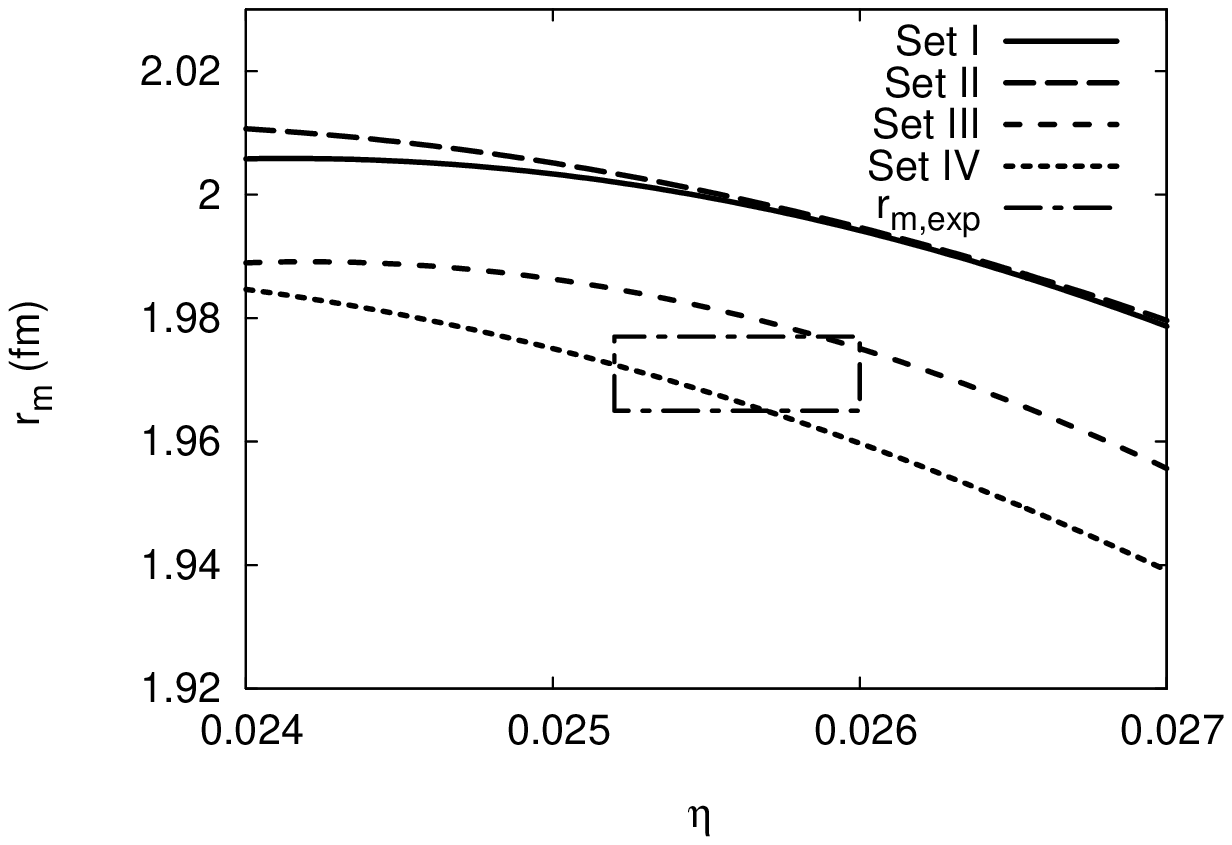,height=5.5cm,width=5.5cm}\\ 
\epsfig{figure=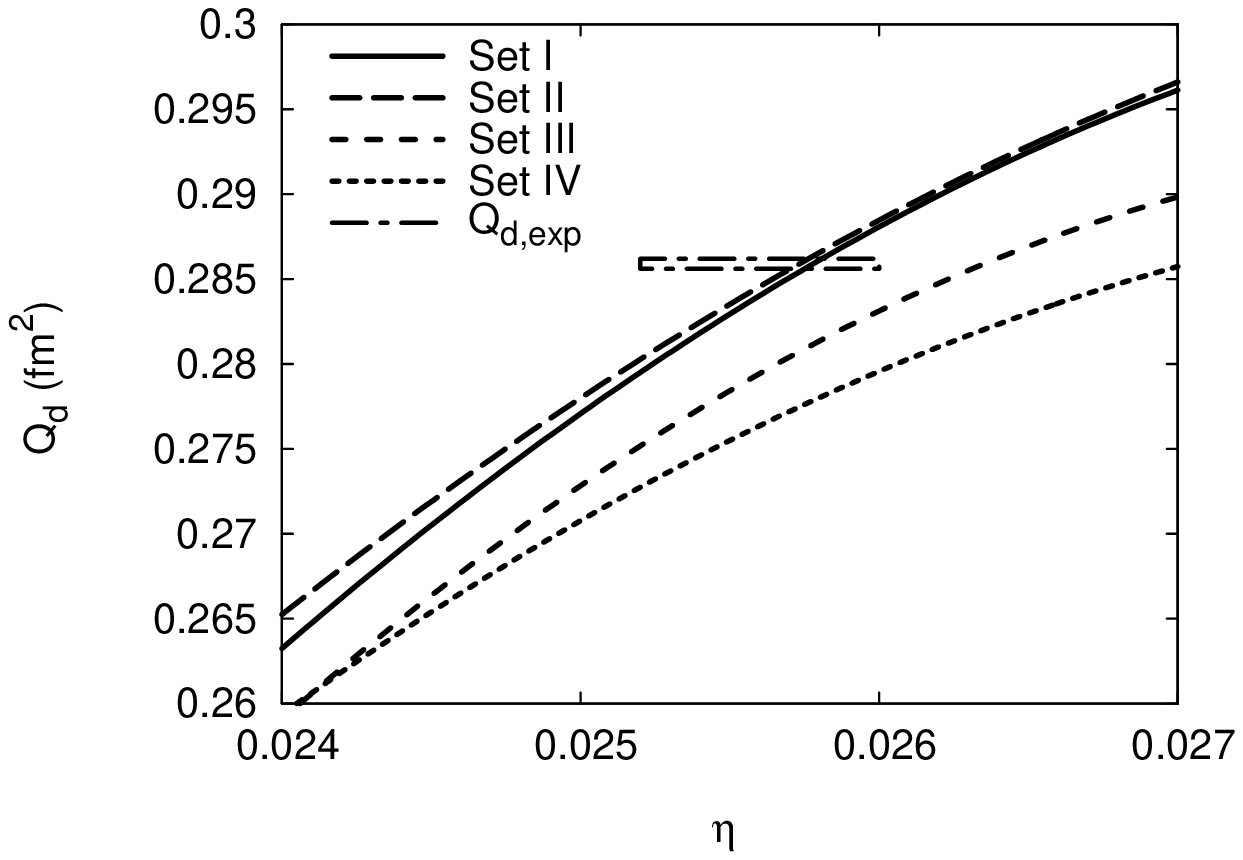,height=5.5cm,width=5.5cm}
\epsfig{figure=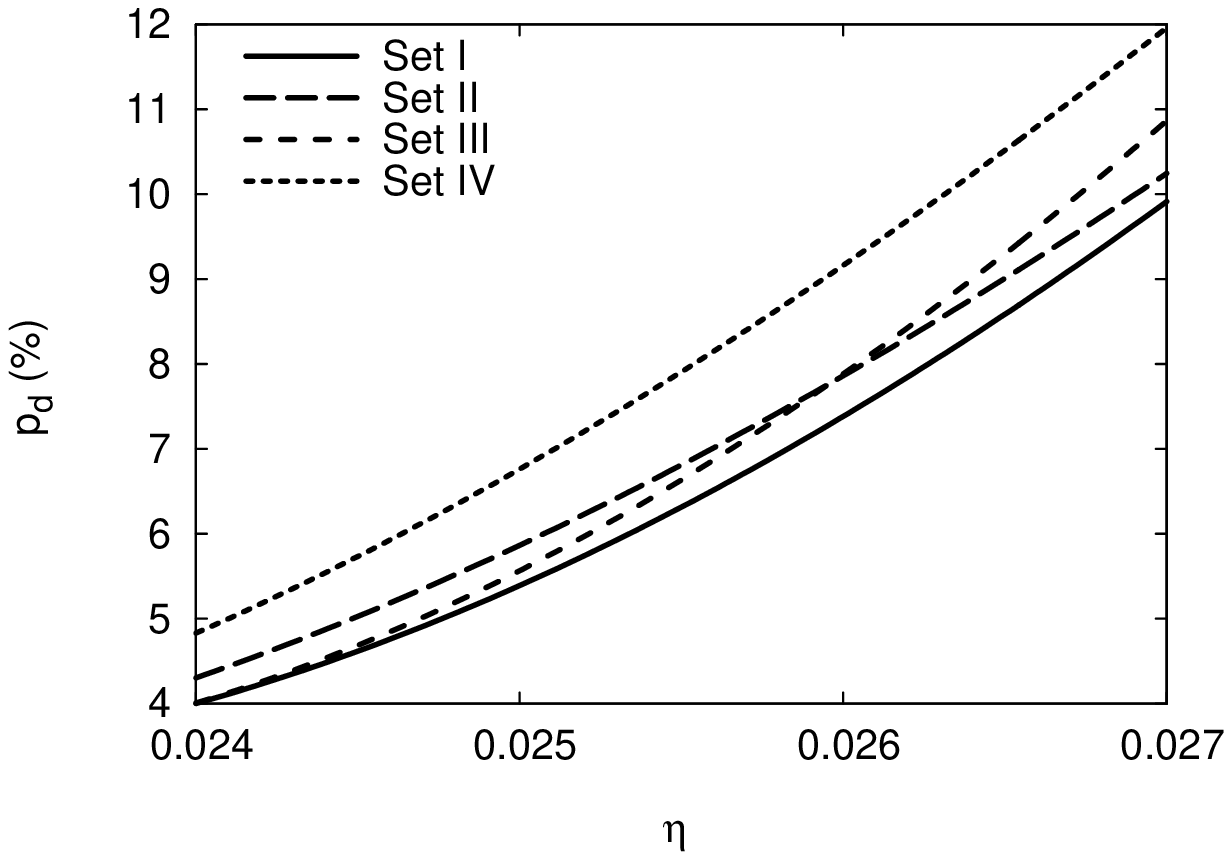,height=5.5cm,width=5.5cm}
\epsfig{figure=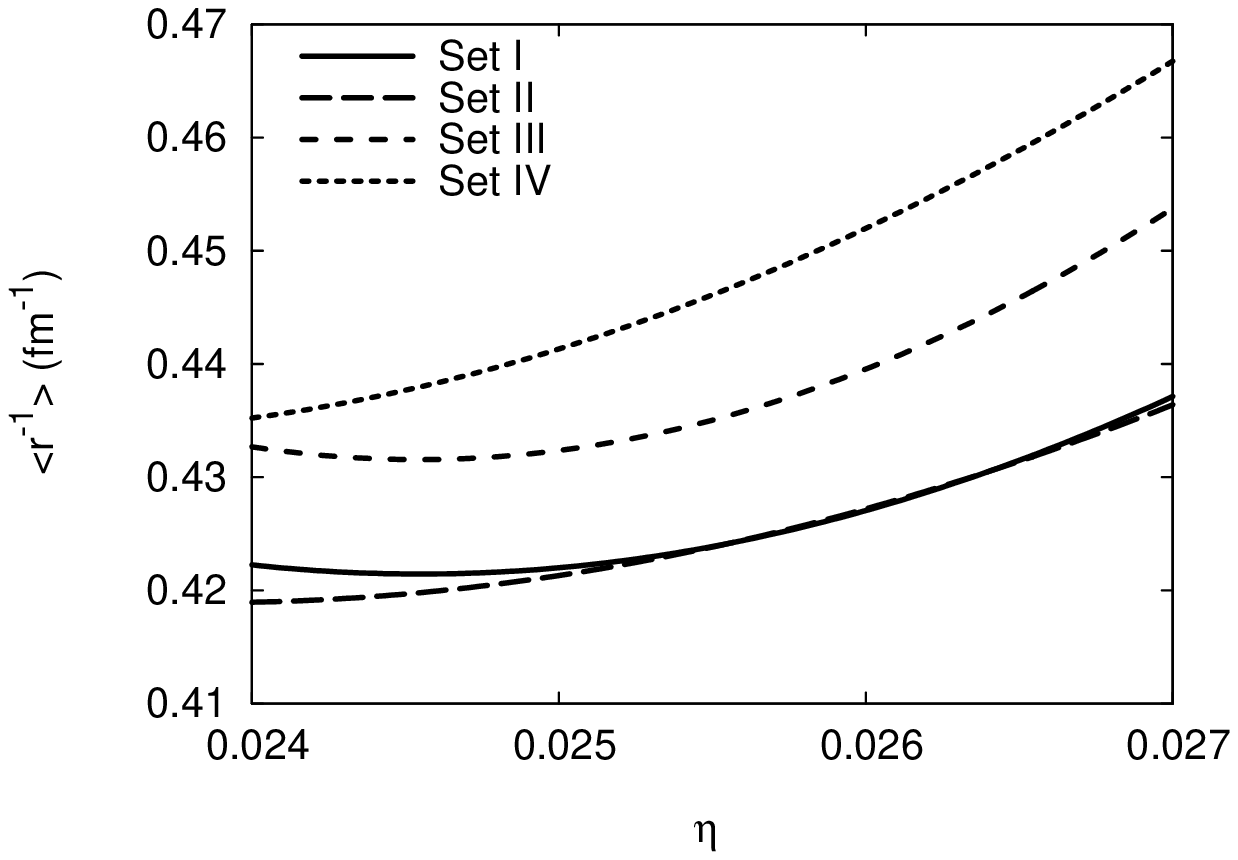,height=5.5cm,width=5.5cm}
\end{center}
\caption{The dependence of the S-wave normalization $A_S$ (in ${\rm
fm}^{-1/2} $, upper left panel) , the matter radius $r_m$ (in ${\rm
fm} $,upper right panel), the quadrupole moment $ Q_d $ (in ${\rm
fm}^2 $, lower left panel) (lower mid
panel) and the inverse radius $\langle r^{-1} \rangle $ (in ${\rm
fm}^{-1}$ lower right panel) for the TPE
potential as a function of the asymptotic $D/S$ ratio $\eta $. The
boxes represent the experimental values. The predicted Quadrupole
moments as well as the matter radius should be corrected for MEC's
(accounting for adding $0.01 {\rm fm}^2 $ and $0.003 {\rm fm}$
respectively) on top of the potential result.  We display the four
sets of chiral coupling constants.}
\label{fig:[eta]}
\end{figure*}

\underline{Set IV }
\begin{eqnarray} 
1 / A_S^2 &=& 3.40962 - 190.713\,\eta + 4198.86\,{\eta}^2 \nonumber \\
r_m^2 / A_S^2 & =& 5.40066 - 49.2912\,\eta + 1232.81\,{\eta}^2 \nonumber \\ 
Q_d / A_S^2 &=& -0.306469 + 33.9354\,\eta - 318.598\,{\eta}^2 \nonumber \\ 
P_D / A_S^2 &=& 1.66525 - 155.233\,\eta + 3681.89\,{\eta}^2
\nonumber \\ \langle r^{-1} \rangle / A_S^2 &=&
3.14658 - 226.20 \,\eta + 4907.35 \,{\eta}^2
\end{eqnarray} 
The numerical coefficients in these expressions depend on the deuteron
binding energy and the TPE potential parameters, $g$, $m$, $f$, $c_1$,
$c_3$ and $c_4$. The results for the deuteron properties are given in
Table~\ref{tab:table3}. The uncertainties are due to changing the
input $\gamma$ and $\eta$ within their experimental uncertainties. We
have checked that the short distance cut-offs $a \sim 0.1-0.2 {\rm fm}
$ generates much smaller uncertainties. The explicit dependence on
$\eta$ is displayed in Fig.~\ref{fig:[eta]}. Again, we find a
discrepancy in the case of Set I with the values quoted in the finite
cut-off calculation. Remarkably, our renormalized results in
coordinate space agree most with the momentum space calculation of
Ref.~\cite{Entem:2001cg} corresponding to Set IV. It is noticeable
that this can be done without explicit knowledge of the counterterms
used in that work in momentum space.  This is precisely one of the
points of renormalization; results can be reproduced by just providing
physical input data, and no particular reference to the method of
solution. Let us remind that the $c_1$, $c_3$ and $c_4$ were fixed
from the perturbative study of NN peripheral waves where the cut-off
sensitivity is rather small.  Nevertheless, some significant
discrepancies do also occur.

For the parameter Set IV~\cite{Entem:2003ft} obtained by a N$^3$LO fit
to NN scattering data, our NNLO calculation almost reproduces exactly
the numbers provided in that work. Furthermore, they turn out to be
compatible with the experimental numbers at the $1\sigma$ level within
the uncertainty induced by the asymptotic $D/S$ ratio~\footnote{One
may object that one should not use N$^3$LO parameters to do a NNLO
calculation, since they are obtained by fitting the same
database. However, if there are finite cut-off effects the situation is
not as clear. Finite cut-off effects are minimized in a N$^3$LO
calculation as shown in Ref.~\cite{Epelbaum:2004fk} where the induced
uncertainties are drastically reduced when going from NNLO to
N$^3$LO. Note that in our calculation there are no sizeable cut-off
induced uncertainties already at NNLO.}

One immediate lesson we learn from inspection of
Table~\ref{tab:table3} is that, regardless of the parameter set, only
the experimental uncertainty in the asymptotic D/S ratio for the
deuteron generates theoretical uncertainties about an order of
magnitude larger then the experimental ones. On top of this, one has
also to take into account other uncertainties, such as the one in
$g_{\pi NN}$ and, of course, those induced by $c_1$, $c_3$ and $c_4$,
which generally will generate larger uncertainties if all these
parameters are regarded as independent (see Sect.~\ref{sec:error}
below). In addition, there are systematic errors related to the
accuracy of the expansion in the potential,
Eq.~(\ref{eq:pot_chpt}). In common with non-perturbative finite
cut-off
calculations~\cite{Rentmeester:1999vw,Epelbaum:1999dj,Epelbaum:2003gr,Epelbaum:2003xx,Entem:2003ft,Rentmeester:2003mf}
they are difficult to estimate {\it a priori} given the
non-perturbative nature of our calculation, but are bound to increase
the error (see, however, our discussion in Sect.~\ref{sec:pert} below
on non-integer power counting).  Given the insensitivity of our
results with the short distance cut-off, the procedure used in
Ref.~\cite{Epelbaum:2003gr,Epelbaum:2003xx} of varying the cut-off
becomes unsuitable in our case.

For the deuteron channel one may conclude that the predictive power of
the chiral expansion has reached a limit at NNLO. So, at present, we
do not expect to make theoretical predictions in the deuteron to be
more accurate than experiment. The inclusion of N$^3$LO and higher
orders may provide better central values but is unlikely to improve
the situation regarding error estimates since new unknown coefficients
in the potential appear and the induced uncertainties will generally
increase.

On the other hand, the slope for $A_S$ and $r_m$, Fig.~\ref{fig:[eta]},
suggests that it would be better to take the asymptotic S-wave
normalization or the matter radius as input, since generated errors
may be comparable or even smaller. For instance, if the matter radius
$r_m$ is taken as input we get instead $ \eta= 0.0253(4) $ a
compatible value with similar errors. However, if we take
$A_S=0.8846(9)$ as input for Set IV we get $\eta = 0.0255(1)$ a
compatible value with the experimental one but with much smaller
errors. The reduction of errors is also confirmed in Sets I, II and
III, although the central values are a bit off. This result opens up
the possibility of making a benchmark determination of the asymptotic
$D/S$ deuteron ratio from the chiral effective theory. Obviously, to
do so, the chiral constants should be known with rather high accuracy,
an illusory expectation at the present moment. In this regard it would
perhaps be profitable to pin down the errors for the chiral constants
from peripheral waves. This point will be analyzed
elsewhere~\cite{Pavon2005}. 

Both the loss of predictive power and the very rare possibility of
making model independent theoretical predictions for purely hadronic
processes using Chiral Perturbation Theory more accurate than
experiment we seem to observe in low energy NN scattering is not new
and has already been documented for low energy $\pi\pi$
scattering~\cite{Nieves:1999zb,Colangelo:2000jc,Colangelo:2001df} and
provides a further motivation to use chiral effective approaches.

\begin{table*}
\caption{\label{tab:table3} Deuteron properties and low energy
parameters in the $^3S_1-^3D_1$ channel for the pionless theory
(Short), the OPE and the TPE potential.  We use the non-relativistic
relation $ \gamma= \sqrt{ 2 \mu_{np} B} $ with $B=2.224575(9)$. For
${\rm OPE^{*}}$ we take $g_A = 1.26$ as input instead of $g_{\pi NN} =
13.083 $ as input. The errors quoted in the TPE reflect the
uncertainty in the non-potential parameters $\gamma$, $\eta$ and
$\alpha_0$ only. Differences from this work are attributed to finite
cut-off effects. Experimental values can be traced
from~\cite{deSwart:1995ui}.}
\begin{ruledtabular}
\begin{tabular}{|c|c|c|c|c|c|c|c|c|c|c|c|c|}
\hline  Set & Ref. & $\gamma $ & $\eta$ & $A_S $
& $r_m $ & $Q_d $ & $P_D $ &  $ \langle r^{-1} \rangle $ & $ \alpha_0 $ &
$\alpha_{02} $ & $ \alpha_2 $ & $r_0 $ \\ \hline
{\rm Short} & - & Input & 0 & 0.6806 & 1.5265 &  0 & 0 & $ \infty $& 4.3177& 0 & 0 & 0   \\ 
{\rm OPE} & \cite{PavonValderrama:2005gu} & Input & 0.02633 & 0.8681(1) & 1.9351(5) & 0.2762(1) &
7.31(1)\% & 0.476(3) & 5.335(1) & 1.673(1) & 6.693(1) & 1.638(1) \\
%
${\rm OPE}^{*}$ &  \cite{PavonValderrama:2005gu} & Input & 0.02555 & 0.8625(2) & 1.9234(5) & 0.2667(1)
& 7.14(1)\% & 0.471(3) & 5.308(1) & 1.612(1) & 6.325(1) & 1.602(2) \\ \hline \hline  
Set I & ${\rm N}^2$LO \cite{Epelbaum:1999dj} &  & 0.0245 & 0.884  & 1.967 & 0.262
& 6.11 \% & -- & 5.420  &  -- & -- & 1.753 \\ 
Set I & ${\rm N}^2$LO \cite{Entem:2001cg} &  & 0.0256 & 0.8846  & 1.9756 & 0.281
& 4.17 \% & -- & 5.417  &  -- & -- & 1.753 \\ 
Set I & This work & Input & Input & 0.900(2) & 1.999(4) & 0.284(4)
& 6(1)\% & 0.424(3) & Input  &  2.3(2) & 3(3) & 1.4(4)\\
\hline 
Set II& \cite{Rentmeester:1999vw}  & - & - &- & - & -
& - & - & - &  - & -& - \\
Set II& This work  & Input & Input & 0.900(2) & 1.999(4) & 0.285(4)
& 7(1)\% & 0.424(4) &  Input  &  2.22(15) & 4(2) & 1.46(19)\\ \hline 
Set III &  ${\rm N}^2$LO \cite{Epelbaum:2003xx} & -- & 0.0256 & 0.873  & 1.972  & 0.272
& 5 \% & - & 5.427  &   - & - & 1.731 \\
Set III &  ${\rm N}^3$LO \cite{Epelbaum:2004fk} & -- & 0.0254 & 0.882  & 1.979  & 0.266
& 3 \% & - & 5.417  &   - & - & 1.745 \\
Set III & This work & Input & Input & 0.891(3) & 1.981(5) & 0.279(4)
& 7(1)\% & 0.436(3) & Input  &  1.88(10) & 5.7(16) & 1.67(8)\\ \hline 
Set IV  & ${\rm N}^3$LO  \cite{Entem:2003ft}  & -- & 0.0256 & 0.8843 & 1.968  & 0.275 
& 4.51 \% & - & 5.417  &  - & - & 1.752  \\ 
Set IV  & This work & Input & Input & 0.884(4) & 1.967(6) & 0.276(3)
& 8(1)\% & 0.447(5) & Input  &  1.67(4) & 6.6(4) & 1.76(3)\\ \hline \hline 
NijmII & \cite{Stoks:1993tb,Stoks:1994wp} & 0.2316 & 0.02521 & 0.8845(8) & 1.9675 & 0.2707 & 
5.635\%  & 0.4502 & 5.418 & 1.647 & 6.505 & 1.753 \\
Reid93 & \cite{Stoks:1993tb,Stoks:1994wp} & 0.2316 & 0.02514 & 0.8845(8) & 1.9686 & 0.2703 & 
5.699\% & 0.4515 & 5.422 & 1.645 & 6.453 & 1.755 \\ \hline 
Exp. & - &  0.2316 &  0.0256(4)  & 0.8846(9)  & 1.971(6)  &
0.2859(3) & - & - &  5.419(7) & - & - &  1.753(8) \\
\end{tabular}
\end{ruledtabular}
\end{table*}

\subsection{Low energy parameters} 
\label{sec:low-energy} 

\begin{figure}[ttt]
\begin{center}
\epsfig{figure=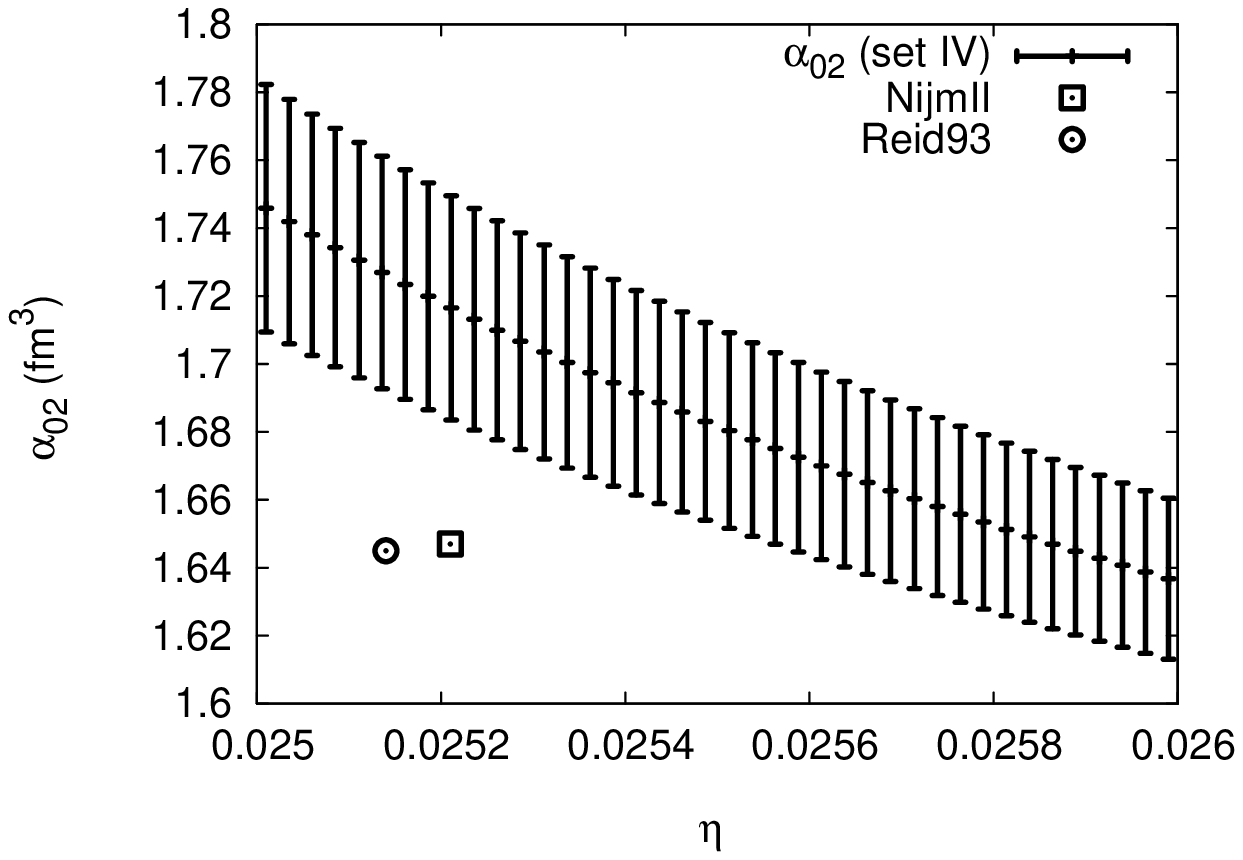,height=5.5cm,width=6.5cm}  
\epsfig{figure=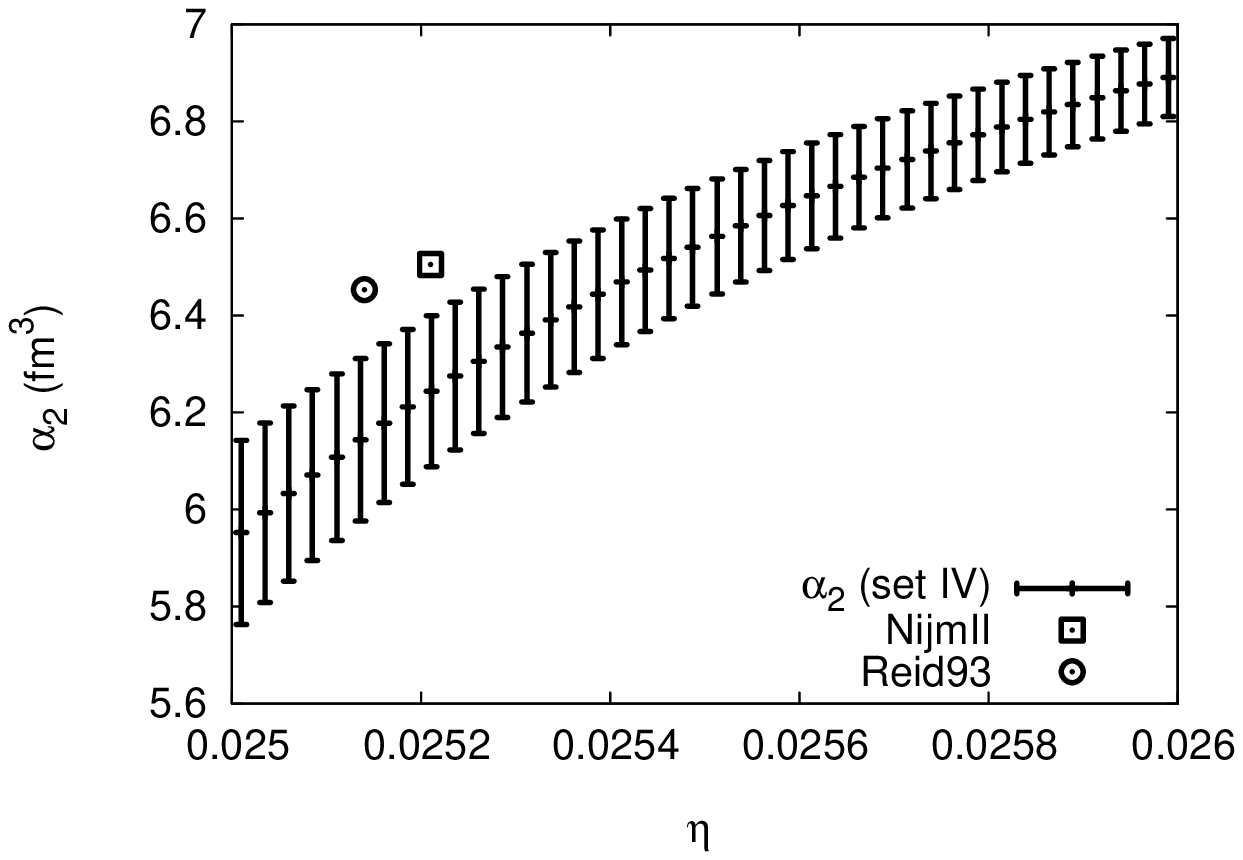,height=5.5cm,width=6.5cm}
\end{center}
\caption{The dependence of the $\alpha_{02}$ scattering length (in
${\rm fm}^3$) and the $\alpha_{2}$ scattering length (in ${\rm fm}^5$)
for the TPE potential as a function of the asymptotic $D/S$ ratio
$\eta $. The boxes represent the Reid93 and Nijm II values determined
in Ref.~\cite{PavonValderrama:2004se}. We use $\alpha_0 = 5.419(7) {\rm
fm} $ to generate the bands.  We use the set IV low energy constants.}
\label{fig:alpha[eta]}
\end{figure}

The zero energy wave functions are taken asymptotically
as~\footnote{We correct an error in Eq.(45) of our previous
work~\cite{PavonValderrama:2005gu} where $\alpha_2$ appears. The
corrected numerical value is $\alpha_2 = 6.693 {\rm fm}^5$.}
\begin{eqnarray}
u_{0,\alpha} (r) & \to & 1- \frac{r}{\alpha_0} \, , \nonumber \\ w_{0,\alpha} (r) &
\to & \frac{3 \alpha_{02}}{\alpha_0 r^2 } \, , \nonumber \\ u_{0,\beta} (r)
&\to & \frac{r}{\alpha_0} \, , \nonumber \\ w_{0,\beta} (r) &\to& 
\left( \frac{\alpha_2}{\alpha_{02}} + \frac{\alpha_{02}}{\alpha_0} \right)
\frac{3}{r^2}- \frac{r^3}{15 \alpha_{02}} \, . 
\label{eq:zero_energy}
\end{eqnarray} 
Using these zero energy solutions one can determine the effective
range. The $^3S_1$ effective range parameter is given by
\begin{eqnarray} 
r_0 &=& 2 \int_0^\infty \left[ \left(1-\frac{r}{\alpha_0} \right)^2 -
u_\alpha (r)^2 - w_\alpha (r)^2 \right] dr \, . \nonumber \\ 
\label{eq:r0_triplet} 
\end{eqnarray} 
In the zero energy case, the vanishing of the diverging exponentials
at the origin imposes a condition on the $\alpha $ and $\beta$ states
which generate a correlation between $\alpha_0$ , $\alpha_{02}$ and
$\alpha_2$. Using the superposition principle of boundary conditions
we may write the solutions in such a way that 
\begin{eqnarray} 
u_{0,\alpha} (r) &=& u_1 (r) - \frac{1}{\alpha_0} u_2 (r) + \frac{3
\alpha_{02}}{\alpha_0} u_3 (r) \nonumber \\ w_{0,\alpha} (r) &=& w_1
(r) - \frac{1}{\alpha_0} w_2 (r) + \frac{3 \alpha_{02}}{\alpha_0} w_3
(r) \nonumber \\ u_{0,\beta} (r) &=& \frac{1}{\alpha_0} u_2 (r) +
\left( \frac{3 \alpha_2}{\alpha_{02}} + \frac{3\alpha_{02}}{\alpha_0}
\right) u_3 (r) -\frac1{15 \alpha_{02}} u_4 (r) \nonumber \\
w_{0,\beta} (r) &=& \frac{1}{\alpha_0} w_2 (r) + \left( \frac{3
\alpha_2}{\alpha_{02}} + \frac{3\alpha_{02}}{\alpha_0} \right) w_3 (r)
-\frac1{15\alpha_{02}} w_4 (r) \nonumber \\
\label{eq:sup_zero}
\end{eqnarray}

where the functions $u_{1,2,3,4}$ and $w_{1,2,3,4}$ are independent on
$\alpha_0$, $\alpha_{02}$ and $\alpha_2$ and fulfill suitable boundary
conditions. The orthogonality constraints for the $\alpha$ and $\beta$
states read in this case
\begin{eqnarray}
u_\gamma u_{0,\alpha}' - u_\gamma' u_{0,\alpha} + w_\gamma
w_{0,\alpha}' - w_\gamma ' u_{0,\alpha} \Big|_{r=r_c} &=& 0 \nonumber \\
u_\gamma u_{0,\beta}' - u_\gamma' u_{0,\beta} + w_\gamma w_{0,\beta}'
- w_\gamma ' u_{0,\beta} \Big|_{r=r_c} &=& 0 \nonumber \\
\label{eq:orth_triplet} 
\end{eqnarray} 
yielding  two relations between $\gamma$, $\alpha_{02}$, $\alpha_2$,
$\eta$ and $\alpha_0$, meaning that two of them are not independent.
Using the superposition principle decomposition of the bound state,
Eq.~(\ref{eq:sup_bound}), and for the zero energy states,
Eq.~(\ref{eq:sup_zero}), we make the orthogonality relation explicit
in $\alpha_0$, $\alpha_{02}$, $\alpha_2$ and $\eta$. If we would use
$\alpha_0$, $\alpha_{02}$, $\alpha_2$ as input parameters the
orthogonality constraint is actually a non-linear eigenvalue problem
for $\gamma$ and $\eta$. The values of $\alpha_{02}$ and $\alpha_2$
are not so well known although they have been determined in potential
models in our previous work~\cite{PavonValderrama:2004se}. In
contrast, $\gamma$, $\eta$ and $\alpha_0$ are well determined
experimentally. Thus, in the deuteron scattering channel we will use
$\gamma$, $\eta$ and $\alpha_0 $ as independent input parameters and
$\alpha_{02}$, $\alpha_2$ as predictions. This same set of independent
parameters was also adopted in Ref.~\cite{deSwart:1995ui} within the
high quality potential model treatment, although the role of the short
distance Van der Waals singularity was not recognized.  Fixing the
experimental value of $\gamma$ we get the following relations for
different parameter choices of $c_1$, $c_3$ and $c_4 $,
\begin{widetext} 
\underline{Set I} 
\begin{eqnarray} 
\alpha_{02}&=&\frac{2.01763 - 0.456461\,\alpha_0 - 44.8947\,\eta +
11.9351\,\alpha_0 \,\eta}{-0.314426 + 13.1555\,\eta} \nonumber 
\\ 
\alpha_2 &=& \frac{-0.023522 + 1.04677\,\eta - 11.6459\,{\eta}^2 + 
    \alpha_0\,\left( 0.008423 - 0.537856\,\eta + 9.39376\,{\eta}^2 \right) }{
    \alpha_0\,\left( -0.023901 + \eta \right)^2} 
+ \frac{\alpha_{02}^2}{\alpha_0}
\end{eqnarray} 

\underline{Set II} 
\begin{eqnarray} 
\alpha_{02}&=&\frac{1.71745 - 0.373228\,\alpha_0 -
  33.4616\,\eta + 8.76639\,\alpha_0\,\eta} {-0.228075 +
  9.865911\,\eta} \nonumber 
\\ 
\alpha_2 &=& \frac{-0.030303 +
  1.18083\,\eta - 11.5032\,{\eta}^2 + \alpha_0\,\left(
  0.009559 - 0.566611\,\eta + 9.71850\,{\eta}^2 \right) }{
  \alpha_0\,\left( -0.023118 + \eta \right)^2} + \frac{\alpha_{02}^2}{\alpha_0}
\end{eqnarray} 

\underline{Set III } 
\begin{eqnarray} 
\alpha_{02} &=& \frac{2.36659 - 0.550871\,\alpha_0 - 59.1666\,\eta +
  15.7488\,\alpha_0 \,\eta} {-0.414962 + 17.2806\,\eta}  \nonumber 
\\ 
\alpha_2 &=& \frac{ -0.018755 +
  0.937802\,\eta - 11.7229\,{\eta}^2 + \alpha_0\,\left(
  0.007437 - 0.505434\,\eta + 9.09925\,{\eta}^2 \right) }{
  \alpha_0\,\left( -0.024013 + \eta \right)^2} + \frac{\alpha_{02}^2}{\alpha_0}
\end{eqnarray} 

\underline{Set IV } 
\begin{eqnarray} 
\alpha_{02} &=& \frac{1.89526 
- 0.418953 \alpha_0 - 41.8369 \eta + 10.8526 \alpha_0 \eta}
{-0.279236 + 12.2978 \eta} \nonumber 
\\ 
\alpha_2 &=& \frac{-0.023751 +
  1.04857\,\eta - 11.5733\,{\eta}^2 + \alpha_0\,\left(
  0.008050 - 0.518604\,\eta + 9.31798\,{\eta}^2 \right) }{
\alpha_0\,\left( -0.023118 + \eta \right)^2}  + \frac{\alpha_{02}^2}{\alpha_0}
\end{eqnarray} 
\end{widetext} 
The numerical coefficients appearing in these equations depend on the
deuteron wave number $\gamma$ and the TPE parameters, $g$,$f$,$m$ and
$c_1$, $c_3$ and $c_4$. The dependence on $\eta$ for fixed values of
$\alpha_0$ within its experimental uncertainty is depicted in
Fig.~\ref{fig:alpha[eta]}. We see that for fixed chiral couplings
$c_1$, $c_3$ and $c_4$, the $\eta $ uncertainty dominates the
errors. Numerical values can be seen at Table~\ref{tab:table3}. Note
the large discrepancy in the effective range $r_0$ for Sets I and II
with the experimental number. Finite cut-off effects are observed in
Set III although the $\eta$ induced uncertainty would make the value
compatible with that estimate. Good agreement is observed again for
Set IV, particularly in the $E_1$ and $^3D_1$ scattering lengths and
the effective range $r_0$, but only the latter provides a clear TPE
improvement as compared to OPE. The quantities $\alpha_{02}$ and
$\alpha_2$ are compatible with typical
expectations~\cite{PavonValderrama:2004se} from the high quality
potential models.

\subsection{Phase Shifts}

\begin{figure*}[]
\begin{center}
\epsfig{figure=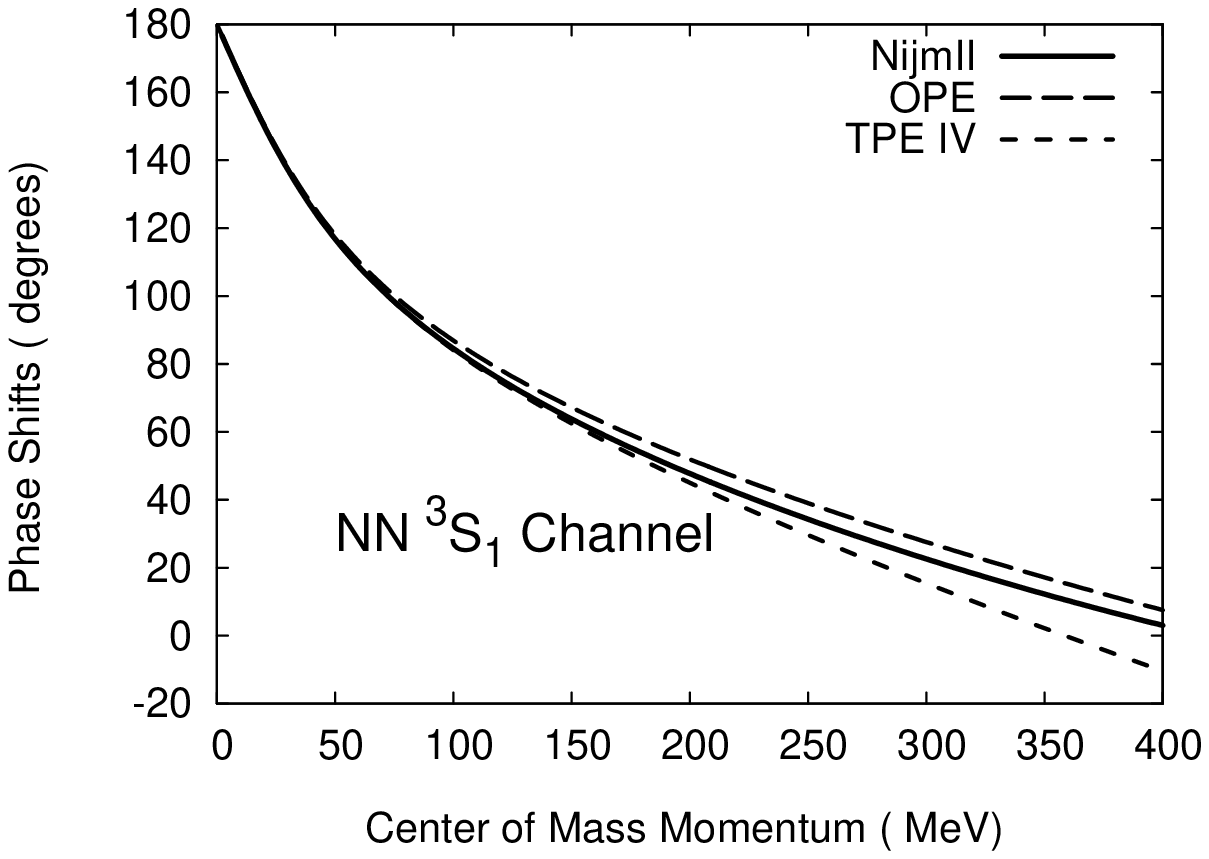,height=5.5cm,width=5.5cm}
\epsfig{figure=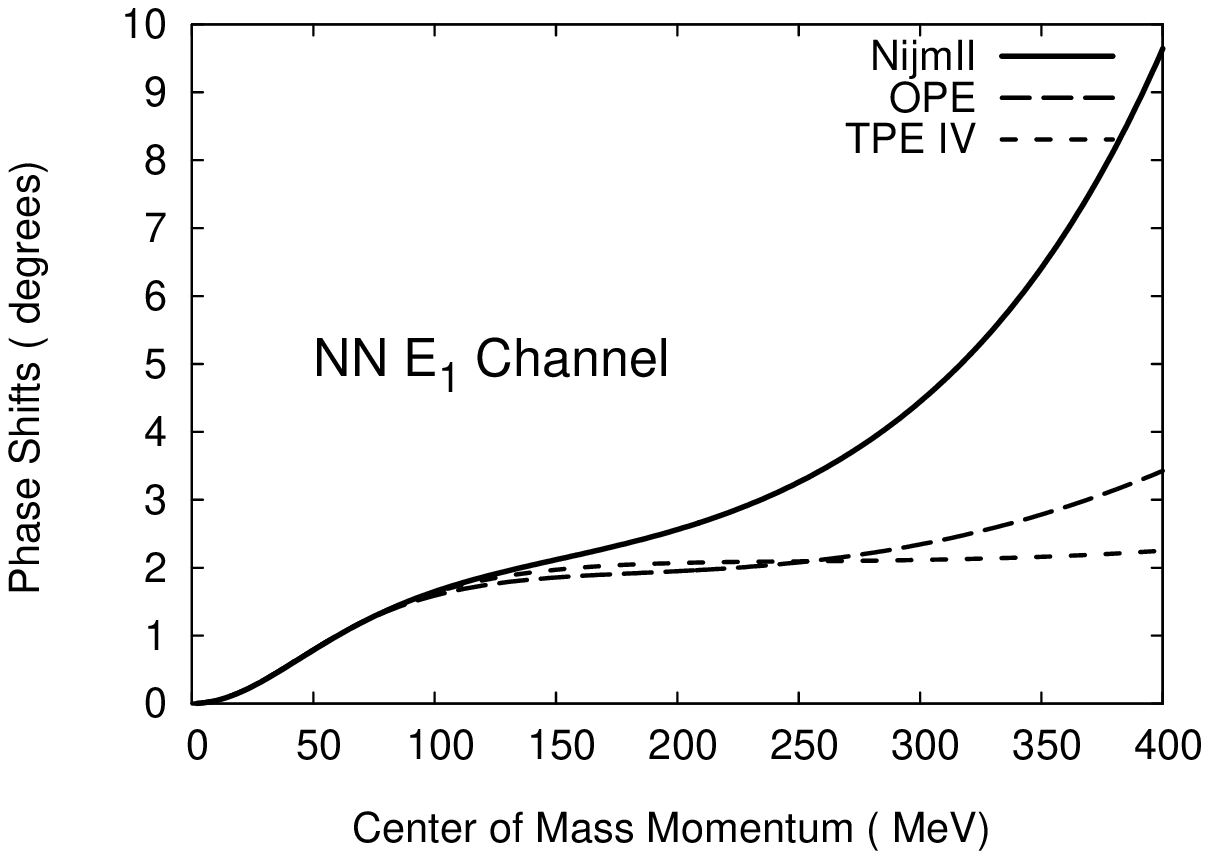,height=5.5cm,width=5.5cm}
\epsfig{figure=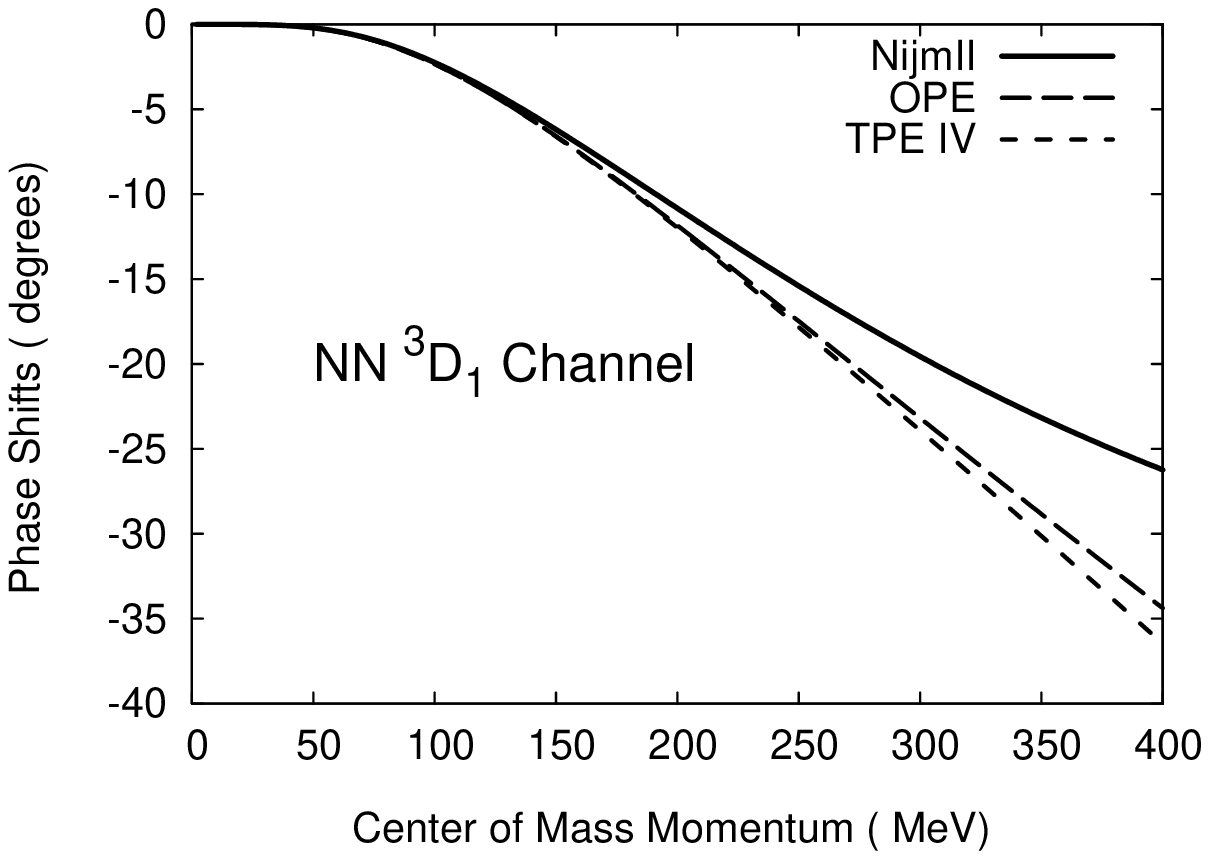,height=5.5cm,width=5.5cm}
\end{center}
\caption{Renormalized Eigen Phase shifts for the OPE and TPE
potentials as a function of the CM np momentum in the triplet
$^3S_1 - {}^3 D_1$ channel compared to the Nijmegen
results~\cite{Stoks:1993tb}. The regular scattering wave functions are
orthogonal to the regular deuteron bound state wave functions
constructed from the OPE with $\gamma=0.231605 {\rm fm}^{-1}$, $
m=138.03 {\rm MeV}$ and $g_{\pi NN}=13.083 $ for the OPE contribution 
to the TPE potential and $g=1.26$ for the TPE contribution to the 
TPE potential. We take Set IV (see main text).}
\label{fig:phase-shifts}
\end{figure*}

For the $\alpha$ and $\beta$ positive energy scattering states we
choose the asymptotic normalization
\begin{eqnarray}
u_{k,\alpha} (r) &\to & \frac{\cos \epsilon}{\sin \delta_1}\Big( \hat
j_0 (kr) \cos \delta_1 - \hat y_0 (kr) \sin \delta_1 \Big) \, , \nonumber \\ w_{k,\alpha}
(r) &\to & \frac{\sin \epsilon}{\sin \delta_1}\Big( \hat j_2 (kr) -
\hat y_2(kr) \sin \delta_1 \Big) \, , \nonumber \\ \\  
u_{k,\beta} (r) & \to & -\frac1{\sin \delta_1}\Big( \hat j_0 (kr) \cos \delta_2 - \hat y_0 (kr)
\sin \delta_2 \Big) \, ,  \nonumber \\ 
w_{k,\beta} (r) &\to & \frac{\tan \epsilon}{\sin \delta_1}\Big( \hat
j_2 (kr) \cos \delta_2 - \hat y_2(kr) \sin \delta_2 \Big) \, , \nonumber \\ 
\label{eq:phase_triplet}
\end{eqnarray} 
where $ \hat j_l (x) = x j_l (x) $ and $ \hat y_l (x) = x y_l (x) $
are the reduced spherical Bessel functions and $\delta_1$ and
$\delta_2$ are the eigen-phases in the $^3S_1$ and $^3D_1$ channels,
and $\epsilon$ is the mixing angle $E_1$. The use of the superposition
principle for boundary conditions as well as the orthogonality
constraints, 
\begin{eqnarray}
u_\gamma u_{k,\alpha}' - u_\gamma' u_{k,\alpha} + w_\gamma
w_{k,\alpha}' - w_\gamma ' u_{k,\alpha} \Big|_{r=r_c} &=& 0 \nonumber \\
u_\gamma u_{k,\beta}' - u_\gamma' u_{k,\beta} + w_\gamma w_{k,\beta}'
- w_\gamma ' u_{k,\beta} \Big|_{r=r_c} &=& 0 \nonumber \\
\label{eq:orth_triplet_k} 
\end{eqnarray} 
analogous to Eq.~(\ref{eq:orth_triplet}), to the deuteron wave
functions. If orthogonality would be applied to the zero energy state
one obtains an explicit relation of $\delta_1$, $\delta_2$ and
$\epsilon$ with the scattering lengths $\alpha_0$, $\alpha_2$ and
$\alpha_{02}$ as a direct generalization to the coupled channel case
the one channel singlet case given by
Eq.~(\ref{eq:phase_singlet}). The explicit expressions are rather
cumbersome and will not be written down here explicitly.  The results
are depicted in Fig.~\ref{fig:phase-shifts} for Set IV. We observe a
clear improvement in the threshold region, in consonance with the low
energy parameters of Table~\ref{tab:table3} and a moderate improvement
over the OPE results in the intermediate energy region. This suggests
that finite cut-off effects may also be built in the phase shifts as
well as the low energy parameters.

\medskip

\section{Error analysis and Determination of chiral couplings from low energy NN data and the deuteron}
\label{sec:error} 

\subsection{Propagating experimental errors in $c_1$, $c_3$ and $c_4$.}

The results in the previous sections clearly show that deuteron and
low energy scattering properties in the $^1S_0$ and $^3S_1-^3D_1$
channels are sensitive to finite cut-off effects and also to the
values of the chiral constants after removal of the cut-off. We will
assume that the values for $c_1$, $c_3$ and $c_4$ are free of
uncertainties. Then, Set IV provides the best description of triplet
data but produces a slightly off value for the effective range in the
singlet channel at the $2\sigma$ confidence level. Note that in the
singlet case the theoretical prediction for $r_0$ does not have a
large source of error as in the triplet case where uncertainties in
$\eta$ dominate the error. On the contrary, Set III provides a
compatible value for the effective range in the singlet channel but
incompatible values for the triplet channel in $A_S $ and $r_0$ at the
$3-4 \sigma $ confidence level on the experimental side. Thus, on this
basis we may reject Set III and accept Set IV.

To improve on this analysis, let us try to include some errors on the
chiral coefficients. The $\pi N$ analysis of
Ref.~\cite{Buettiker:1999ap} (Set I) and the $NN$ fit of
Ref.\cite{Rentmeester:1999vw} (Set II) yield some
errors. Ref.~\cite{Entem:2003ft} (Set IV) does not quote errors but we
will take the educated guess of a $5 \%$ error for $c_3$ and a $30 \%
$ error for $c_4 $~\cite{Machleidt-2005}. We can propagate then by a
Monte-Carlo simulation implementing also the errors in $g_{\pi
NN}=13.1 \pm 0.1 $, $\alpha_{0,s}= -23.77\pm 0.05 $, $\alpha_{0,t} =
5.419 \pm 0.007 $, $\eta= 0.0256 \pm 0.0004 $. We assume for
simplicity that all these quantities are fully uncorrelated. This will
in general enhance the errors, as compared to the case where
correlations in $c_1$, $c_3$ and $c_4$ with $\pi N$ would be taken
into account. Perhaps, the best thing would be to consider a
simultaneous analysis of both $NN$ and $\pi N $ low energy data to
build in correlations. Obviously, we do not expect good central values
for the observables judging from Table~\ref{tab:table3}. But there is
still the possibility of large error bars.

\begin{table}
\caption{\label{tab:table_errors} Singlet $^1S_0 $ and triplet
$^3S_1-^3D_1$ scattering and deuteron properties with error estimates
using the chiral TPE potential. We make a MonteCarlo calculation of
the input parameters $g_{\pi NN}=13.1 \pm 0.1 $, $\alpha_{0,s}= -23.77\pm
0.05 $, $\alpha_{0,t} = 5.419 \pm 0.007 $, $\eta= 0.0256 \pm 0.0004 $ and
the chiral constants $c_1$, $c_3$ and $c_4$.  The quoted values span
an interval where $68 \%$ of the output is contained.}
\begin{ruledtabular}
\begin{tabular}{|c|c|c|c|c|}
 & Set I & Set II &  Set IV & Exp. \\ 
\hline  $ c_1 $  & $-0.81(15)$ & $-0.76(7) $ & $-0.81 $ &    \\ 
        $ c_3 $  & $-4.69(1.34)$ & $-5.08(24) $      & $ -3.20(16) $  &   \\ 
        $ c_4 $  & $ 3.40(4)$ & $ 4.78(10) $    &  $ 5.40(1.65) $ &   \\ 
\hline  \hline 
        $ r_{0,s} $  & $2.92^{+0.08}_{-0.04}$ & $2.97^{+0.03}_{-0.02}$ & $2.86^{+0.04}_{-0.03}$     & $2.77 \pm 0.05$   \\ \hline  
        $ r_{0,t} $  & $1.36^{+0.33}_{-0.75}$ & $1.48^{+0.14}_{-0.25}$ & $1.76^{+0.03}_{-0.06}$     & $1.753 \pm 0.008$   \\ \hline  
        $ A_s $  & $0.899^{+0.008}_{-0.009}$ & $0.900^{+0.003}_{-0.004}$      &  $0.884^{+0.005}_{-0.008}$ & $0.8849 \pm 0.0009$    \\   \hline  
        $ Q_d $  & $0.284^{+0.005}_{-0.007}$ & $0.284^{+0.005}_{-0.004}$    & $0.276^{+0.004}_{-0.004}$ &  $0.2859 \pm 0.0003$ \\ \hline  
        $ r_m $  & $1.998^{+0.015}_{-0.019}$ & $1.998^{+0.007}_{-0.007}$    & $1.965^{+0.011}_{-0.014}$ & $1.971 \pm 0.006$ \\ \hline 
       $ P_d $  & $6.6^{1.0}_{-0.9}$ & $7.1^{+0.9}_{-0.9}$    & $8.3^{+1.4}_{-1.5}$  &  -- \\ \hline 
       $ \alpha_{02} $  & $2.26^{+0.51}_{-0.39}$ & $2.20^{+0.23}_{-0.16}$   & $1.67^{+0.13}_{-0.13}$ &--   \\ \hline 
       $ \alpha_{2} $  & $3.6^{+3.3}_{-6.9}$ & $4.0^{+1.6}_{-2.9}$   & $ 6.71^{+0.48}_{-0.83}$ & --  \\ \hline 
\end{tabular}
\end{ruledtabular}
\end{table}
The outcoming distributions in the low energy and deuteron parameters
are somewhat asymmetric. Actually, for a given set of $c_1$, $c_3$ and
$c_4$ distributions we observe the appearance of upper bounds in the
$^3S_1 $ effective range, namely $r_{0,t} \le 1.79, 1,75, 1.81 {\rm
fm}$ for Sets I,II and IV respectively where the out-coming
distributions become more dense.  The results of the error propagation
are summarized in Table~\ref{tab:table_errors}. Thus, we see that the
values of the chiral coefficients deduced from low energy $\pi
N$~\cite{Buettiker:1999ap} are globally inconsistent, at the $1\sigma$
level, with the low energy NN threshold parameters after uncertainties
are taken into account. The same remark applies to Set
II~\cite{Rentmeester:1999vw}.  Again, the loss of predictive power
becomes manifest for all the sets although Set IV provides the best
central values and the smallest errors.  The situation for the
quadrupole moment is noteworthy since the difference to the potential
value is attributed to Meson Exchange Currents (MEC) and relativistic
effects, which provide a correction of about $0.01 {\rm fm}^2$ (see
Ref.[25] in Ref.~\cite{Entem:2001cg} and also
Ref.~\cite{Phillips:2003jz}). As we see, this is about the size of the
error deduced from our analysis. In the case of the deuteron matter
radius the situation is even worse since MEC's contributions are much
smaller $0.003 {\rm fm}$~\cite{Phillips:2003jz} while our predicted
errors are larger. It would be extremely interesting to reanalyze the
problem with the present deuteron wave
functions~\cite{Phillips-Pavon}.

\subsection{Determination of  $c_1$, $c_3$ and $c_4$.}

Another possibility is to attempt a direct fit to the data. The standard
approach is to fit the partial waves to a NN
database~\cite{Stoks:1993tb,Stoks:1994wp}.  The problem with such an
approach is that, unfortunately, there is no error assignment on the
phase shifts and hence a reliable assessment of errors cannot be
made. Actually, besides the work of
Ref.~\cite{Rentmeester:1999vw,Rentmeester:2003mf} where a full partial
wave analysis was undertaken, other
works~\cite{Entem:2003ft,Epelbaum:2003xx,Epelbaum:2004fk} assume fixed
values for $c_1$ , $c_3$ and $c_4$ without attempting any error
analysis based on input uncertainties.  But even if data for the phase
shifts with errors were known one expects the quality of the fit to
worsen as the energy is increased as we think that the chiral approach
to NN interaction should work best at low energies. If the data were
known with uniform uncertainty one would fit until $\chi^2/DOF$
exceeds one providing an energy window. In such a fit all points are
equally weighted while we know that the description at low energies,
where the theory works best, will be compromised by the highest
possible energy within such an energy window.

\begin{figure*}[ttt]
\begin{center}
\epsfig{figure=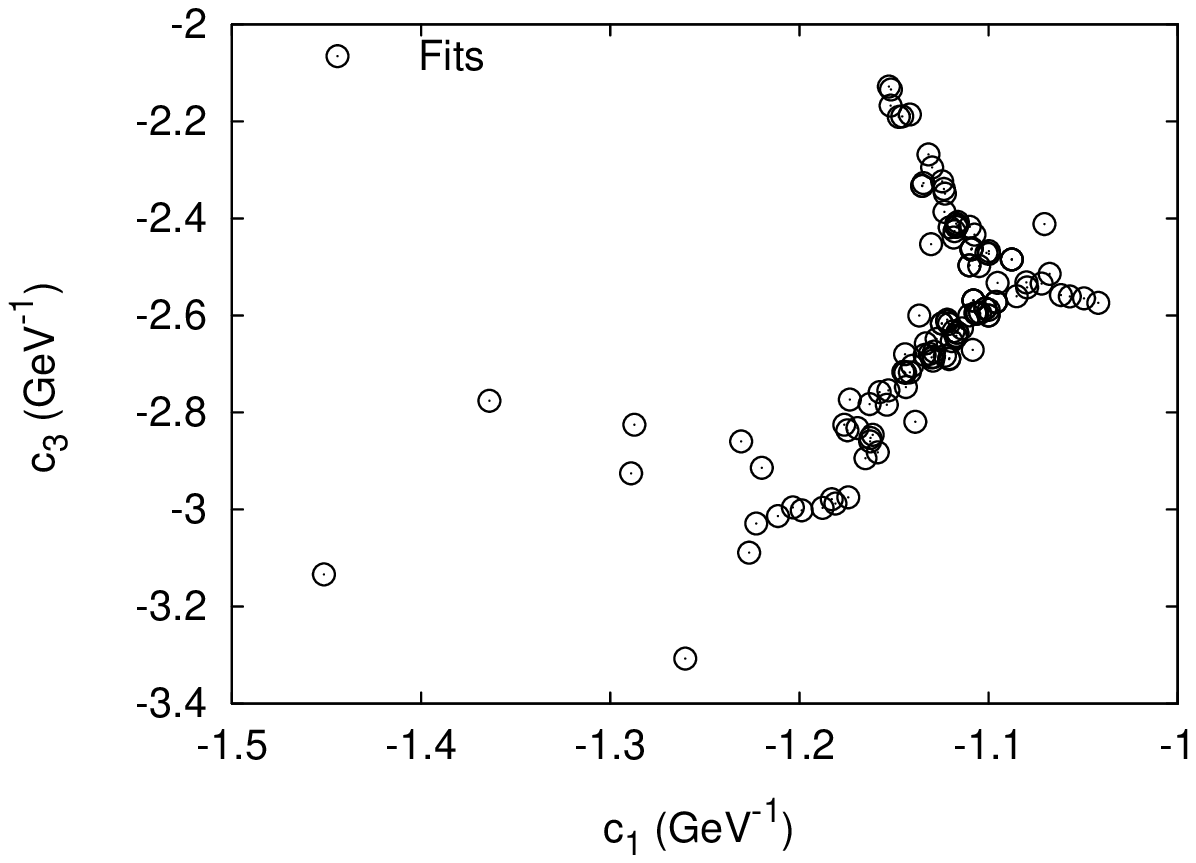,height=5.5cm,width=5.5cm}
\epsfig{figure=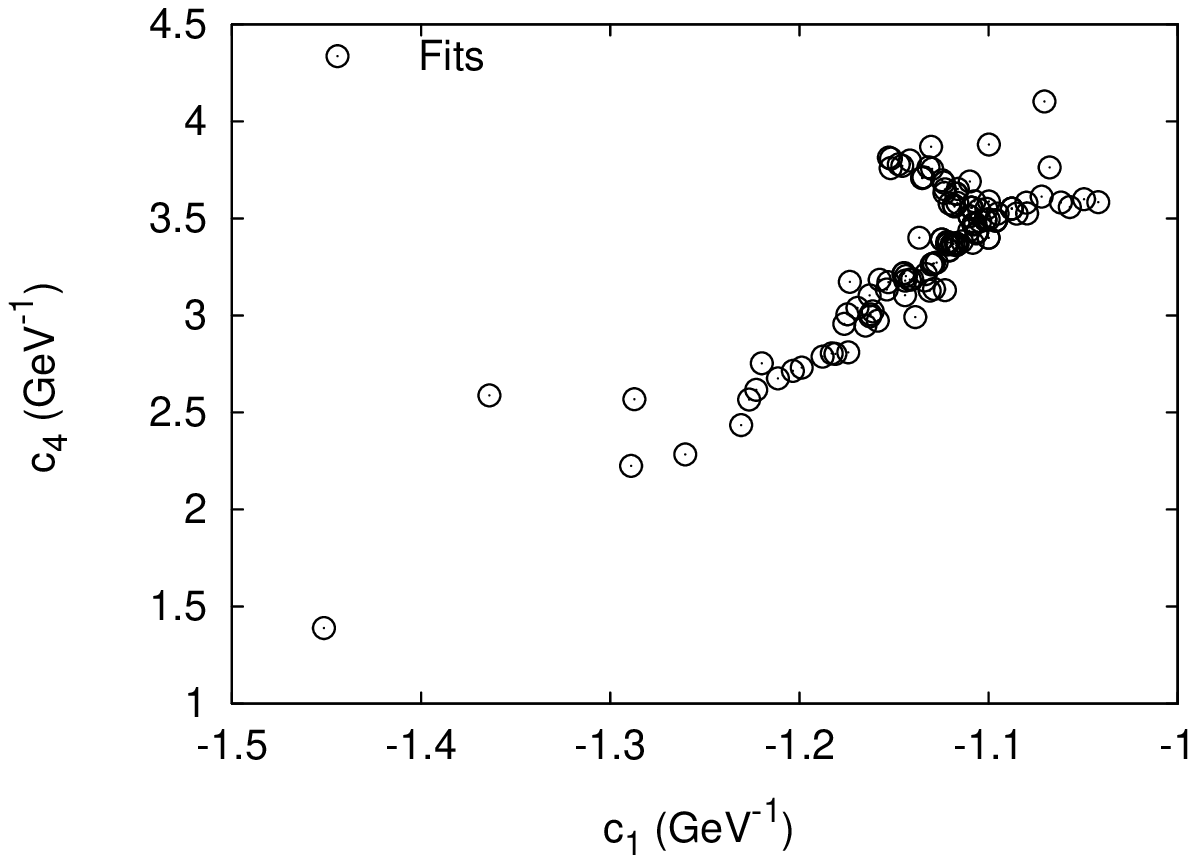,height=5.5cm,width=5.5cm}
\epsfig{figure=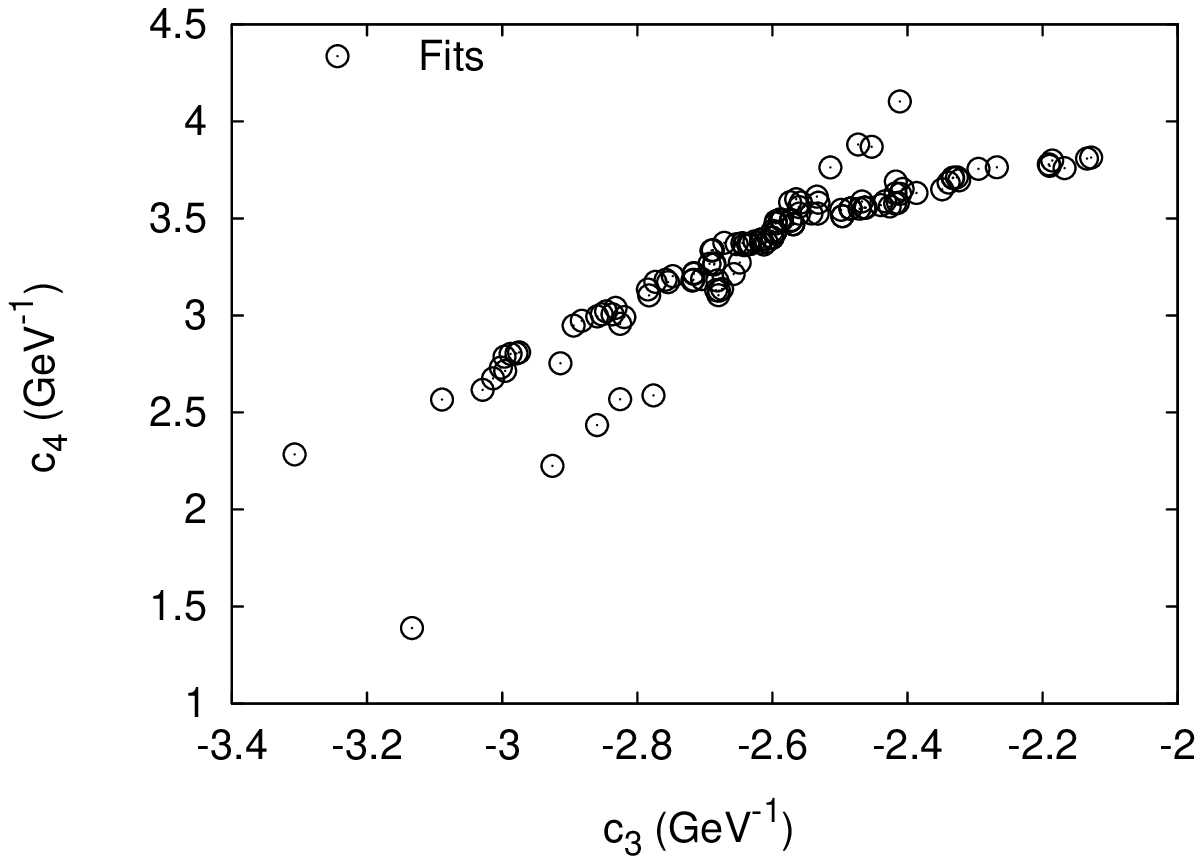,height=5.5cm,width=5.5cm}
\end{center}
\caption{Correlation plots for the low energy constants $c_1$, $c_3$
and $c_4$ obtained from the renormalized TPE potential by fitting the
singlet and triplet effective ranges $r_{0,s} = 2.77 \pm 0.05 $ and $r_{0,t} =
1.753 \pm 0.08 $ and the deuteron s-wave asymptotic normalization
$A_S=0.8846 \pm 0.0009 $. The dispersion in the data reflects the
dispersion in all input parameters, $g_{\pi NN}=13.11 \pm 0.08 $,
$\alpha_{0,s}= -23.77\pm 0.05 $, $\alpha_{0,t} = 5.419 \pm 0.007 $, $\eta=
0.0256 \pm 0.0004 $ which provide $\chi^2 < 1$.}
\label{fig:correlations}
\end{figure*}

Along the previous line of reasoning we propose, instead, to fit
directly the low energy threshold parameters which central values and
errors are well known and widely accepted. The basic ingredient is to
use purely hadronic information in the process to avoid any
contamination due to electromagnetic effects. Specifically, we make a
Monte-Carlo sampling of the input data assuming that as primary data
they are gaussian distributed and uncorrelated. For any of the samples
we make a $\chi^2$ fit to the values $r_{0,s}$, $r_{0,t}$ and $A_S$, i.e. we
minimize
\begin{eqnarray}
\chi^2 = \left( \frac{r_{0,s} - r_{0,s}^{\rm exp}}{\Delta r_{0,s}}\right)^2  
+\left( \frac{r_{0,t} - r_{0,t}^{\rm exp}}{\Delta r_{0,t}}\right)^2  
+\left( \frac{A_S - A_S^{\rm exp}}{\Delta A_S}\right)^2 \nonumber \\  
\end{eqnarray}
and determine then the optimal values of $c_1$ , $c_3$ and $c_4$. We
only accept values where $\chi^2 < 1$, and the resulting distribution
of chiral constants $c_1$, $c_3$ and $c_4$ is given in
Figs.~\ref{fig:correlations}. As we see, there is a very strong,
almost linear, correlation between $c_3$ and $c_4$. This can be easily
understood in terms of the short distance dominance of the singlet
effective range, since for the pure Van der Waals contribution, and in
the limit of large scattering length $r_{0,s} \sim R = ( M
C_6)^{1/4}$, with $C_6$ given in
Eq.~(\ref{eq:pot_sing_short}). Deviations from linearity are induced
from the larger relative error of $r_{0,s}$ ($1\%$) as compared to
$r_{0,t} $ and $A_S$ ($0.1\% $). This is different from the large
scale partial phase-shift analysis obtained in
Ref.~\cite{Rentmeester:2003mf} where a very small correlation between
$c_3$ and $c_4$ of about 0.2 was found. We have checked that
cutting-off data with decreasing values of $\chi^2 $, excludes the
points where the distribution is sparse, so that the dense part indeed
reflects the uncertainties in the input data. The fact that the three
coefficients seem to be on a line is just a consequence of solving by
minimization a system of three equations and three unknowns.

\medskip

We use $g_{\pi NN}=13.083 $, $\alpha_{0,s}= -23.77\pm 0.05 $,
$\alpha_{0,t} = 5.419 \pm 0.007 $, $\eta= 0.0256 \pm 0.0004 $ and fit $c_1$
, $c_3$ and $c_4$ to the values $r_{0,s}=2.77 \pm 0.05 $, $r_{0,t}= 1.753 \pm
0.08 $ and $A_S= 0.8846 \pm 0.0009 $. Our final result for a
sample with 125 points with $\chi^2 < 1 $ is
\begin{eqnarray}
c_1 &=& -1.13_{-0.04}^{+0.02} ({\rm stat})\,  {\rm GeV}^{-1} \, , \nonumber \\
c_3 &=& -2.60_{-0.23}^{+0.18} ({\rm stat})\, {\rm GeV}^{-1} \, , \nonumber \\
c_4 &=& + 3.40_{-0.40}^{+0.25} ({\rm stat})\, {\rm GeV}^{-1} \, .  
\end{eqnarray}
The central value is the mean and the errors have been obtained by the
standard method of excluding the $16\%$ left and right extreme values
of the variables, so as to have $68\%$ confidence level between the
upper and lower values. Cutting-off data with $\chi^2 < 0.5$ does not
change significantly the result.

At the $2\sigma$ level, our values for $c_1$, $c_3$ and $c_4$ are
compatible with the analysis of low energy $\pi N $ scattering of
Ref.~\cite{Buettiker:1999ap}, $c_1 = -0.81 \pm 0.15 $, $c_3 = -4.69
\pm 1.34 $ and $c_4=3.40 \pm 0.04 $, but incompatible with the NN full
partial wave analyses~\cite{Rentmeester:1999vw,Rentmeester:2003mf}
where an energy dependent boundary condition at $a=1.4 {\rm fm} $ was
used. It is difficult to say whether other determinations for the
chiral couplings based on NN scattering are incompatible with ours,
since no error estimates have been provided.

\subsection{Estimate of the systematic errors}

As we have mentioned, any approach based on power counting of the
potential cannot make an {\it a priori} estimate of the accuracy of
the calculation. Nevertheless, we can have an idea by simply varying
the input parameters. 

At LO we may use either $g_A=1.26 $ as input or $g_{\pi NN}=13.1 $,
since the difference is the Goldberger-Treiman discrepancy, which
should be a higher order correction. The effect can be seen by
comparing OPE with OPE$^*$ in Table \ref{tab:table3}. When compared to
the TPE result, for e.g. Set IV, the error is underestimated this way.
At NNLO we use the same procedure in the TPE piece. Again, the
difference should be higher orders.  Numerically this is equivalent to
include $g_{\pi NN}=13.1 \pm 0.1 $ and consider this systematic error
as an statistical one, which has already been taken into account.

We can estimate the systematic error in the chiral constants by
varying the input used to determine the c's. To correlate the singlet
and triplet channels we must keep $r_{0,s}$ and $\alpha_{0,s}$. So we
can interchange the inputs $A_S$, $r_{0,t} $ with the outputs $r_m $,
$Q_d$, $\alpha_{02}$ and $\alpha_2$. This yields a total of 15
possible combinations. Another question concerns the assessment an
error to the fitted variables whenever there is no direct experimental
quantity, since this choice weights the determination of the c's. This
is the case for $\alpha_{02}$ and $\alpha_2$, where we make the
educated guess of taking the difference between the Reid93 and NijmII
values as determined in Ref.~\cite{PavonValderrama:2004se} as an
estimate of the error. The situation with $Q_d$ is a bit special and
we exclude it from the analysis~\footnote{The discrepancy of potential
models to the experimental value $\sim 0.01 {\rm fm}^2$, attributed to
MEC's and relativistic effects~\cite{Phillips:2003jz}, is about two
orders of magnitude larger than the error in the experimental number
$\sim 0.0003 {\rm fm}^2$ and the discrepancy between potential models
$\sim 0.0004 {\rm fm}^2$. It is not clear whether the discrepancy can
be pinned down with similar errors~\cite{Phillips-Pavon}.}. The
results are listed in Table~\ref {tab:table_systematic}. We see that
this estimate on the systematic error provides a larger fluctuation
than direct propagation of the input errors for $c_1$. Symmetrizing the
errors we get 
\begin{eqnarray}
c_1 &=& -1.2 \pm 0.2 \, ({\rm syst})\, {\rm GeV}^{-1} \, ,
\nonumber \\ c_3 &=& -2.6 \pm 0.1 \, ({\rm syst}) \, {\rm
GeV}^{-1} \, , \nonumber \\ c_4 &=& + 3.3 \pm 0.1 \, ({\rm syst})
\, {\rm GeV}^{-1} \, .
\end{eqnarray}
If we attempt a fit to all observables assigning $\Delta Q_d = 0.01{\rm
fm^2} $ and $\Delta \alpha_{02} = 0.4 {\rm fm} $ we get $ c_1 = -0.9$,
$c_3 = -2.71$ and $ c_4 = 3.85$ with a large $ \chi^2 /DOF = 3 $
basically due to the small errors. Obviously a more realistic estimate
of the errors would be desirable.

\begin{table}[bbb]
\caption{\label{tab:table_systematic} 
Central values for the chiral constants $c_1$, $c_3$ and $c_4$ 
depending on the input. we only include $\chi^2 < 1$.}
\begin{ruledtabular}
\begin{tabular}{|c|c|c|c|c|}
Fitted & $c_1$  & $c_3$ &  $c_4$ & $\chi^2 $ \\ \hline 

$r_{0,s}, A_S, r_{0,t} $      & -1.09 & -2.61 & 3.36 & 0.06  \\
$r_{0,s}, A_S, r_m $          & -1.23 & -2.57 & 3.34 & 0.3 \\
$r_{0,s}, r_{0,t}, r_m $      & -1.45 & -2.54 & 3.26 & 0.2 \\
$r_{0,s}, r_{0,t}, \alpha_2 $ & -1.09 & -2.64 & 3.17 & 0.4 \\
$r_{0,s}, r_m, \alpha_2 $     & -1.03 & -2.70 & 3.26 & 0.03 
\end{tabular}
\end{ruledtabular}
\end{table}

\section{The role of chiral Van der Waals forces} 
\label{app:a}

As we have pointed out, our approach is not the conventional one of
adding short distance counterterms following a given {\it a priori}
power counting regardless on the approximation where the long distance
potential has been constructed. Instead, the potential power counting
dictates the form of the short distance physics by demanding a finite
limit when the regulator is removed.  In order to stress the
differences with previous approaches it is interesting to see how much
of the phase shifts is determined from the short distance chiral
potential {\it without} adding a short range contribution to the
effective range. In the standard approach this can be achieved by
adding a counterterm $C_2$ in the $S-$wave channels. In
Ref.~\cite{PavonValderrama:2005gu} we showed that both perturbatively
and non-perturbatively the orthogonality constraints for the OPE
potential imply $C_2=0$.  Here we will see that the bulk of the
$S-$wave interaction can be explained mainly in terms of the chiral
Van der Waals force when renormalization is carried out, without any
additional short distance contribution or counterterm.

For a pure Van der Waals potential of the form
\begin{eqnarray}
U= -\frac{R^4}{r^6} \, , 
\end{eqnarray} 
the zero energy wave function can be analytically
computed~\cite{Frank:1971} in terms of Bessel functions
$J_\nu(x)$. Normalizing to the asymptotic form $u_0 (r) \to 1 - r
/\alpha_0 $ we get
\begin{eqnarray} 
u_0 (r) &=& \Gamma \left( \frac54 \right) \sqrt{\frac{2 r}{R}}
J_{\frac14} \left( \frac{R^2}{2 r^2} \right) \nonumber \\ &-& 
\Gamma \left(\frac34 \right)
\sqrt{\frac{r R}{2}} J_{-\frac14} \left( \frac{R^2}{2 r^2} \right) \frac1{\alpha_0}
\, .
\end{eqnarray}  
The effective range can also be computed analytically~\cite{Gao98,GF1999}
from Eq.~(\ref{eq:r0_singlet}) yielding
\begin{eqnarray} 
 r_0 &=& \frac{-4\,R^2}{3\,\alpha_0} +
\frac{4\,R^3\,{\Gamma(\frac{3}{4})}^2}{3\,\alpha_0^2\,\pi } +
\frac{16\,R\,{\Gamma(\frac{5}{4})}^2}{3\,\pi } \, , \nonumber \\ &=&
1.39473\,R - \frac{1.33333\,R^2}{\alpha_0} +
\frac{0.637318\,R^3}{\alpha_0^2} \, ,
\label{eq:r0_vdw}
\end{eqnarray} 
in agreement with the general low energy theorem of Eq.~(\ref{eq:r0_univ}).
Taking the values of Table~\ref{tab:table_vdw} for $R= (MC_6)^{1/4}$
one gets in the singlet $^1S_0$ channel
\begin{eqnarray} 
r_{0,s} &=& 2.39811 -
\frac{3.9418}{\alpha_{0,s}}+\frac{3.23959}{\alpha_{0,s}^2} \, \qquad ({\rm
Set \, I} ) \, ,\nonumber \\ 
r_{0,s} &=& 2.49192 -
\frac{4.25624}{\alpha_{0,s}}+\frac{3.63486}{\alpha_{0,s}^2} \, \qquad ({\rm
Set \, II} ) \, , \nonumber \\ 
r_{0,s} &=& 2.2227 -
\frac{3.8625}{\alpha_{0,s}}+\frac{2.57944}{\alpha_{0,s}^2} \, \qquad ({\rm
Set \, III} ) \, , \nonumber \\ 
r_{0,s} &=& 2.29099 -
\frac{3.59753}{\alpha_{0,s}}+\frac{2.82459}{\alpha_{0,s}^2} \, \qquad ({\rm
Set \, IV} ) \, . \nonumber \\ 
\end{eqnarray} 
The numerical agreement at the ten percent level of the $\alpha_{0,s}$
independent term with the full chiral TPE result,
Eq.~(\ref{eq:r0_todos})), is striking~\footnote{The formula
(\ref{eq:r0_vdw}) can also be used as a numerical test of the
integration method and of the numerical solution of the differential
equations. This is a non-trivial condition due to the rapid
oscillations of the wave function at the origin. We have checked that
it is accurately reproduced.}.  On the other hand, first order
perturbation theory in the OPE potential yields (see Sect.~ A of
Ref.~\cite{PavonValderrama:2005gu}) in the form of
Eq.~(\ref{eq:r0_univ}) the result
\begin{eqnarray} 
r_{0,s} &=& \frac{g_{\pi NN}^2 }{8 M \pi} \left( 1 - \frac{8}{3 \alpha_{0,s} m } +
\frac{2}{\alpha_{0,s}^2 m^2 }\right) \, , \nonumber \\ &=& 1.4369 -
\frac{5.4789}{\alpha_{0,s}} + \frac{5.8758}{\alpha_{0,s}^2} \, . 
\label{eq:r0_ope_pert}
\end{eqnarray} 
Note that the coefficient in $1/\alpha_{0,s}^2$ is slightly better described by
the OPE perturbative value than the full OPE result (see
Eq.~(\ref{eq:r0_todos})), a not unreasonable result since this coefficient is
sensitive to the longest range part of the interaction. Likewise, the
bulk of the $\alpha_0$-independent coefficient is given {\it just} by
the most singular contribution to the full chiral potential.  As we
see, for large scattering lengths the effective range scales with the
Van der Waals singlet radius $R_s= (MC_{6,^1S_0})^{1/4} $ and not with
the pion Compton wavelength $1/m$, confirming the dominance of the
short distances singularity in the singlet channel.

\begin{figure*}[ttt]
\begin{center}
\epsfig{figure=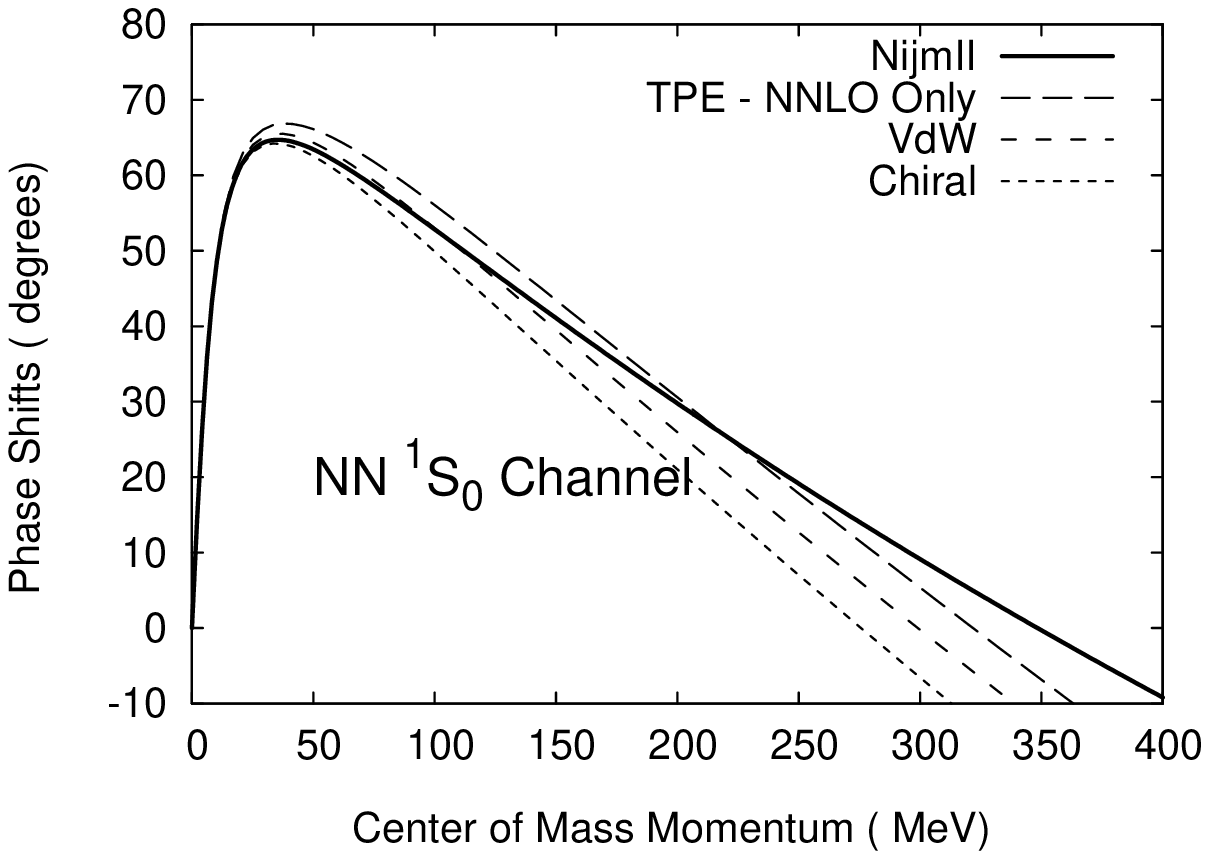,height=6.5cm,width=6.5cm}
\epsfig{figure=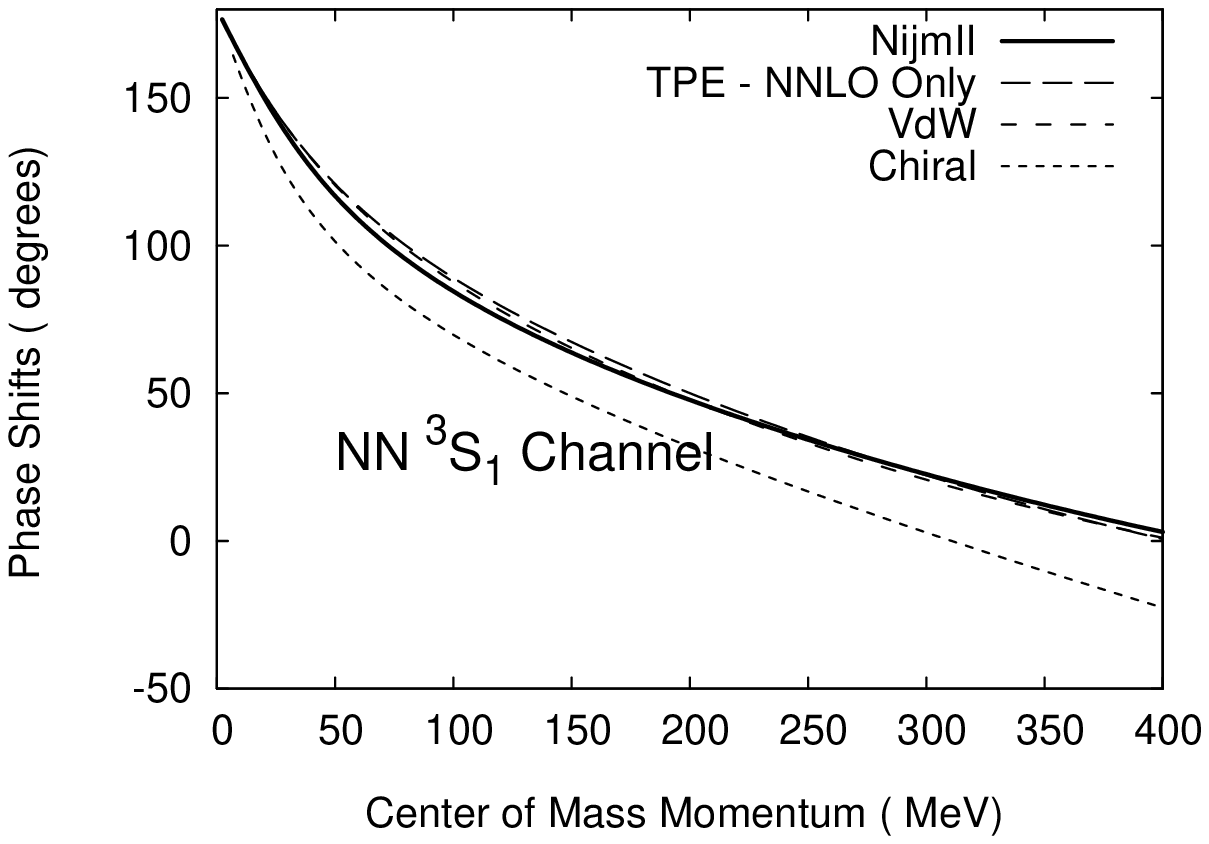,height=6.5cm,width=6.5cm}
\end{center}
\caption{Renormalized Eigen Phase shifts in the $^1S_0$ and
$^3S_1-^3D_1$ channel for the pure chiral Van der Waals $C_6/r^6$
potential (Vdw), and the pure NNLO terms compared to the
renormalized phase shifts with the same parameters from
Table~\ref{tab:table_vdw} for Set IV. We also compare to the Nijmegen
database~\cite{Stoks:1993tb}.}
\label{fig:vdw}
\end{figure*}

\begin{figure}[ttt]
\begin{center}
\epsfig{figure=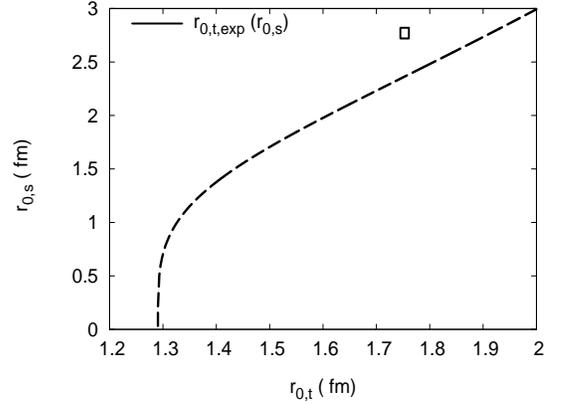,height=5.5cm,width=7.5cm}
\end{center}
\caption{Van der Waals correlation between the singlet and triplet
effective ranges using the experimental singlet and triplet scattering
lengths. The point represents the experimental values. }
\label{fig:r0s-r0t}
\end{figure}

For the triplet channel, the equation cannot be solved analytically,
and the effective range has a correction due to the D-wave (see
Eq.~(\ref{eq:r0_triplet}) ). Moreover, the scattering length is a factor
five times smaller than in the singlet case, so that we do not expect
in principle such a dramatic agreement. If we neglect the mixing with
the D-wave and take the $R_t = (M C_{6,^3S_1})^{1/4}$ of
Eq.~(\ref{eq:vdw_triplet}) we get 
\begin{eqnarray} 
r_{0,t} &=& 2.50174 - \frac{4.28983}{\alpha_{0,t}} +\frac{3.67797}{\alpha_{0,t}^2} \, \qquad ({\rm
Set \, I} ) \, , \nonumber \\  
r_{0,t} &=&  2.58537  - \frac{4.58143}{\alpha_{0,t}} + \frac{4.05928}{\alpha_{0,t}^2} \, \qquad ({\rm
Set \, II} ) \, , \nonumber \\  
r_{0,t} &=&  2.35089  - \frac{3.78809}{\alpha_{0,t}} + \frac{3.05196}{\alpha_{0,t}^2} \, \qquad ({\rm
Set \, III} ) \, , \nonumber \\  
r_{0,t} &=& 2.40877 - \frac{3.97691}{\alpha_{0,t}} +
\frac{3.28297}{\alpha_{0,t}^2} \, \qquad ({\rm
Set \, IV} ) \, , \nonumber \\ 
\end{eqnarray} 
which, using the triplet scattering length value, $\alpha_{0,t}=5.42 $
yields $r_{0,t} = 1.83,1.87, 1.75, 1.78 $ respectively, in remarkable
agreement with the experimental value. An estimate of the mixing
effect can be made by using the largest van der Waals eigen radius
$R_+ = (M C_{6,+})^{\frac14} $ obtained by diagonalizing the
interaction at short distances.  From Table~\ref{tab:table_vdw},
Eq.~(\ref{eq:r0_vdw}) and the experimental value of the scattering
length we get $r_0 = 2.00, 2.07,1.95, 2.05 {\rm fm}$ for Sets I,II,III
and IV respectively, accounting for about $85 \%$ percent of the full
value. Instead, perturbation theory for OPE,
Eq.~(\ref{eq:r0_ope_pert}) yields $r_0 = 0.62 {\rm fm}$, and full OPE
$r_0=1.64{\rm }$.  Actually, using the relation
\begin{eqnarray} 
 MC_{6,^1S_0} - M C_{6,^3S_1} &=& R_s^4 - R_t^4 \nonumber \\ &=&
 \frac{3 g^2}{64 \pi^2 f^4} (4 - 9 g^2 )
\end{eqnarray} 
we get an explicit correlation between $\alpha_{0,s}$, $r_{0,s}$,
$\alpha_{0,t}$ and $r_{0,t}$ regardless on the numerical values of the
chiral constants $c_3$ and $c_4$.  In the range of physical parameters
this looks like a linear correlation (see Fig.~\ref{fig:r0s-r0t})
between the singlet and triplet effective ranges. For $r_{0,t}=1.75 $
one gets $r_{0,s}=2.34$.

To check further the dominance of chiral Van der Waals interactions,
we plot in Fig.~\ref{fig:vdw} the phase shifts for a variety of
situations including the pure Van der Waals contributions, as well as
the contribution of the NNLO only, which reduces to the previous case
at short distances but decays exponentially as $ \sim e^{-2 mr }$ at
long distances. The plots confirm, again, our estimations
based on the pure Van der Waals potential of the effective range for
the $s-$waves, and this is the reason why the triplet $s-$wave is
better reproduced than the singlet case for Set IV. Obviously, by
adjusting the effective range changing the chiral parameters $c_3$ and
$c_4$ we could obtain a much better description of the data.

The results of this study show that the singularity of the chiral Van
der Waals force is not a feature that should be avoided, but instead
provides a very simple way to describe the scattering data for the
$s-$waves.

Finally, it is interesting to note, that central waves based on taking
the chiral limit of the potential are less accurately described than
the phase-shifts obtained from the pure Van der Waals contribution. In
this limit the singlet $^1S_0$ channel contains in addition to the Van
der Waals term a $1/r^5$ contribution stemming from the NLO TPE
contribution. The triplet $^3S_1-^3D_1$ TPE contribution has a similar
structure in addition to the OPE tensor $1/r^3$ singular short
distance contribution.

\section{The TPE Potential at NLO: A missing link ? }
\label{app:b} 

In the previous sections we analyzed the renormalization of the NNLO
potential. In this Section we analyze the NLO in the singlet $^1S_0$
and triplet $^3S_1-^3D_1$ channels and the problem that arises in the
latter.  We argue that similar trends are observed in finite cut-off
calculations. We also suggest several scenarios on how the problem may
be overcome.

\subsection{Convergence in the singlet $^1S_0$ channel} 
\label{sec:conv-1S0}

In the singlet $^1S_0$ channel the potential at short distances
behaves as~\cite{Kaiser:1997mw,Friar:1999sj,Rentmeester:1999vw}
\begin{eqnarray} 
U_{^1S_0} \to  \frac{M C_{5,^1S_0}}{r^5} \, , 
\end{eqnarray} 
where 
\begin{eqnarray}
M C_{5,^1S_0} &=& \frac{M(1 + 10 g^2 - 59 g^4) }{256 \pi^3 f^4} \, 
\end{eqnarray} 
The singlet coefficient is negative and, according to the discussion
in Sect.~\ref{sec:short}, one has an undetermined short distance phase
which can be fixed by using the scattering length as input. The
effective range in the singlet channel is given by
\begin{eqnarray}
r_0 = 2.122 - \frac{4.889}{\alpha_0} + \frac{5.499}{\alpha_0^2} \, \qquad ({\rm
NLO} ) \, , 
\label{eq:r0_NLO}
\end{eqnarray} 
which compared with the LO and NNLO results, Eq.~(\ref{eq:r0_todos}),
shows a convergence rate.  To show that this trend to convergence is
not fortuitous we display in Table~\ref{tab:convergence} the threshold
parameters of the effective range expansion $k \cot \delta =
-1/\alpha_0 + r_0 k^2 /2 + v_2 k^4 + v_3 k^6 + v_4 k^8 $ depending on
the terms kept in the expansion of the potential given by
Eq.~(\ref{eq:pot_chpt}). As we see there is a clear trend to
convergence, although the higher order threshold parameters display a
slower convergence rate since they are increasingly
sensitive to the shorter range regions. This trend is confirmed in
Fig.~\ref{fig:LO-NLO-NNLO} for the phase shift. Obviously, there is
scale separation in the singlet potential, and higher order potentials
although more singular at the origin yield contributions in the right
direction.

\begin{table}[bbb]
\caption{\label{tab:convergence} Convergence of the threshold
parameters of the effective range expansion $k \cot \delta =
-1/\alpha_0 + r_0 k^2 /2 + v_2 k^4 + v_3 k^6 + v_4 k^8 $ in the
singlet $^1S_0$ channel depending on the successive inclusion of terms
in the potential $U=U_{\rm LO} + U_{\rm NLO}+ U_{\rm NNLO} + \dots $.
LO means LO alone (and taking $g_{\pi NN} = 13.083 $ and $g_A=1.26$),
NLO means LO+NLO and so on. The only input is the scattering length
$\alpha_0$ besides the potential parameters. For the NNLO case we use
Set IV for the chiral constants $c_1$, $c_3$ and $c_4$ given in
Table~\ref{tab:table_vdw}. }
\begin{ruledtabular}
\begin{tabular}{|c|c|c|c|c|c|}
\hline $^1S_0 $ & LO & NLO  &  NNLO  & Exp.  &  Nijm II \\ \hline
$\alpha_0( {\rm fm}) $ & Input & Input  & Input  & -23.74(2) & -23.73 \\ 
$r_0( {\rm fm}) $ & 1.44 & 2.29  & 2.86  & 2.77(5) & 2.67 \\ 
$v_2 ( {\rm fm}^3) $ & -2.11 & -1.02  & -0.36  & -- & -0.48 \\ 
$v_3 ( {\rm fm}^5) $ & 9.48 & 6.09  & 4.86  & -- & 3.96 \\ 
$v_4 ( {\rm fm}^7) $& -51.31 & -35.16  & -27.64  & -- & -19.88 
\end{tabular}
\end{ruledtabular}
\end{table}

\subsection{The problem in the triplet $^3S_1-^3D_1$ channel}
\label{sec:triplet_problem} 

The triplet $^3S_1-^3D_1$ potential at short
distances has the behaviour~\cite{Kaiser:1997mw,Friar:1999sj,Rentmeester:1999vw}
\begin{eqnarray} 
U_{^3S_1}(r) &\to& \frac{M C_{5,^3S_1}}{ r^5 } \, , \nonumber \\ 
U_{E_1}(r) &\to& \frac{M C_{5,E_1}}{ r^5 } \, , \nonumber \\ 
U_{^3D_1}(r) &\to& \frac{M C_{5,^3D_1}}{ r^5 } \, ,  
\end{eqnarray} 
where 
\begin{eqnarray} 
M C_{5,^3S_1} &=& \frac{3 M( -1 -10 g^2 + 27 g^4) }{256
\pi^3 f^4} \, ,  \nonumber \\ M C_{5,E_1} &=& -\frac{15 M g^4 }{64 \sqrt{2}
\pi^3 f^4} \nonumber \\ M C_{5,^3D_1} &=& \frac{3 M( -1 -10 g^2 +37
g^4) }{256 \pi^3 f^4} \, ,  
\end{eqnarray} 

\begin{figure*}[]
\begin{center}
\epsfig{figure=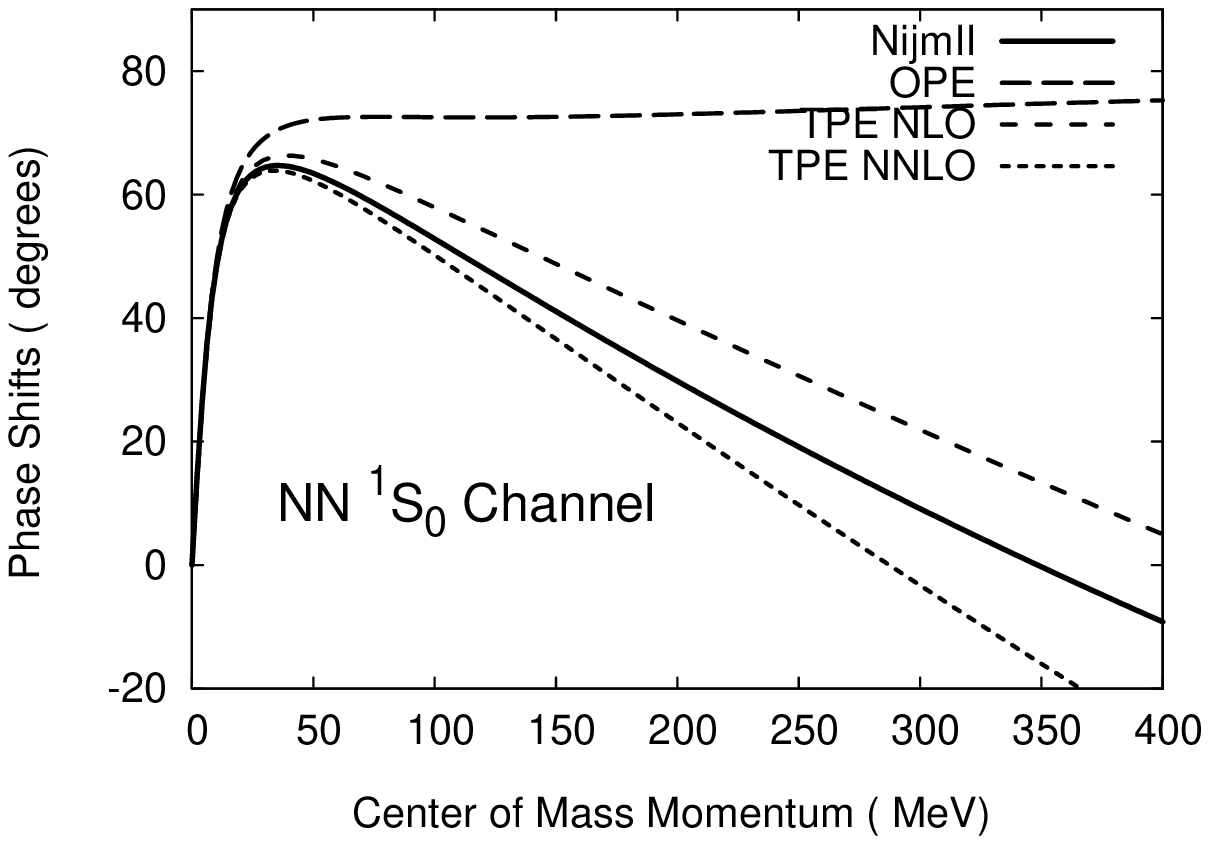,height=6.5cm,width=6.5cm}
\epsfig{figure=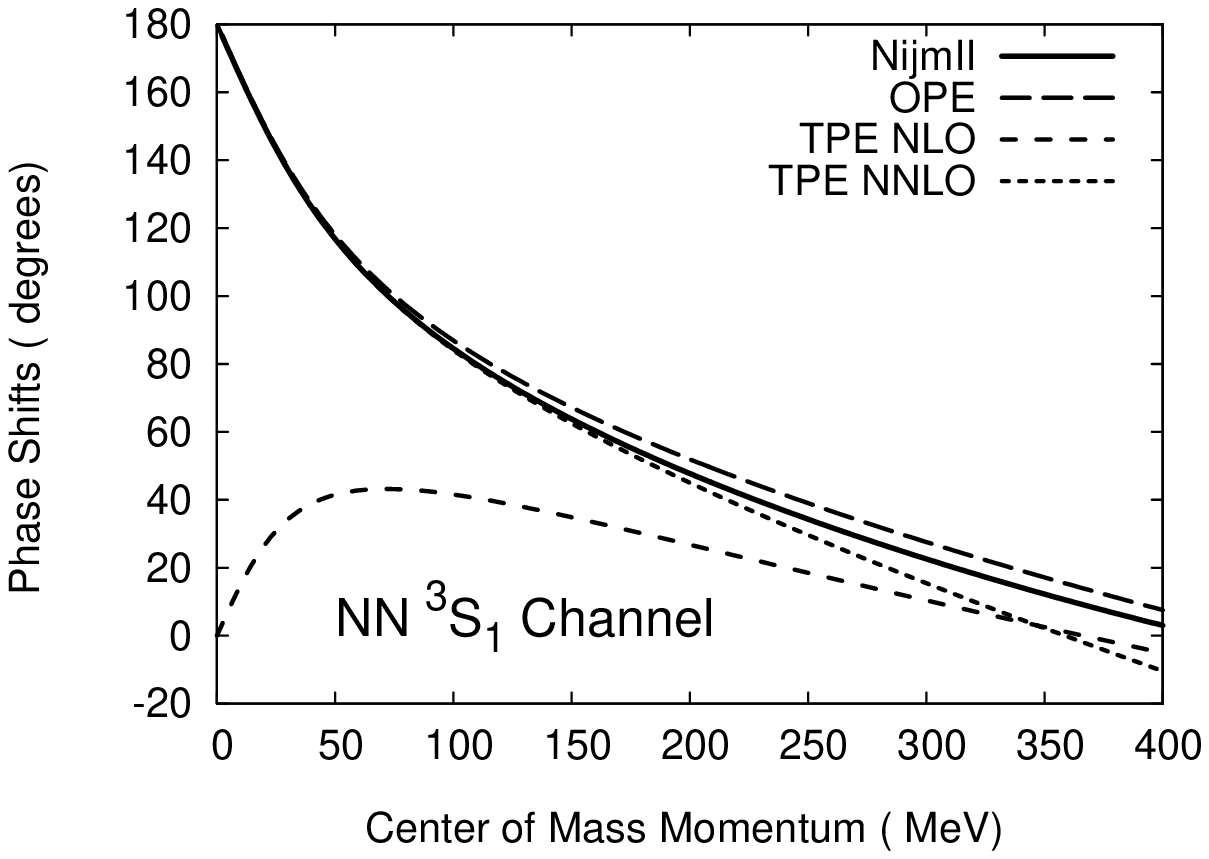,height=6.5cm,width=6.5cm}\\ 
\epsfig{figure=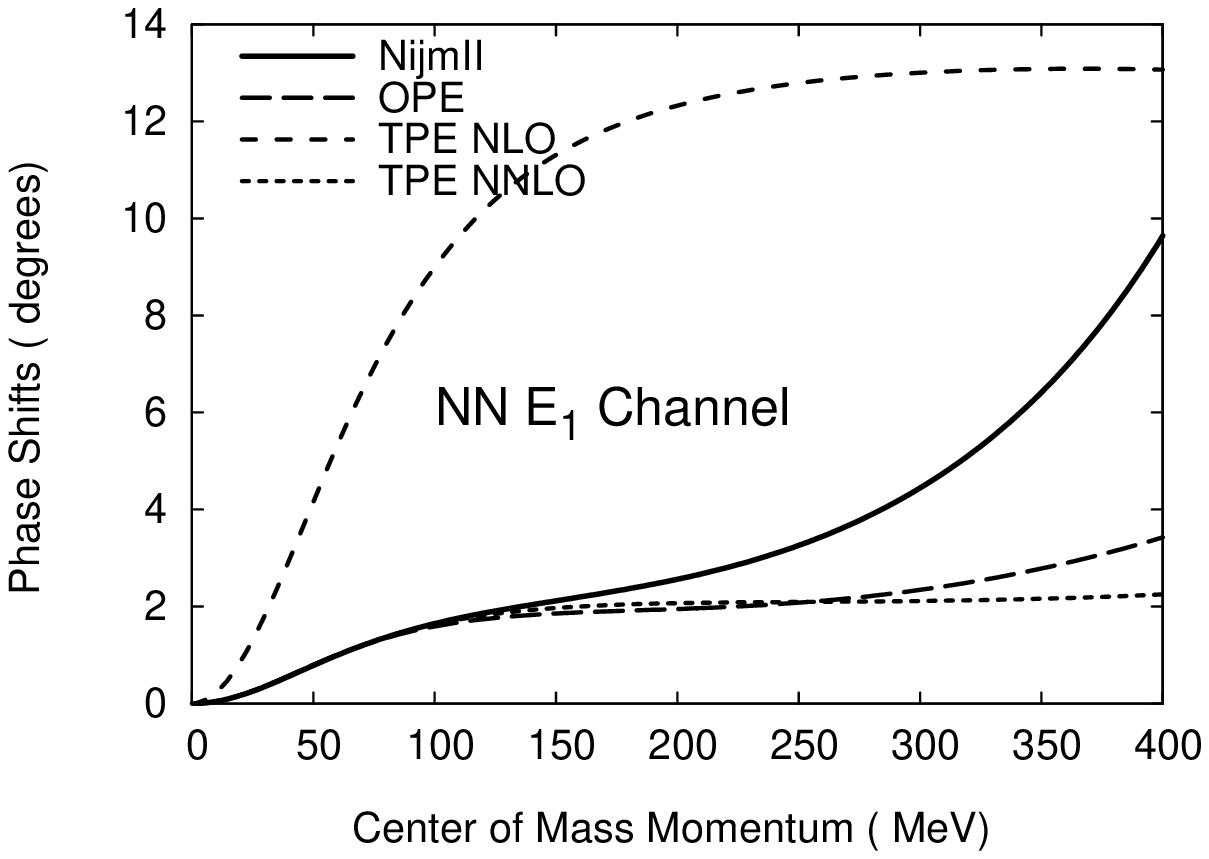,height=6.5cm,width=6.5cm}
\epsfig{figure=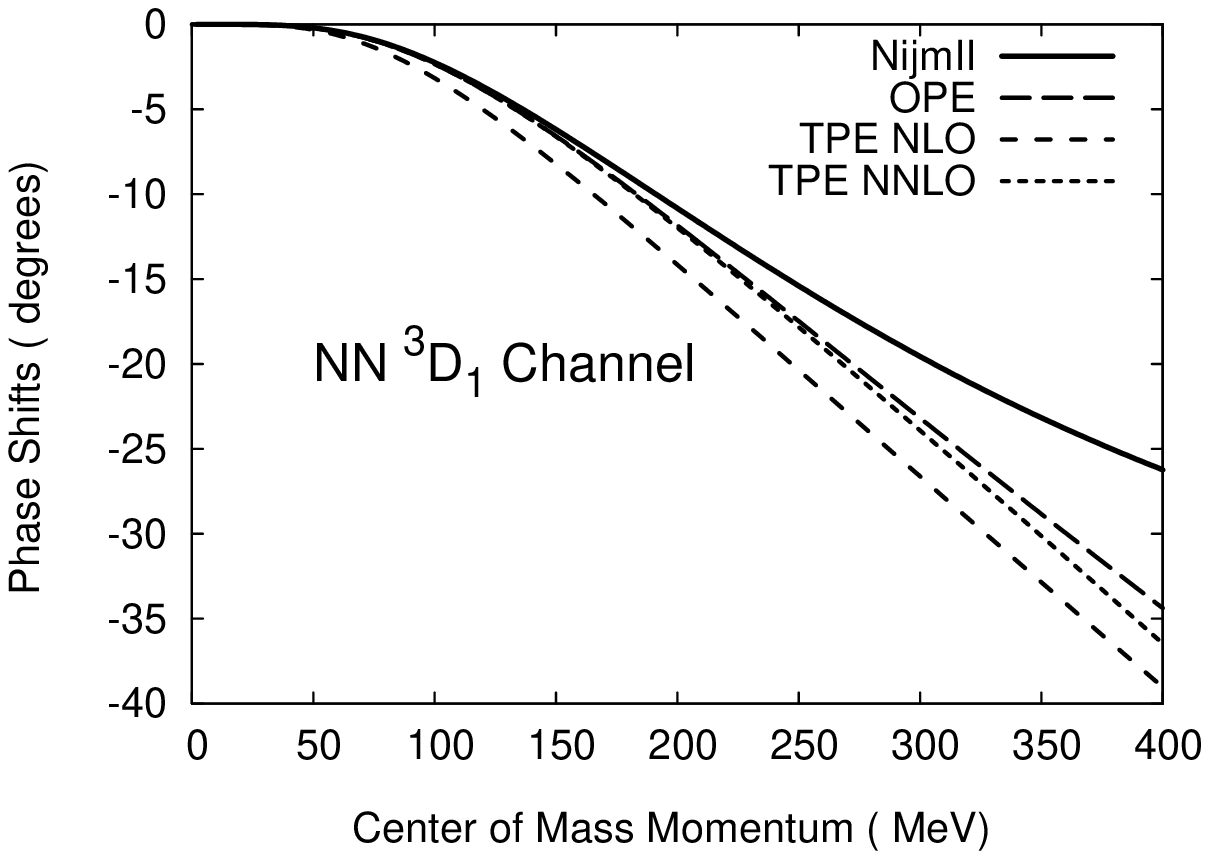,height=6.5cm,width=6.5cm}
\end{center}
\caption{Renormalized Eigen phase shifts at LO, NLO and NNLO as a
function of the CM np momentum $k$ in the singlet $^1S_0$ and triplet
$^3S_1-^3D_1$ channels compared to the Nijmegen
results~\cite{Stoks:1993tb} for different parameter sets.}
\label{fig:LO-NLO-NNLO}
\end{figure*}

On the other hand, the diagonalized
triplet coefficients are
\begin{eqnarray}
M C_{5,+} &=& \frac{3 M(-1 - 10 g^2 +17 g^4) }{256 \pi^3 f^4} \, ,\nonumber \\ 
M C_{5,-} &=& \frac{3 M( -1 -10 g^2 + 47 g^4) }{256 \pi^3 f^4} \, ,
\end{eqnarray} 
and the mixing angle is given by $\tan \theta = \sqrt{2}$, differing
by $-\pi$ as compared to the OPE
case~\cite{PavonValderrama:2005gu}. For $ 0.5356 < g < 0.8217 $ one
would have an attractive-repulsive situation ( see
Sect.~\ref{sec:short}), as in the OPE
case~\cite{PavonValderrama:2005gu} and in such a case one could take
either the deuteron binding energy or the $^3S_1$ scattering
length. However, for the physical value $g=1.26$ one has two short
distance repulsive eigenchannels, and hence one must take the
exponentially decaying regular solutions at the origin.  Let us remind
the according to Sect.~\ref{sec:short} {\it finite renormalized
results can only be obtained by precisely choosing the regular
solution at the origin}.  In this case there are no short distance
phases, and the scattering lengths, as well as the phase shifts are
completely determined from the potential. The (finite) renormalized
results are depicted in Fig.~\ref{fig:LO-NLO-NNLO}. As we see, the
singlet $^1S_0$ phase-shift shows a very reasonable trend, since NLO
and improves on the LO, and it is improved by the NNLO potential. We
remind that in the three cases the scattering length is exactly the
same. However, not completely unexpectedly, the triplet channel
results worsen the LO ones. In the next subsection we show that if,
demanding the standard Weinberg counting requires the irregular
solution at the origin, hence yielding to divergent renormalized
results.

\subsection{Finite cut-offs and the Weinberg counting} 
\label{sec:finite} 

The special status of the NLO calculation as compared to the LO and
NNLO ones has been recognized in previous studies in momentum
space~\cite{Epelbaum:1999dj} where regularization was implemented by
using a sharp cut-off $\Lambda$. As noted by these authors, the
allowed cut-off variations at NLO are {\it smaller} ($\sim 380-600$
MeV ) than at LO ($\sim 700-800$ MeV) or NNLO ($\sim 800-1000$) but
the reasons have not been made clear. Let us focus on the triplet
$^3S_1 - ^3D_1 $ channel.  Within our coordinate space renormalization
scheme this trend can be easily understood. At LO one fixes only one
parameter, say $\alpha_0$, and because one has attractive and
repulsive potentials at short distances the system will naturally be
driven into the exponential regular solution at the origin. Obviously,
if one would fix some other parameter independently, say $r_0$ (or
equivalently using a counterterm $C_2$, and not the one predicted by
the regular solutions, one would be driven instead to the irregular
solution, not allowing to remove the cut-off in practice. In such a
situation one would be forced to keep the cut-off finite at the scale
where the repulsive core sets in.  However, at LO the Weinberg power
counting does not allow to fix this additional parameter and one can
comfortably reach higher cut-off values. On the contrary, at NLO one
has two repulsive eigenpotentials and one cannot fix any low energy
parameter arbitrarily. Otherwise one would be attracted to the
irregular solutions at short distances. On the other hand, they are
attractive at long distances, so that one would expect a stability
region where the potential becomes flat before turning into a
repulsive core in both eigenchannels. This is exactly what one
observes in the NLO calculation of Ref.~\cite{Epelbaum:1999dj}. The
occurrence of such a plateau is to some extent fortuitous since it is
associated to the critical points of the potential, and not with some
a priori estimate on the validity range of the NLO potential. Finally,
in the NNLO calculation because both eigenpotentials have an
attractive character one can again increase the cut-off since there
are no irregular solutions in the problem where one can be attracted
to. It is very rewarding that our coordinate space analysis of short
distance singularities anticipates when these features can be
expected.  On the other hand this does not imply that finite cut-off
calculations are necessarily wrong, simply that the observed features
when the cut-off approaches the limit can be understood.

\begin{figure*}[]
\begin{center}
\epsfig{figure=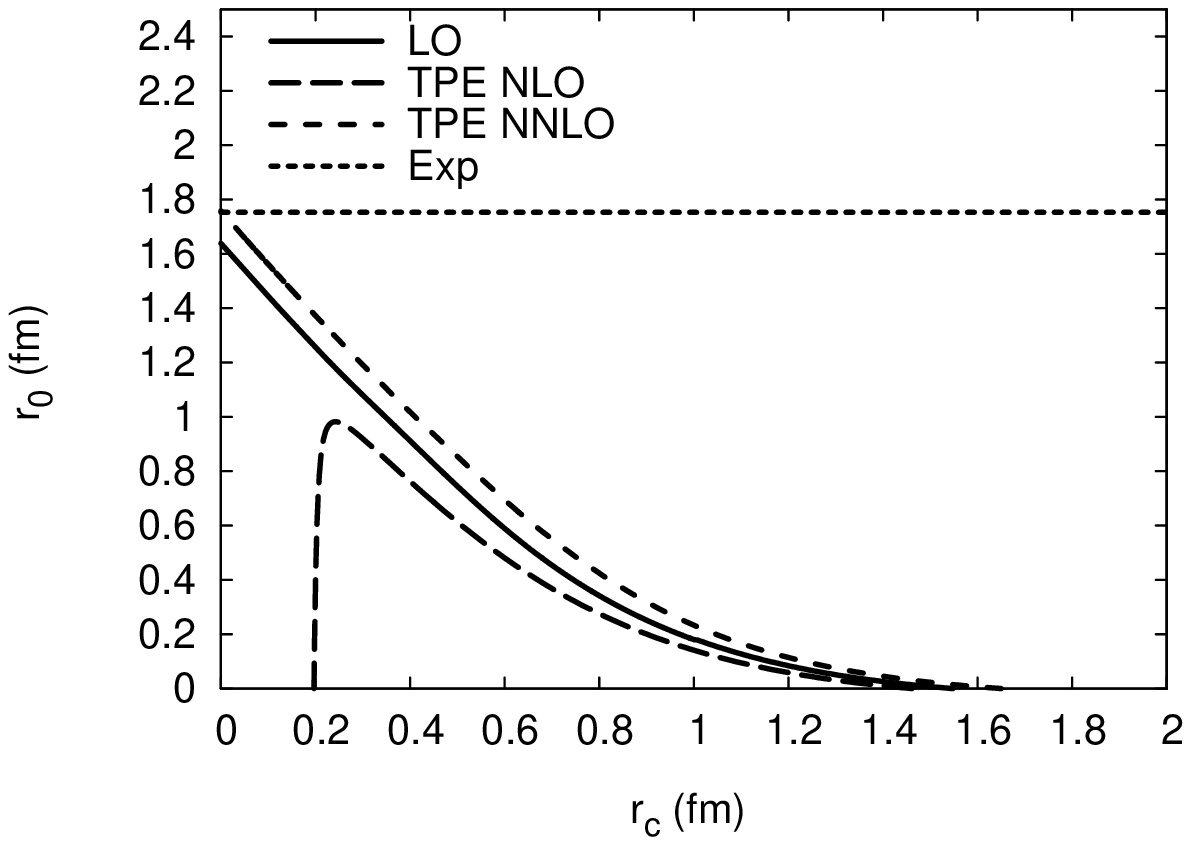,height=5.5cm,width=6.5cm}  
\epsfig{figure=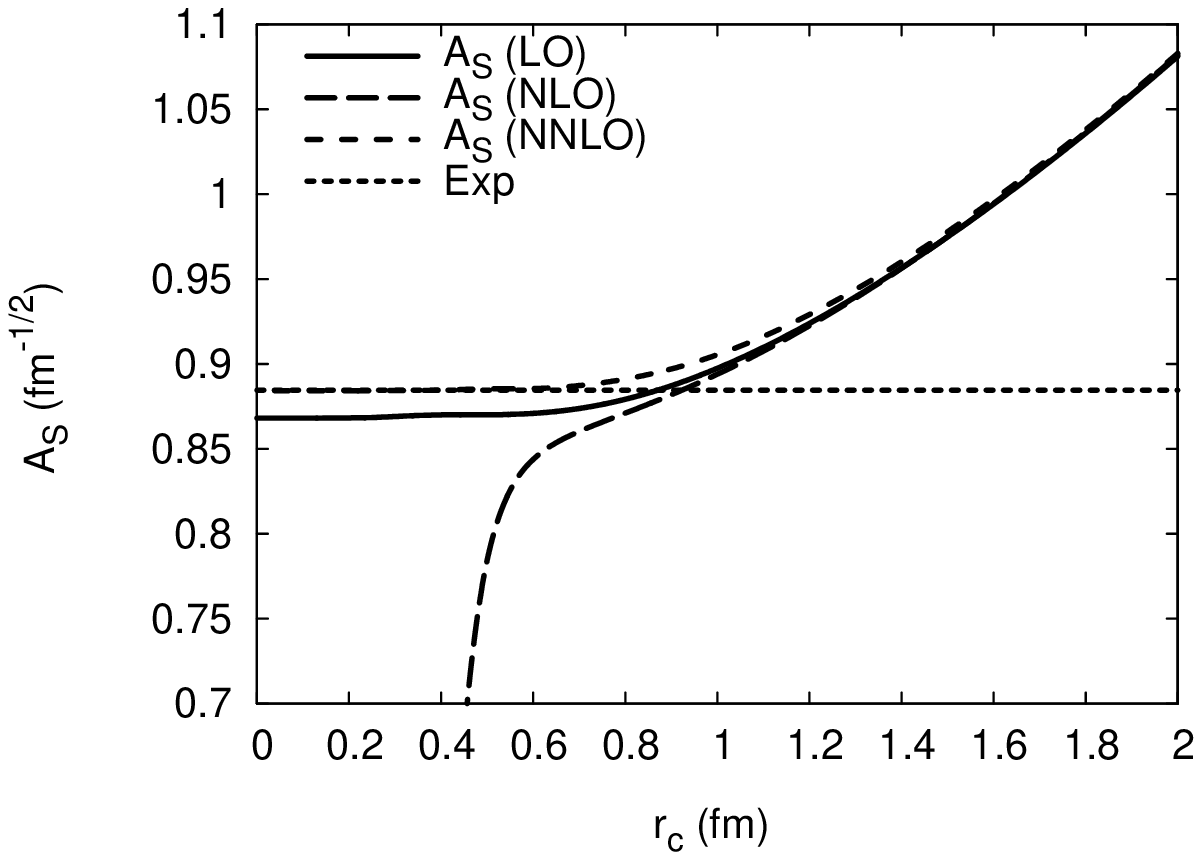,height=5.5cm,width=6.5cm} 
\end{center}
\caption{Cut-off dependence of the effective range, $r_0$ (in fm) and
the deuteron wave function renormalization, $A_S$ (in ${\rm
fm}^{-1/2}$), in the triplet $^3S_1-^3D_1$ channel at LO, NLO and NNLO
in the Weinberg counting. At LO we fix the deuteron binding energy
(one counterterm), at NLO and NNLO we fix the deuteron biding energy,
the asymptotic $D/S$-mixing, $\eta$ and the scattering length
$\alpha_0$ (three counterterms). We use Set IV of chiral coupling
constants.}
\label{fig:w_counting}
\end{figure*}

The previous discussion can be illustrated in our approach by looking
at the short distance cut-off dependence of the effective range, $r_0$
(in fm) and the deuteron wave function renormalization, $A_S$ (in
${\rm fm}^{-1/2}$), in the triplet $^3S_1-^3D_1$ channel at LO, NLO
and NNLO in the standard Weinberg counting as presented in
Fig.~\ref{fig:w_counting}. As described in
Sect.~\ref{sec:power_counting} any counterterms can be mapped into a
given renormalization condition. Once these conditions are fixed we
can ask whether other properties are finite or not. Thus, at LO we fix
the deuteron binding energy (one counterterm), at NLO and NNLO we fix
the deuteron biding energy, the asymptotic $D/S$-mixing, $\eta$, and
the scattering length $\alpha_0$ (three counterterms). In all cases it
is clear that by lowering the cut-off at LO and NNLO of the
approximation one nicely approaches the experimental values. This
raises immediately the question whether there is a given value of the
cut-off where NLO improves over LO. As we see such a region does not
exist.  In addition, although there is a nice and clear trend in both
LO and NNLO for distances below $0.5 {\rm fm}$ for $A_S $ down to the
origin, this is not so at NLO.  So, in this case it is not true that
low energy properties are independent on short distance details, in
contrast to the standard EFT wisdom. Moreover,
Fig.~\ref{fig:w_counting} shows explicitly the conflict between the
Weinberg counting and the remotion of the cut-off at NLO because $r_0
$ and $A_S $ diverge due to the onset of the irregular solution, as
anticipated in our study of short distance solutions (see
Sect.~\ref{sec:short}). We have checked that this is a general feature
on both deuteron and scattering properties. On the contrary, LO and
NNLO have a rather smooth limit because in these two cases Weinberg
power counting on the short distance counterterms turns out to be
compatible with the choice of the regular solution at the
origin. Thus, {\it in the $^3S_1-^3D_1$ channel the renormalized
solution at NLO in the Weinberg counting is divergent while LO and
NNLO are convergent}. In conclusion, the present analysis shows in a
somewhat complementary manner as done in
Sect.~\ref{sec:triplet_problem} that indeed the NLO is problematic, at
least non-perturbatively. In Sect.~\ref{sec:pert} we will see that the
problem is not solved if the NLO contribution is computed within a
perturbative framework using the exact OPE-distorted wave basis as a
lowest order approximation.

\subsection{The role of relativity and the $\Delta$ resonance in the renormalization problem} 
\label{sec:delta}

The requirement of renormalizability may be regarded as a radical
step, and renormalized LO calculations demand violating dimensional
power counting on the counterterms~\cite{Nogga:2005hy} in non central
waves such as $^3P_0$, due to an attractive $1/r^3$ singularity. To
reach a finite limit the authors of Ref.~\cite{Nogga:2005hy} propose
to promote counterterms which in Weinberg's power counting are of
higher order. However, in their proposal it is intriguing that they
choose to promote just one counterterm in coupled channels, while they
could have used a coupled channel counterterm, i.e. three counterterms
in total. In the boundary condition approach we know from the start
how many independent parameters must be {\it exactly} taken to reach a
finite and unique limit, the reference to power counting is only
specified at the level of the potential. Note that the power counting
in the potential fixes its short attractive-repulsive singular
character, and this is the origin of the conflict of assuming an {\it
a priori} power counting for the counterterms. Finiteness requires
that some forbidden counterterms must be allowed
(promoted)~\cite{Nogga:2005hy} but also that some allowed counterterms
must be forbidden (demoted).  In such a framework, our NLO
calculations in the $^3S_1-^3D_1 $ channel lead to finite but
nonsensical results due to the repulsive-repulsive $1/r^5$ singularity
(See Sect.~~\ref{app:b}). On the other hand, if one fixes as required
by the power counting the scattering length, the limit does not exist
because one is driven to the exponentially diverging solution at the
origin (For instance, Eq.~(\ref{eq:r0_triplet}) gives $r_0 \to -\infty
$). How then can we reconcile finiteness with fixing of the parameters
?. As we pointed out already, the singular short distance behaviour of
the chiral potential is in fact a long distance feature which changes
dramatically when changing the long distance physics. Actually, one
may reverse the argument and use renormalizability as a selective
criterium for admissible long distance potentials. In the following we
want to provide at least two possible scenarios how this might happen,
i.e. , how modifying the potential at long distances by introducing
physically relevant information the short distance behaviour of the
potential changes.

In the first place, the chiral potential, Eq.~(\ref{eq:pot_chpt}), was
derived in the heavy baryon expansion. The short distance character
may change when not taking such a limit since the combination $M r$
does make the order of limits ambiguous. A proper treatment of
relativistic effects requires inclusion of antinucleons in loops, and
a satisfactory EFT treatment of relativistic effects remains a
challenging open problem because the standard crossing vs. unitarity
non-perturbative divorce. On top of this one should use a satisfactory
relativistic two body equation, which necessarily makes the problem
fully non-local in coordinate space. Nevertheless, there exist
``relativistic'' potentials where {\it some} of the terms of higher
power in 1/M than the TPE obtained in heavy-baryon ChPT are
kept~\cite{Higa:2003jk,Higa:2003sz,Higa:2004cr} which have $1/r^7$ Van
der Waals short distance behaviour with attractive-repulsive eigen
potentials~\cite{Higa2005} meaning that as in the OPE case one has one
free parameter. A calculation using these incomplete ``relativistic''
potentials will be presented elsewhere~\cite{HPR2005}.

A second scenario is related to the role played by the $\Delta$
resonance~\footnote{We thank D. Phillips for drawing our attention to
this point.} not included in the present analysis.  As pointed out in
Ref.~\cite{Kaiser:1998wa}, the $\Delta$ provides the bulk of the
chiral constants, yielding $-c_3= 2c_4 = g_A^2 / 2 \Delta$, with
$\Delta = 293 {\rm MeV}$ the nucleon-delta mass splitting, yielding
$c_3 = -2.7 {\rm GeV}^1 $ and $c_4 = 1.35 {\rm GeV}^{-1} $. The
difference to the parameters of Ref.~\cite{Entem:2002sf} may be due to
some other resonances. On the other hand, in terms of scales one has
$\Delta \sim 2 m_\pi $, which might be regarded as a small
parameter. This obviously does not mean that $\Delta$ vanishes in the
chiral limit. In the standard chiral counting of the potential,
Eq.~(\ref{eq:pot_chpt}), the combinations $\bar c_1 = M c_1 $, $\bar
c_3 = M c_3 $ and $ \bar c_4 = M c_4 $ are considered to be zeroth
order, but according to the previous argument they could be regarded
to be enhanced by one negative power. Thus the nominally NNLO terms
containing $c_3$ and $c_4$ might become NLO contributions, and hence
changing the repulsive-repulsive $1/r^5$ singularity into an
attractive-attractive $1/r^6$ one. On the other hand, the $c_3$ and
$c_4$ contributions of the standard NNLO dominate the short distance
Van der Waals contributions. Actually, much of the NNLO potential is
built from these terms all over the range. According to this reasoning
our NNLO calculation may be closer to a NLO one where the $N\Delta$
splitting is regarded as small parameter. In fact, taking NLO+$\Delta$
with $-c_3= 2c_4 = g_A^2 / 2 \Delta$ and $\eta=0.0256$ one gets $A_S =
0.8869 {\rm fm}^{-1/2}$, $Q_D = 0.2762 {\rm fm}^2$, $r_m=1.9726 {\rm
fm}$ and $P_d=0.06$ in overall agreement with
Table~\ref{tab:table3}. It would be rather interesting to look for
further consequences of this $\Delta$-counting at higher orders. The
importance of the $\Delta$ in the NN problem has been stressed in
several works already on power counting
grounds~\cite{VanKolck:1993ee,Ordonez:1995rz,Epelbaum:1999dj,Pandharipande:2005sx}
but the crucial role played on the renormalization problem, i.e., the
fact that the cut-off can be completely removed has not been
recognized. Our discussion suggests that the momentum space cut-off
could also be confortably removed in this $\Delta-$counting, unlike
the delta-less NLO.

The two possible scenarios outlined above do not prove that the
requirement of renormalizability is necessarily right, but suggest
that looking into the short distance singular behaviour of long
distance chiral potentials together with the mathematical requirement
of finiteness may provide a significant physical insight into the NN
problem. In the language of Ref.~\cite{Nogga:2005hy} where promotion
of counterterms on the basis of the renormalizability requirement has
been stressed, we are perhaps led also to the demotion of counterterms
(like for relativistic potentials), or alternatively the promotion of
terms in the potential (like in the $\Delta$ counting described
above).

\subsection{Van der Waals forces, the molecular analogy  and the chiral quark model} 

The previous arguments show that it is possible to change the
attractive/repulsive character of the potential at short distances by
organizing the calculation of the potential in a different manner, but
does not give a clue on why this actually happens. Remarkably, the
analogy with atomic neutral systems subjected to Van der Waals forces
illustrated in Sect.~\ref{app:a} goes further and provides valuable
insight into the problem. In low energy molecular physics where one
works in a Born-Oppenheimer approximation, all atomic constituents,
electrons and nuclei interact through the Coulomb force arising from
one photon exchange. At long distances between distant electrons the
potential is a dipole-dipole interaction
\begin{eqnarray}
V_{\rm dip} (R) = e^2 \sum_{A,B} \left[ \frac{\vec r_A \cdot \vec
  r_B}{R^3} - 3 \frac{( \vec r_A \cdot \vec R ) (\vec r_B \cdot \vec R
  )}{R^5} \right]
\end{eqnarray} 
where the sum runs over electrons belonging to different atoms. In
second order perturbation theory the atom-atom energy at a separation
distance $R$ reads,
\begin{eqnarray} 
V_{AA} &=& \langle AA | V_{\rm dip} | A A \rangle \nonumber \\ &+& \sum_{AA \neq A^* A^*}
\frac{|\langle AA | V_{\rm dip} | A^* A^* \rangle|^2}{E_{AA}-E_{A^*
A^*}}+ \dots 
\end{eqnarray} 
where $|AA \rangle $ and $|A^* A^* \rangle $ is the electron wave
function corresponding to a pair of separated clusters in their atomic
ground state and excited states respectively. The first order
contribution vanishes for atoms with no permanent dipole moment.  The
mutual electric polarization causes the Van der Waals interaction
between the two atoms, $C_6/R^6$ and because it is second order
perturbation theory it is obvious that the $C_6 $ contribution to the
potential will always be attractive. However, it is not clear that
higher order terms would always be attractive.  It is remarkable that
the theorem of Thirring and Lieb~\cite{LiebThirring86} establishes
that the Coulomb force between constituents implies all terms in the
expansion being attractive, without appealing to the
dipole-approximation. Thus, according to this result the long distance
force will always be singular and attractive at short distances, and
that is exactly what one needs.  In such a situation making a long
distance expansion of the potential, $U = -R_6^4/r^6 - R_8^6 / r^8 +
\dots $ and computing the scattering phase shifts by fixing always the
same scattering length, along the lines pursued in this paper, makes
much sense.  Moreover, one expects the results for the phase shifts
to be convergent if there is scale separation between the
corresponding Van der Waals radii $R_6 \gg R_8 \gg \dots $. Our
experience with several atomic systems confirms these
expectations~\cite{Alvaro}.

The argument in the NN system is a straightforward generalization of
the molecular system above. It is well known that there are no colour
hidden states between colour neutral systems, so that at long
distances one may assume only exchange of colourless objects. The
longest range object will be the pion, and the mutual (chiral)
polarizability will cause attraction between the nucleons, exactly in
the same way as for atom-atom interactions. If we use as an example
the chiral quark model, assume for simplicity non-relativistic
constituent quarks one obtains the OPE for quarks.  To second order
perturbation theory we get the NN potential in the Born-Oppenheimer
approximation
\begin{eqnarray}
V_{NN} &=& \langle NN | V_{\rm OPE} | NN \rangle \nonumber \\  &+& \sum_{HH \neq NN
}\frac{ |\langle NN | V_{\rm OPE} | HH \rangle |^2}{ E_{NN}-E_{HH} }+
\dots 
\end{eqnarray} 
where $V_{NN}$ represents the potential in the $NN$ operator basis.
This yields {\it exactly} when $H H=N \Delta$ the results found in
Ref.~\cite{VanKolck:1993ee,Ordonez:1995rz,Kaiser:1998wa} and naturally
explains why the contribution from one $\Delta$ intermediate state is
attractive at short distances. Although this analogy with molecular
systems is very suggestive, the generalization to all orders along the
lines of the Lieb-Thirring theorem within a QCD context remains at
present an optimistic speculation.

\section{Renormalized Perturbation Theory versus non-integer power counting} 
\label{sec:pert} 

\subsection{Perturbations on boundary conditions} 

In all our calculations we have taken a long distance potential
calculated perturbatively, and scattering amplitudes have been
computed non-perturbatively by fully iterating a potential computed in
perturbation theory, as initially suggested by
Weinberg~\cite{Weinberg:1990rz}. We will call this form of solution
non-perturbative for brevity. This requires a non-perturbative
treatment of the renormalization problem, which naturally implies that
the short distance renormalization conditions (or counterterms) are
determined by the most singular contribution of the long distance
potential at the origin.  For the NN chiral potential it turns out
that the higher the order the more singular the potential. As a
consequence we have seen in Sect.~\ref{sec:finite} that for instance
Weinberg counting at NLO in the $^3S_1-^3D_1$ channel is incompatible
with renormalization and finiteness due to the short distance
repulsive character of the NLO potential. Although naively this looks
counterintuitive, it is important to realize that there are also
situations, like LO and NNLO, where the regularity condition of the
wave function conspires against the singularity so that the net effect
is well behaved in the scattering amplitudes and deuteron properties.
Our results in Sect.~\ref{sec:singlet} and \ref{sec:triplet}, suggest
that the pattern obtained when comparing LO and NNLO looks quite
converging numerically, although there appears to be no way of making
an {\it a priori} estimate of the corrections.

\begin{figure}[]
\begin{center}
\epsfig{figure=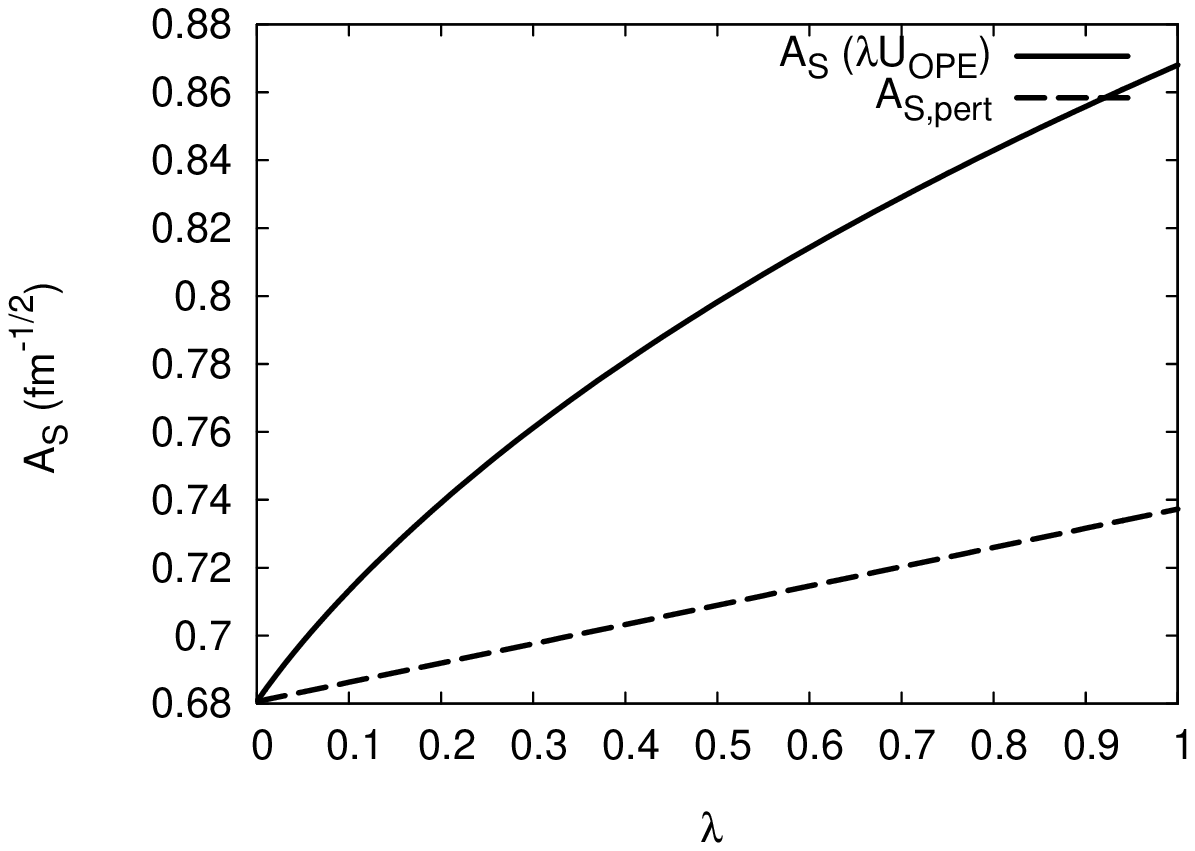,height=5cm,width=6.5cm} \\  
\epsfig{figure=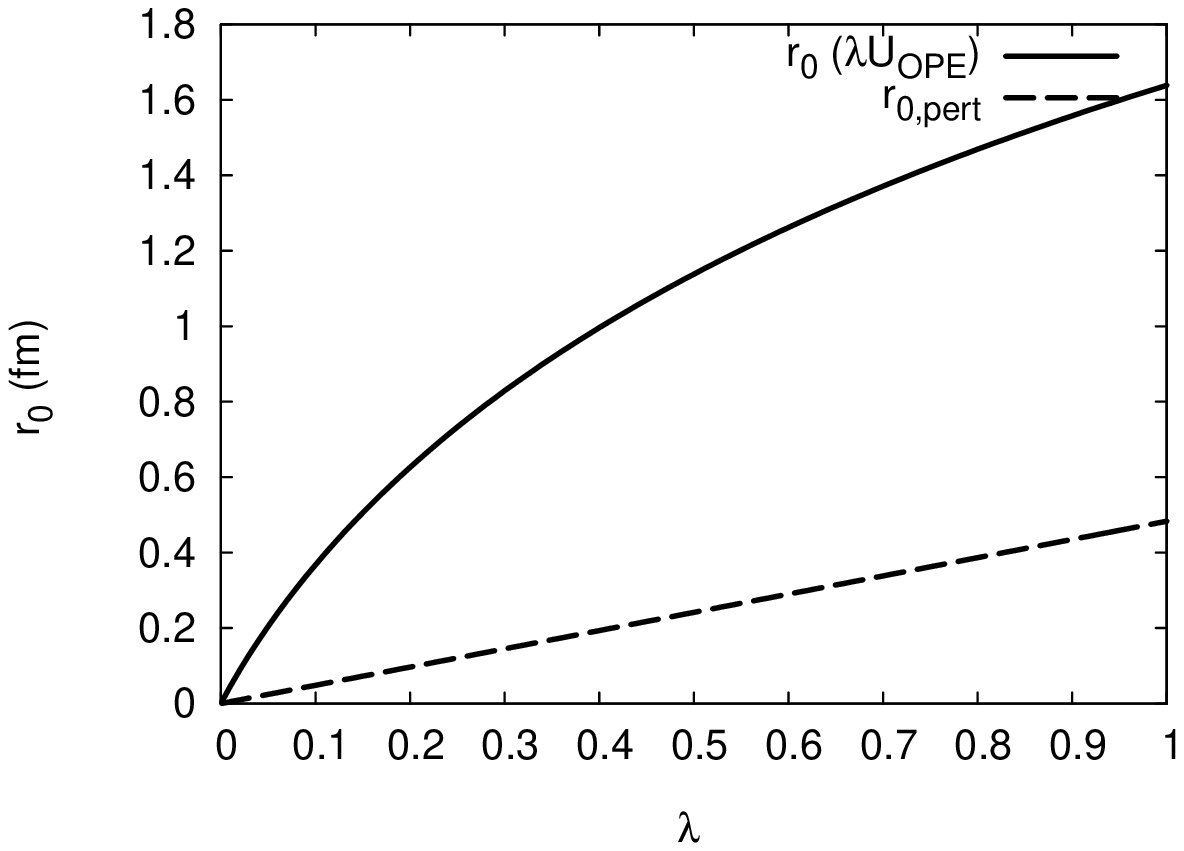,height=5cm,width=6.5cm}
\end{center}
\caption{Dependence of the s-wave function normalization $ A_S
(\lambda)$ in the deuteron and the effective range $r_0 (\lambda) $ in
the $^3S_1-^3D_1$ channel when one scales the OPE potential $ U_{\rm
LO} \to \lambda U_{\rm LO } $. In all cases we fix the deuteron
binding energy to its experimental value.}
\label{fig:Asr0-3S1.lambda}
\end{figure}

Perturbative treatments might circumvent this difficulty since they
have the indubitable benefit of allowing an {\it a priori} estimate of
the systematic error via dimensional power counting. This causes no
problem in the calculation of the long distance potential. However, we
anticipate already that singular potentials are indeed singular
perturbations, and power counting may not work as one naively expects
for the full amplitudes. Kaplan, Savage and
Wise~~\cite{Kaplan:1998we,Kaplan:1998sz} suggested such a perturbative
scheme some years ago, where the lowest order approximation was a
contact theory and OPE and higher order corrections could be computed
in perturbation theory. This is equivalent to consider $m M /f^2 $ to
be first order and $m^2 /f^2 $ second order, so that a calculation
involving the chiral constants would be N$^3$LO in that
counting. Unfortunately, the expansion turned out to be non-converging
at NNLO~\cite{Fleming:1999ee}. In our coordinate space formulation,
this approach corresponds to assume for the $S$ waves a boundary
condition fixing the scattering length
$\alpha_0$~\cite{PavonValderrama:2005gu} (See Appendix A of that work)
and making long distance potential perturbations. In our previous work
we verified that perturbation theory could only account for a
contribution to the deuteron and $^3S_1-^3D_1$ scattering observables
at first order. Unfortunately, the second order was divergent, while
non-perturbatively, i.e., exactly solving the Schr\"odinger equation
for the OPE potential, the results were not only finite but also
numerically quite close to experiment. This deserves some explanation.
In Fig.~\ref{fig:Asr0-3S1.lambda} we show the results in the deuteron
channel when we scale the OPE potential $ U_{\rm OPE} \to \lambda
U_{\rm OPE } $ for the s-wave function normalization $ A_S (\lambda)$
and the effective range $r_0 (\lambda) $ as a function of the scaling
parameter $\lambda$ by keeping the deuteron binding energy fixed to
its experimental value.  The non-perturbative result is compared to
the first order perturbation theory used in
Ref.~\cite{PavonValderrama:2005gu}. Clearly, perturbation theory fails
even for weak coupling. The experimental value for $r_0$ could be
obtained by adding a counterterm $C_2$ as done by Kaplan, Savage and
Wise~~\cite{Kaplan:1998we,Kaplan:1998sz}. A non-vanishing $C_2$ not
only violates the orthogonality of the zero energy and deuteron wave
functions for a long distance local potential but also introduces a
new parameter, reducing the predictive power. Moreover, the
non-perturbative inclusion of this $C_2$ counterterm with the OPE
potential yields divergent results (see the discussion in
Sect.~\ref{sec:finite}). In fact, much of the strength of $C_2 $ is
naturally provided by the short distance $1/r^3$ singularity of the
OPE potential.

\subsection{Perturbations on the OPE potential} 

Recently, Nogga,Timmermans and van Kolck (NTvK) have
suggested~\cite{Nogga:2005hy} treating the OPE effects
non-perturbatively, i.e. to all orders, while TPE and higher as well
as $\Delta$ contributions should be computed in perturbation theory (see
also Refs.~\cite{Griesshammer:2005ga,Birse:2005um,Birse:2005pm} for
related ideas). In this section, we analyze such a proposal
disregrading the Delta. Our main conclusions will not change although
numbers could be modified. In such a situation the perturbative
expansion is equivalent to consider the $m M /(4 \pi f^2) $ to be
zeroth order while $m^2 /(4 \pi f)^2 $ is taken to be second order and
$m/M$ is first order, so that the potential can written as
\begin{eqnarray} 
U^{(0)} &=& U_{1 \pi}^{(0)} \, , \nonumber \\ 
U^{(2)} &=& U_{1 \pi}^{(2)} + U_{2 \pi}^{(2)} \, , \nonumber \\   
U^{(3)} &=& U_{1 \pi}^{(3)} + U_{2 \pi}^{(3)} \, . 
\label{eq:pot_pert}
\end{eqnarray} 
Non-perturbatively $U_{1 \pi } = U_{1 \pi}^{(0)}+ U_{2 \pi}^{(0)}+
U_{3 \pi}^{(0)}+ \dots $ amounts to take $g_{\pi NN} =13.1 $ in the
OPE piece, hence accounting for the Goldberger-Treiman discrepancy. In
perturbation theory, we must take $U_{1 \pi}^{(0)} $ with $g_A=1.26 $
and include OPE corrections to higher order.  Note that the missing
first order implies substantial simplifications in the perturbative
treatment. Indeed NNLO can be done within first order perturbation
theory since going to second order perturbation theory considers
${U^{(2)}}^2$ which is N$^3$LO. In the remaining of this section we
analyze some aspects of such proposal by scaling the strength of the
perturbation and show that the appearance of non-analytical behaviour
is intrinsic to singular potentials, yielding to perturbative
divergences. As we will see, finite perturbative calculations
including chiral TPE to NNLO would require 4 counterterms for the
singlet $^1S_0$ and 6 counterterms for the triplet $^3S_1-^3D_1$
channel. Our non-perturbative results, i.e., fully iterated NNLO
potentials, are based on just 1 and 3 counterterms respectively.

\subsection{Singlet $^1S_0$ channel in distorted OPE waves}

\begin{figure}[]
\begin{center}
\epsfig{figure=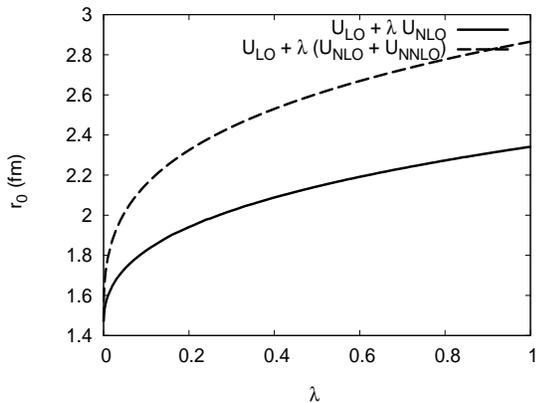,height=5.5cm,width=7.5cm}
\end{center}
\caption{Dependence of the effective range $r_0 (\lambda) $ in the
$^1S_0 $ channel when the scaled potentials $ U = U_{\rm LO } +
\lambda U_{\rm NLO} $ and $ U = U_{\rm LO } + \lambda ( U_{\rm NLO} +
U_{\rm NNLO} ) $ are considered. In all cases we fix the scattering
length to its experimental value.}
\label{fig:r0-1S0.lambda}
\end{figure}

Let us examine first the $^1S_0$ channel, and consider the effect of
the NLO and NNLO TPE potentials on top of the LO OPE potential in long
distance perturbation theory. The fact that they are taken second and
third order respectively means that the effect will be additive at
NNLO in the scattering properties. For the total potential in
Eq.~(\ref{eq:pot_pert}) we write the wave function as
\begin{eqnarray}
u_k (r)=u^{(0)}_k (r)+ u^{(2)}_k (r) + u^{(3)}_k (r) + \dots 
\dots
\end{eqnarray}
and the phase shift becomes
\begin{eqnarray}
\delta_0 = \delta^{(0)}_0 + \delta^{(2)}_0 + \delta^{(3)}_0 + \dots
\end{eqnarray}
At LO the scattering length $\alpha_0^{(0)}$ is a free parameter which
we fix to the physical value, $\alpha_0^{(0)}=\alpha_0$. As we did in
our non-perturbative treatment in Sect.~\ref{sec:conv-1S0}, we will
keep the scattering length fixed to its experimental value at any
order of the approximation, so that differences may be only
attributable to the potential. In the normalization of
Eq.~(\ref{eq:norm}) the correction to the phase shift is just given by
\begin{eqnarray}
\delta_0^{(2)} = - k \sin^2 \delta_0^{(0)} \int_{r_c}^\infty U^{(2)}
(r) u_{k}^{(0)}(r)^2 dr \, ,
\end{eqnarray} 
and a similar expression for $\delta_0^{(3)}$ which can be deduced by
the standard Lagrange's identity. Here, a short distance cut-off $r_c$
has been assumed because at short distances the NLO potential diverges
as $U_{\rm NLO} \sim 1/r^5 $. The previous formula yields a change
also in the scattering length, so that we may eliminate the cut-off
radius by subtracting off the zero energy contribution by fixing
$\alpha_0^{(2)}=0$. It is convenient to recast the result in the form
of an effective range expansion in the OPE distorted wave basis,
\begin{eqnarray} 
k \cot \delta_0 + \frac1{\alpha_0} &=& k \cot \delta^{(0)}_0 +
\frac1{\alpha_0^{(0)}} \nonumber \\ &+& \int_{r_c}^\infty dr U^{(2)} (r)
\left[ u_{k}^{(0)} (r)^2 - u_0^{(0)} (r)^2 \right] \nonumber \\  
&+& \int_{r_c}^\infty dr U^{(3)} (r)
\left[ u_{k}^{(0)} (r)^2 - u_0^{(0)} (r)^2 \right] \, , \nonumber \\ 
\end{eqnarray}
which guarantees $\alpha_0^{(2)}=\alpha_0^{(3)}=0$, due to the one
subtraction. If we expand in powers of the energy the LO wave function
we get
\begin{eqnarray}
u_{k}^{(0)} (r) = u_0^{(0)} (r) + k^2 u_{2}^{(0)} (r) + k^4
u_{4}^{(0)} (r) + \dots \, . 
\end{eqnarray} 
where 
\begin{eqnarray}
-u_0^{(0)}{''}(r) + U(r) u_0^{(0)}  (r) &=& 0 \nonumber \\ 
-u_{2}^{(0)}{''}(r) + U(r) u_2^{(0)}  (r) &=& u_{0}^{(0)} (r)  \nonumber  \\
-u_{4}^{(0)}{''}(r) + U(r) u_4^{(0)}  (r) &=& u_{2}^{(0)} (r)  \nonumber  \\ 
\dots && 
\label{eq:singlet:le}
\end{eqnarray} 
These equations can be solved recursively. Thus, the NLO correction to
the effective range is given by
\begin{eqnarray} 
r_0^{(2)} = 4 \int_{r_c}^\infty U^{(2)}
(r)  u_{2}^{(0)} (r) u_0^{(0)} (r) dr \, . 
\end{eqnarray}
To estimate the short distance contribution we use the OPE exchange
potential in the form $U_{\rm LO} = - e^{-m r} / (R_s r) $, with $R_s
= 16 f^2 \pi / g^2 m^2 M $ the characteristic length $^1S_0$-channel
scale. Note that the OPE potential in the $^1S_0$ channel is Coulomb
like at short distances for which the complete regular plus irregular
solution is known.  One could then use a short distance expansion of
the general analytical Coulomb solution. This facilitates guessing the
solution at short distances for the zeroth energy wave function. The
higher energy wave functions can be computed straightforwardly,
yielding for $r \to 0$
\begin{eqnarray}
u_0^{(0)} (r) &\sim&  c_0 \left[1 + m r -\frac{3r}{2R_s} - \frac{r}{R_s} \log\left(\frac{r}{R_s} \right) \right] + c_1 r \nonumber \\ 
u_2^{(0)} (r) &\sim&  -c_0 r R_s + {\cal O} (r^3 ) \nonumber \\
u_4^{(0)} (r) &\sim&   \frac1{3!} c_0 r^3  R_s + {\cal O} (r^5 ) \nonumber \\
u_6^{(0)} (r) &\sim&  -\frac{1}{5!} c_0 r^5 R_s + {\cal O} (r^7 ) \nonumber \\
u_8^{(0)} (r) &\sim&   \frac{1}{7!} c_0 r^7 R_s + {\cal O} (r^9 )
\label{eq:LO_short}
\end{eqnarray} 
as can be readily checked by solving the Schr\"odinger equation in
powers of energy, Eq.~(\ref{eq:singlet:le}). The coefficients $c_1$
and $c_0 $ correspond to the linearly independent regular and
irregular solutions respectively and are determined by matching to the
integrated in asymptotic condition $u_0^{(0)} (r) \to 1- r/\alpha_0 $
at large distances and at zero energy. Obviously, the irregular
solution contributes, $c_0 \neq 0$, because $\alpha_0 $ is taken to be
independent of the potential, and hence terms proportional to the
coefficient $c_1$ are subleading. Thus, we get
\begin{eqnarray} 
r_0^{(2)} \sim 4 \int_{r_c} dr \frac{ M C_5 }{r^5} (-c_0^2 R_s) r \, .  
\end{eqnarray}
Thus, we conclude that the first order perturbative result is badly
divergent. This is very puzzling since the non-perturbative
calculation in Sect.~\ref{sec:conv-1S0} yields a finite number (see
Eq.~(\ref{eq:r0_NLO})), and suggests non-analytical dependence on the
coupling constant. To enlighten the situation, let us scale the NLO
potential by a factor $\lambda$, $U_{\rm NLO} \to \lambda U_{\rm NLO}
$, and compute non-perturbatively the effective range as a function of
the scaling parameters, $r_0 (\lambda) $, with the obvious conditions
$r_0 (0) = r_0^{(0)}$ and $r_0 (1)=r_0^{\rm NLO}$. The result is
presented in Fig.~\ref{fig:r0-1S0.lambda}. The infinite slope at the
origin can be clearly seen. Numerically, we find that for small
$\lambda \ll 0.1 $, the correction to the effective range behaves as
$r_0 - r_0^{(0)} \sim \sqrt{\lambda}$, whereas for $\lambda \sim 1 $,
it behaves as $r_0 - r_0^{(0)} \sim \lambda^{\frac13}$. This
fractional power counting $\lambda^\alpha$, with $0 < \alpha < 1 $ is
evident from the universal low energy theorem, Eq.~(\ref{eq:r0_univ}),
for a potential with a {\it single } scale, $U(r) = F(r/R)/R^2 $, and
appears when the short distance regulator is removed~\footnote{In
Ref.~\cite{Beane:2000wh} a potential well was used as a short distance
regulator which was not removed, and hence the non-analiticity was not
seen.}. An explicit example is provided by Eq.~(\ref{eq:r0_vdw}) when
a pure Van der Waals potential acts as a perturbation to a boundary
condition ( a contact theory with only $\alpha_0$), since the strength
of the potential is $ \lambda M C_6 = R^4 $, but $r_0 $ contains $R
\sim \lambda^{1/4}$, $R^2 \sim \lambda^{1/2}$ and $R^3 \sim
\lambda^{3/4}$.  It would be interesting to predict {\it a priori}
this non-perturbative non-integer power counting analytically for
potentials with multiple scales as we have done here
numerically~\cite{Alvaro}~\footnote{Fractional power counting has also
been reported to occur also in the EFT analysis of the three body
problem for the pionless
theory~\cite{Griesshammer:2005ga,Birse:2005pm}.}

Obviously, to prevent the perturbative divergence one could subtract
an energy dependent contribution, and provide the effective range as
an input parameter~\footnote{This is equivalent to use a short
distance energy dependent boundary condition on the solution and hence
to violate the orthogonality conditions discussed in
Sect.~\ref{sec:short}.}. Then one would get
\begin{eqnarray} 
k \cot \delta &=& k \cot \delta^{(0)} + \frac12 \left(r_0 -
r_0^{(0)} \right) k^2 \nonumber \\ &+& \int_{r_c}^\infty dr \left[ U^{(2)}
(r) + U^{(3)} (r) \right] \times \nonumber \\ 
&& \Big[ u_{k}^{(0)} (r)^2 - u_0^{(0)} (r)^2  - 2 k^2 u_0^{(0)} (r) u_2^{(0)} (r)^2 \Big]
\end{eqnarray}
Note that this equation requires assuming $ r_0 - r_0^{(0)} = {\cal O}
(\lambda) $, while non-perturbatively we find $ r_0 - r_0^{(0)} =
{\cal O} (\lambda^\frac12 )$. Now the NLO and NNLO corrections to the
$v_2$ parameter would come as a prediction,
\begin{eqnarray} 
v_2^{(2)} &+& v_2^{(3)} = \int_{r_c}^\infty dr \left[ U^{(2)} (r) +
 U^{(3)} (r) \right] \times \nonumber \\ && \left[ 2 u_{4}^{(0)}
(r) u_0^{(0)} (r) + u_2^{(0)} (r)^2 \right] \, , 
\label{eq:deltav2}
\end{eqnarray}
which is also divergent since the leading behaviour of the integrand
is $ \sim 1/r^3 $ at NLO and $ \sim 1/r^4 $ at NNLO for small $r$, see
Eq.~(\ref{eq:LO_short}). Thus, a further subtraction would be needed,
predicting the correction to $v_3$,
\begin{eqnarray} 
v_3^{(2)} &+& v_3^{(3)} = \int_{r_c}^\infty dr \left[ U^{(2)} (r) +
 U^{(3)} (r) \right] \times \nonumber \\
&& \left[ 2 u_{6}^{(0)}
(r) u_0^{(0)} (r) + 2 u_2^{(0)} (r) u_4^{(0)} (r)\right] \, , \nonumber \\  
\label{eq:deltav3}
\end{eqnarray}
which is logarithmically divergent due to
Eq.~(\ref{eq:LO_short}). Finally, if a fourth subtraction is implemented 
a convergent prediction is obtained for $v_4$ at NLO and NNLO, 
\begin{eqnarray} 
v_4^{(2)} &+& v_4^{(3)} = \int_{r_c}^\infty dr 
\left[ U^{(2)} (r) + U^{(3)} (r) \right] \times \nonumber \\ 
&& \left[ 2 u_{8}^{(0)}
(r) u_0^{(0)} (r) + 2 u_2^{(0)} (r) u_4^{(0)} (r) + u_4^{(0)} (r)^2
\right] \, . \nonumber \\ 
\label{eq:deltav4}
\end{eqnarray}
These four subtractions, needed to make a renormalized {\it
perturbative} prediction of the $^1S_0$ phase shift at NNLO , actually
correspond to having 4 counterterms, i.e. fixing $\alpha_0, r_0,v_2$
and $v_3$. This result disagrees with the standard Weinberg counting
(two counterterms at NLO and NNLO in the $^1S_0$ channel). Moreover,
besides the loss of predictive power as compared to the
non-perturbative result where only one counterterm is needed the
deduced renormalized value for $v_4$ is worsened in perturbation
theory, since $v_4^{(0)}=-50.74 {\rm fm}^7 $, $v_4^{(2)}= -10.45 {\rm
fm}^7$ and $v_4^{(3)}= -2.88 {\rm fm}^7$.  The situation is summarized
in Table~\ref{tab:1S0_pert} where we show our numerical results
obtained in perturbation theory as explained above and they are
compared to the NijmII and Reid93 potential model calculations (See
e.g. Ref.~\cite{PavonValderrama:2004se}). Although these are not
directly experimental data it is noteworthy that they differ by a few
percent while the perturbative calculation is about a factor 3 larger.
The integrals for $v_4$ are rather well converging and the matching
between the numerical solution and the short distance solutions,
Eq.~(\ref{eq:LO_short}), is quite stable in the region around $r \sim
0.1 {\rm fm}$.

\begin{table}[]
\caption{\label{tab:1S0_pert} Threshold parameters of the effective
range expansion $k \cot \delta = -1/\alpha_0 + r_0 k^2 /2 + v_2 k^4 +
v_3 k^6 + v_4 k^8 $ in the singlet $^1S_0$ channel in OPE distorted
waves perturbation theory.  We take $U^{(0)} = U_{1 \pi }^{(0)} $ with
$g_A=1.26$. For the NNLO case we use Set IV for the chiral constants
$c_1$, $c_3$ and $c_4$ given in Table~\ref{tab:table_vdw}. }
\begin{ruledtabular}
\begin{tabular}{|c|c|c|c|c|c|c|}
\hline $^1S_0 $ & LO & NLO$_{\rm pert}$  &  N$^2$LO$_{\rm pert}$ & Exp.  &  Nijm II & Reid93 \\ \hline
$\alpha_0( {\rm fm}) $ & Input & Input  & Input  & -23.74(2) & -23.73 & -23.74\\ 
$r_0( {\rm fm}) $  & 1.383     &  Input & Input  & 2.77(5) & 2.67 & 2.75 \\ 
$v_2 ( {\rm fm}^3) $ & -2.053 & Input  &  Input  & -- & -0.48 & -0.49 \\ 
$v_3 ( {\rm fm}^5) $ & 9.484 & Input  & Input  & -- & 3.96 & 3.65 \\ 
$v_4 ( {\rm fm}^7) $& -50.74 & -61.19  & -64.07  & -- & -19.88 & -18.30 
\end{tabular}
\end{ruledtabular}
\end{table}

Thus, in this particular example of the $^1S_0$ channel one sees that
our non-perturbative approach based on the choice of the regular
solutions at the origin predicts the phase shift and hence all low
energy parameters from $\alpha_0$ and the potential as displayed in
Table~\ref{tab:convergence}. A perturbative treatment of the amplitude
based on OPE distorted waves requires to fix $\alpha_0, r_0,v_2$ and
$v_3$ at NNLO.  The phenomenological success and converging pattern
observed when the potential is considered at LO,NLO and NNLO is solved
non-perturbatively is very encouraging. The price to pay is to face
non-analytical behaviour which implies a non-integer power
counting. The trend observed here can be generalized to other
channels. A more thorough discussion of this issue will be presented
elsewhere~\cite{Alvaro}.

\subsection{Triplet  $^3S_1-^3D_1$ channel in distorted OPE waves}

We turn now to the triplet $^3S_1-^3D_1$ channel. The reasoning is a
straightforward, although tedious, coupled channel generalization of
the $^1S_0$ case, with the additional feature that the short distance
behaviour is dominated by a $1/r^3$ singularity (instead of 
$1/r$), and so the short distance behaviour is different. It is
convenient to introduce the potential matrix as
\begin{eqnarray}
{\bf U} (r) &=& 
\begin{pmatrix}
U_{^3S_1} (r) & U_{E_1} (r) \\ U_{E_1} (r) & U_{^3D_1} (r)
\end{pmatrix}  \, , 
\end{eqnarray}
and the matrix wave function,
\begin{eqnarray}
{\bf u}_k (r) &=& {\bf A} 
\begin{pmatrix}
u_{k,\alpha}(r)  & u_{k,\beta}(r)  \\ 
w_{k,\alpha}(r)  & w_{k,\beta}(r)  
\end{pmatrix}  \, , 
\end{eqnarray}
with ${\bf A}$ a constant energy dependent matrix, subject to a
slightly different normalization than Eq.~(\ref{eq:phase_triplet}),
\begin{eqnarray}
{\bf u}_k (r)\to \frac1k \hat {\bf j} (kr) {\bf D}^{-1} \hat {\bf M} -
  \hat {\bf y} (kr) {\bf D}
\end{eqnarray}
Here, $ {\bf \hat M} $ is the effective range matrix defined by its
relation to the unitary ${\bf S}$-matrix,
\begin{eqnarray}
{\bf D} {\bf S} {\bf D}^{-1} = \left( {\bf \hat M} + {\rm i} k {\bf
D}^2 \right) \left({\bf \hat M} - {\rm i} k {\bf D}^2 \right)^{-1}
\, ,
\end{eqnarray} 
and ${\bf D} = {\rm diag} ( 1, k^2 ) $. The reduced Bessel functions
matrices are given by $\hat {\bf j} = {\rm diag} ( \hat j_0 , \hat j_2
) $ and $\hat {\bf y} = {\rm diag} ( \hat y_0 , \hat y_2 ) $ with $
\hat j_l (x) = x j_l (x) $ and $\hat y_l (x) = x y_l (x) $. At low
energies, one has the effective range expansion(see e.g.
\cite{PavonValderrama:2004se} and references therein),
\begin{eqnarray}
\hat {\bf M} = -({\bf a})^{-1} + \frac12 {\bf r} k^2 + {\bf v} k^4 + 
\dots  
\end{eqnarray} 
Here, we have introduced the scattering length matrix,
\begin{eqnarray}
{\bf a} &=& 
\begin{pmatrix}
\alpha_{0}  & \alpha_{02}  \\ 
\alpha_{02} & \alpha_2   
\end{pmatrix}  \, , 
\end{eqnarray}
the effective range matrix 
\begin{eqnarray}
{\bf r} &=& 
\begin{pmatrix}
r_{0}  & r_{02}  \\ 
r_{02} & r_2   
\end{pmatrix}  \, , 
\end{eqnarray}
and so on. These parameters have been determined in
\cite{PavonValderrama:2004se} from the potentials of
Ref.~\cite{Stoks:1994wp}. Proceeding similarly as in the one channel
case, one gets, after one subtraction at zero energy the effective
range function in perturbation theory,
\begin{eqnarray}
\hat {\bf M} &+& ({\bf a})^{-1} = \hat {\bf M}^{(0)} + ({\bf
a}^{(0)})^{-1} \nonumber \\ &+& \int_{r_c}^\infty dr \left[ {\bf
u}_k^{(0) \dagger} {\bf U}^{(2)} {\bf u}_k^{(0)} - {\bf
u}_0^{(0) \dagger} {\bf U}^{(2)} {\bf u}_0^{(0)} \right]
\end{eqnarray} 
The condition $\alpha_{0}^{(0)} = \alpha_0 $ must be imposed, since
$\alpha_{02}^{(0)} $ and $\alpha_2^{(0)} $ are predicted from
$\alpha_0^{(0)}$ (at LO one only needs one counterterm). This formula
implies that one introduces two new conditions to fix now
$\alpha_{02}$ and $\alpha_2$ to their experimental value.Along similar
lines as done before, we analyze the finiteness of the previous
expression by computing the effective range matrix.  To this end we
expand the coupled channel wave function in powers of momentum
\begin{eqnarray}
{\bf u}_{k}^{(0)} (r) = {\bf u}_0^{(0)} (r) + k^2 {\bf
u}_{2}^{(0)} (r) + k^4 {\bf u}_{4}^{(0)} (r) + \dots
\end{eqnarray} 
to get 
\begin{eqnarray}
{\bf r}^{(2)} = \int_{r_c}^\infty dr \left[ {\bf u}_2^{(0) \dagger}
{\bf U}^{(2)} {\bf u}_0^{(0)} + {\bf u}_0^{(0) \dagger} {\bf
U}^{(2)} {\bf u}_2^{(0)} \right]
\end{eqnarray} 
The LO OPE short distance behaviour of the triplet wave functions has
been worked out in our previous work~\cite{PavonValderrama:2005gu}. It
is convenient to define the triplet length scale
\begin{eqnarray} 
R_t = \frac{3 g_A^2 M }{32 \pi f_\pi^2}
\label{eq:R_def} 
\end{eqnarray} 
which value $R_t= 1.07764 {\rm fm}$. One has the general structure  
\begin{eqnarray}
u (r) &=& \frac1{\sqrt{3}}\left(\frac{r}{R_t}\right)^{3/4} \Big[ -C_{1R}
  f_{1R} (r) e^{+ 4 \sqrt{2} \sqrt{\frac{ R_t}{r}}} \nonumber \\ &-&
  C_{2R} f_{2R}(r) e^{- 4 \sqrt{2} \sqrt{\frac{ R_t}{r}}} + \sqrt{2}
  C_{1A} f_{1A}(r) e^{- 4 i \sqrt{\frac{ R_t}{r}}} \nonumber \\ &+&
  \sqrt{2} C_{2A} f_{2A}(r) e^{ 4 i\sqrt{\frac{ R_t}{r}}} \Big]
  \nonumber \\ w (r) &=& \frac1{\sqrt{3}}
  \left(\frac{r}{R_t}\right)^{3/4} \Big[ \sqrt{2} C_{1R} g_{1R} (r) e^{+
  4 \sqrt{2} \sqrt{\frac{ R_t}{r}}} \nonumber \\ &+& \sqrt{2} C_{2R}
  g_{2R}(r) e^{- 4 \sqrt{2} \sqrt{\frac{ R_t}{r}}} + C_{1A} g_{1A}(r)
  e^{- 4 i \sqrt{\frac{ R_t}{r}}} \nonumber \\ &+& C_{2A} g_{2A}(r) e^{
  4 i\sqrt{\frac{ R_t}{r}}} \Big] \nonumber \\
\label{eq:LO_short_wf}
\end{eqnarray} 
where the constants $C_{1R}$, $C_{2R}$, $C_{1A}$ and $C_{2A}$ depend
on the energy and the OPE potential parameters. The regular solution
is selected when one takes $C_{1R}=0$. The functions appearing in this
formula are of the form 
\begin{eqnarray}
f(r) &=& \sum_{n=0}^\infty a_n
\left(\frac{r}{R_t}\right)^{n/2} \nonumber \\ 
g(r) &=& \sum_{n=0}^\infty b_n
\left(\frac{r}{R_t}\right)^{n/2} \nonumber \\
\end{eqnarray} 
For the present calculation we only need the power behaviour (see
Appendix B of Ref.~\cite{PavonValderrama:2005gu})
\begin{eqnarray}
{\bf u}_{0}^{(0)} (r) &\sim& r^{3/4} \nonumber \\ {\bf u}_{2}^{(0)}
(r) &\sim& r^{3/4+5/2} \nonumber \\ {\bf u}_{4}^{(0)} (r) &\sim&
r^{3/4+5}
\end{eqnarray} 
which shows that, again, the first order correction to the effective
range matrix is logarithmically divergent because the NLO potential
diverges as $1/r^5$ and ${\bf U}^{(2)} {\bf u}_{0} {\bf u}_{2} \sim
1/r $~\footnote{There is a subtlety here. The terms containing the
regular exponential at the origin are convergent, regardless on the
power of $r$ in the denominator. Naively logarithmically divergent
integrals would become convergent when combined with oscillating
functions. However, these functions appear {\it squared} so that the
logarithmic divergence prevails.}. As previously, the situation could
be amended by adding 3 new counterterms to fix the effective range
matrix ${\bf r}$ and then ${\bf v}$ would come as a prediction. So, at
NLO in perturbation theory one needs a total of 6 counterterms to
generate a coupled channel finite amplitude. When adding the NNLO
contribution this number of counterterms remains the same since ${\bf
U}^{(3)} {\bf u}_{0}^{(0)} {\bf u}_{4}^{(0)} \sim r^{1/2} $ and ${\bf
U}^{(3)} [{\bf u}_{2}^{(0)}]^2 \sim r^{1/2} $.

\begin{figure}[]
\begin{center}
\epsfig{figure=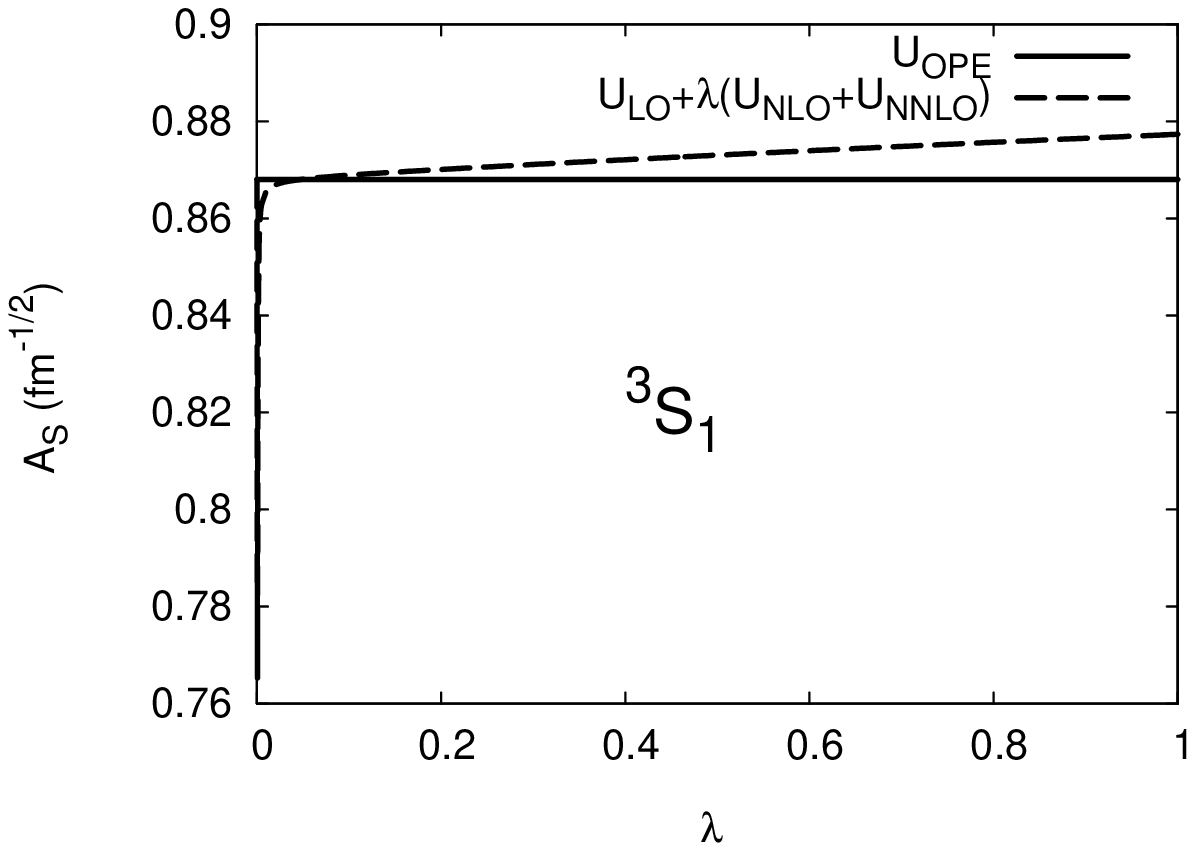,height=5.5cm,width=7.5cm} 
\epsfig{figure=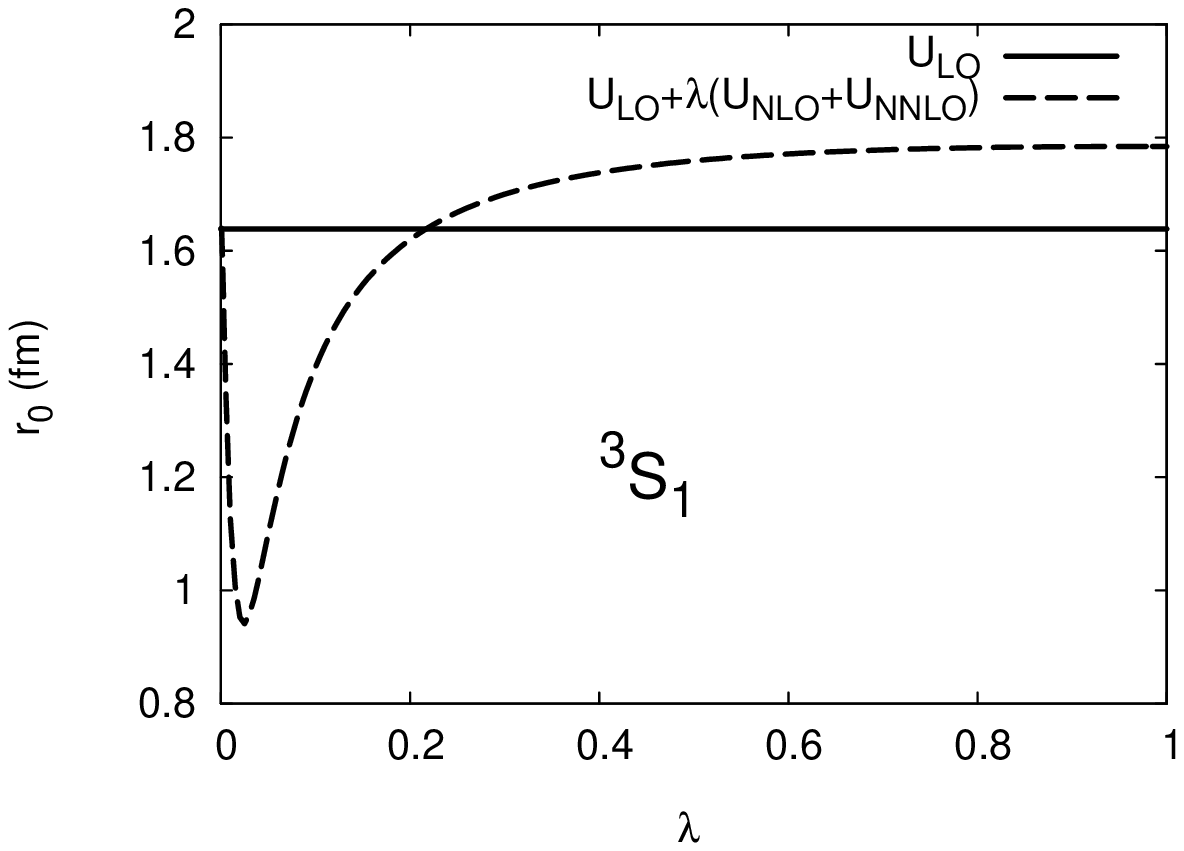,height=5.5cm,width=7.5cm}
\end{center}
\caption{Dependence of the s-wave function normalization $ A_S
(\lambda)$ in the deuteron and the effective range $r_0 (\lambda) $ in
the $^3S_1-^3D_1$ channel when one scales the TPE potential $ U =
U_{\rm LO } + \lambda (U_{\rm NLO}+ U_{\rm NNLO}) $. In all cases we
fix the deuteron binding energy, the asymptotic $D/S$-ration, $\eta$
and the s-wave scattering length, $\alpha_0$ are fixed to their
experimental values.}
\label{fig:Asr0-3S1.lambda.NNLO}
\end{figure}

An illustration of non-analytical non-perturbative behaviour in the
$^3S_1-^3D_1$-channel can be looked up in
Fig.~\ref{fig:Asr0-3S1.lambda.NNLO}. There, the behaviour of the
s-wave function normalization $ A_S (\lambda)$ in the deuteron and the
effective range $r_0 (\lambda) $ in the $^3S_1-^3D_1$ channel when the
TPE potential is scaled as $ U = U_{\rm LO } + \lambda (U_{\rm NLO}+
U_{\rm NNLO}) $ and the deuteron binding energy, the asymptotic
$D/S$-ration, $\eta$ and the s-wave scattering length, $\alpha_0$ are
fixed to their experimental values.

\subsection{The deuteron  in distorted OPE waves } 

To conclude our analysis of perturbation theory we study now the
deuteron bound state. According to Fig.~\ref{fig:Asr0-3S1.lambda.NNLO}
there appears some tiny non-analyticity for very small couplings in
the asymptotic s-wave normalization $A_S$. Note that there is an
apparent linear behaviour with the exception of the very small
$\lambda$ region, making one suspect that the result might be obtained
in perturbation theory. We will see below by an explicit perturbative
calculation that {\it this is not so}. We have checked that this trend
also occurs for other quantities such as the quadrupole moment, $Q_d$,
the matter radius, $r_m$, and the D-state probability, $P_D$. Here, we
show, as it has been done above for the scattering problem, that this
can be traced to a first order divergent renormalized result.

We define the two component deuteron state,
\begin{eqnarray} 
{\bf u}_\gamma (r) = 
\begin{pmatrix}
u_\gamma (r) \\ w_\gamma (r) \end{pmatrix}  
\end{eqnarray}  
In perturbation theory, we expand the potential 
\begin{eqnarray}
{\bf U}(r) = {\bf U}^{(0)}(r)  + {\bf U}^{(2)} (r) +  {\bf U}^{(3)} (r) + \dots
\end{eqnarray} 
and thus the deuteron wave function for fixed energy (or $\gamma$), 
\begin{eqnarray} 
{\bf u}_\gamma (r) = {\bf u}_\gamma^{(0)} (r) + {\bf u}_\gamma^{(2)}
(r) +  {\bf u}_\gamma^{(3)}
(r) + \dots
\end{eqnarray} 
where $( u_\gamma^{(0)} (r) , w_\gamma^{(0)} (r) ) $ correspond
to the lowest order solutions of the problem and $( u_\gamma^{(2)} (r)
, w_\gamma^{(2)} (r) )$ and $( u_\gamma^{(3)} (r)
, w_\gamma^{(3)} (r) )$ satisfy
\begin{eqnarray}
-{\bf u}_\gamma^{(0)}(r) + \left[ {\bf U}^{(0)} (r) + \gamma^2 \right]
 {\bf u}_\gamma^{(0)}(r) &=& 0 \nonumber \\ -{\bf u}_\gamma^{(2)}(r) +
 \left[ {\bf U}^{(0)} (r) + \gamma^2 \right] {\bf u}_\gamma^{(2)}(r)
 &=& -{\bf U}^{(2)} (r) {\bf u}_\gamma^{(0)} (r) \nonumber \\  -{\bf u}_\gamma^{(3)}(r) +
 \left[ {\bf U}^{(0)} (r) + \gamma^2 \right] {\bf u}_\gamma^{(3)}(r)
 &=& -{\bf U}^{(3)} (r) {\bf u}_\gamma^{(0)} (r) \nonumber \\ 
\label{eq:pert_first} 
\end{eqnarray} 
We look for normalized solutions, so that perturbatively
\begin{eqnarray}
1 &=&\int_0^\infty dr {\bf u}_\gamma^{(0) \dagger} (r) {\bf
u}_\gamma^{(0)} (r) \nonumber \\ 0 &=&\int_0^\infty dr \left( {\bf
u}_\gamma^{(2) \dagger} (r) {\bf u}_\gamma^{(0)} (r)+ {\bf
u}_\gamma^{(0) \dagger}  (r) {\bf u}_\gamma^{(2)} (r) \right) \nonumber \\ 0 &=&\int_0^\infty dr \left( {\bf
u}_\gamma^{(3) \dagger} (r) {\bf u}_\gamma^{(0)} (r)+ {\bf
u}_\gamma^{(0) \dagger}  (r) {\bf u}_\gamma^{(3)} (r) \right) \nonumber \\ 
\label{eq:ort1}
\end{eqnarray} 
The zeroth order equation was solved in our previous
work~\cite{PavonValderrama:2005gu}, where it was shown that $\gamma$
was a free parameter, which means $\gamma^{(0)}=\gamma$, and the
regular solution at the origin was selected (see
Eq.~(\ref{eq:LO_short_wf})) to ensure normalizability at the origin.
We will keep always the same fixed value at any order of the
approximation, so that $\gamma^{(2)}=\gamma^{(3)}=0$. To analyze the
NLO and NNLO problem analytically we proceed by the variable
coefficients method.  The zeroth order equation is a homogenous linear
system with four linearly independent solutions,
\begin{eqnarray}
{\bf u}_i^{(0)} (r) =  \begin{pmatrix}
u_i (r) \\ w_i (r) \end{pmatrix} \qquad i=1,2,3,4   
\end{eqnarray}  
The first order equation is an inhomogeneous linear system, which
solution can be written as
\begin{eqnarray} 
u_\gamma^{(2)} (r) + u_\gamma^{(3)} (r) &=& \sum_{i=1}^4 c_i (r) u_i  (r)  \nonumber \\ 
w_\gamma^{(2)}(r) + w_\gamma^{(3)} (r) &=& \sum_{i=1}^4 c_i (r) w_i  (r)  
\end{eqnarray} 
The variable coefficients satisfy 
\begin{eqnarray} 
\sum_{i=1}^4 c_i ' (r) u_i (r)  &=& 0 \nonumber \\ 
\sum_{i=1}^4 c_i ' (r) w_i (r)  &=& 0 \nonumber \\  
\sum_{i=1}^4 c_i ' (r) u_i ' (r)  &=& F_u (r)  \nonumber \\ 
\sum_{i=1}^4 c_i ' (r) w_i ' (r)  &=& F_w (r) 
\label{eq:lin_ode}
\end{eqnarray} 
where we have defined the driving term 
\begin{eqnarray} 
{\bf F} (r) &=&  \begin{pmatrix} F_u (r) \\ F_w (r) \end{pmatrix} \nonumber \\ &=&
-{\bf U}^{(2)} (r) {\bf u}_\gamma^{(0)} (r)  -{\bf U}^{(3)} (r) {\bf u}_\gamma^{(0)} (r)  
\end{eqnarray} 
which at short distances  behaves as 
\begin{eqnarray}
{\bf F}(r) \sim r^{3/4} C_5 r^{-5}  + r^{3/4} C_6 r^{-6}  
\end{eqnarray} 
whereas at large distances one has  
\begin{eqnarray}
{\bf F}(r) \sim e^{-\gamma r} e^{-m r}   
\end{eqnarray} 
To proceed further, we choose the following linearly independent
solutions fulfilling the asymptotic boundary condition at infinity
\begin{eqnarray}
u_1 (r) &\to & e^{-\gamma r} \, , \nonumber \\
w_1 (r) &\to & 0 \, , \nonumber \\ 
u_2 (r) &\to& 0 \,  ,  \nonumber \\ 
w_2 (r) & \to & 
e^{-\gamma r} \left( 1 + \frac{3}{\gamma r} + \frac{3}{(\gamma r)^2}
\right) \, , \nonumber \\
u_3 (r) &\to & e^{\gamma r} \, , \nonumber \\ 
w_3 (r) &\to & 0 \, , \nonumber \\ 
u_4 (r) &\to& 0 \, , \nonumber \\ 
w_4 (r) & \to & 
e^{\gamma r} \left( 1 - \frac{3}{\gamma r} + \frac{3}{(\gamma r)^2}
\right) \, , \nonumber \\
\end{eqnarray} 
Any of these solutions has a short distance behaviour of the general
form given in Eq.~(\ref{eq:LO_short_wf}). So that, {\it all these
solutions are necessarily singular} at the origin. Using Krammer's
rule the solutions to the linear differential system,
Eq.~(\ref{eq:lin_ode}), which are regular at infinity read
\begin{eqnarray}
c_1(r) = \frac1{W} \int_0^r dr'\left| \begin{array}{cccc} 0 & u_2 &
u_3 & u_4 \\ 0 & w_2 & w_3 & w_4 \\ F_u  & u_2 ' & u_3 '& u_4 ' \\
F_w  & w_2 ' & w_3 ' & w_4 '
\end{array}   \right| 
\end{eqnarray} 
\begin{eqnarray}
c_2 (r) = \frac1{W} \int_0^r dr \left| \begin{array}{cccc}
u_1 & 0 & u_3 & u_4  \\  
w_1 & 0 & w_3 & w_4  \\  
u_1 ' & F_u & u_3 '& u_4 ' \\  
w_1 ' & F_w ' & w_3 ' & w_4 '  
\end{array}   \right| 
\end{eqnarray} 
\begin{eqnarray}
c_3 (r) = -\frac1{W}
\int_r^\infty dr' \left| \begin{array}{cccc}
u_1 & u_2 & 0 & u_4  \\  
w_1 & w_2 & 0 & w_4  \\  
u_1 ' & u_2 ' & F_u& u_4 ' \\  
w_1 ' & w_2 ' & F_w & w_4 '  
\end{array}   \right| 
\end{eqnarray} 
\begin{eqnarray}
c_4 (r) = -\frac1{W} \int_r^\infty dr' \left| \begin{array}{cccc}
u_1 & u_2 & u_3 & 0  \\  
w_1 & w_2 & w_3 & 0  \\  
u_1 ' & u_2 ' & u_3 '& F_u \\  
w_1 ' & w_2 ' & w_3 ' & F_w   
\end{array}   \right| 
\end{eqnarray} 
where $W$ is the Wronskian 
\begin{eqnarray}
W= \left| \begin{array}{cccc}
u_1 & u_2 & u_3 & u_4  \\  
w_1 & w_2 & w_3 & w_4  \\  
u_1 ' & u_2 ' & u_3 ' & u_4 ' \\  
w_1 ' & w_2 ' & w_3 ' & w_4 '   
\end{array}   \right| = - 4 \gamma^2  \, . 
\end{eqnarray} 

At asymptotically large distances we have  
\begin{eqnarray}
u^{\rm (2)} (r) &\to& c_S^{(2)} e^{-\gamma r} \nonumber \\ w^{\rm (2)} (r)
&\to& c_D^{(2)} \eta^{(0)} e^{-\gamma r} \left( 1 + \frac{3}{\gamma r} +
\frac{3}{(\gamma r)^2} \right) \nonumber \\ 
\label{eq:long_pert}
\end{eqnarray} 
and similarly for the N$^2$LO correction.  Note that the normalization
condition, Eq.~(\ref{eq:ort1}), implies a linear relation between
$c_S^{(2)}$ and $c_D^{(2)}$ as well as $c_S^{(3)}$ and
$c_D^{(3)}$. The total D/S ratio obtained by including the zeroth
order contribution is given by
\begin{eqnarray}
\eta = \eta^{(0)}\frac{1 +
c_D^{(2)}+ c_D^{(3)} }{1+c_S^{(2)}+ c_S^{(3)} } 
\label{eq:eta_pert} 
\end{eqnarray} 
If we fix $\eta$ we get a relation between $c_S$ and $c_D$. The
coefficients $c_S^{(2)}$ and $c_D^{(2)} $ are given by
\begin{eqnarray}
c_S^{(2)} + c_S^{(3)} 
&=& c_1 (\infty)  \nonumber \\ 
\eta^{(0)} \left( c_D^{(2)}  + c_D^{(3)} \right) &=& c_2 (\infty)
\label{eq:cS_cD_wronsk}  
\end{eqnarray} 
The long distance behaviour of the integrands is well behaved since,
up to inessential powers in $r$, one has
\begin{eqnarray}
c_1' (r) &\sim &  e^{- 2 m r} \nonumber \\
c_2' (r) &\sim &  e^{- 2 m r} \nonumber \\
c_3' (r) &\sim &  e^{- (2 m + 2\gamma) r}
\nonumber \\ c_4' (r) &\sim & e^{- (2 m+ 2
\gamma) r}
\end{eqnarray} 
However, the leading short distance behaviour of the integrand is
given
\begin{eqnarray}
c_i '(r) \sim r^{3/4} ( e^{4 \sqrt{2 R/r}} )^2 \left( \frac{C_5}{r^5}  + 
\frac{C_6}{r^6}  \right)  r^{3/4}
e^{\pm i 4 \sqrt{ R/r}}
\end{eqnarray} 
So, we expect the coefficients $c_S$ and $c_D$ to diverge if the short
distance cut-off is removed, $r_c \to 0$. It is unclear how this
divergence might be avoided. Unlike the scattering problem in
perturbation theory, where energy dependent (and hence orthogonality
violating) subtractions are needed, it would be difficult to accept a
bound state not normalized to unity unless one includes, besides $pn$
other Fock state components, such as $pn \pi $. This short distance
analysis holds also when the $1/r^6$ $\Delta$-contributions are taken
into account perturbatively.

Thus, perturbation theory on the distorted OPE basis for the deuteron
makes sense only as a finite cut-off theory. In the
appendix~\ref{sec:second} we develop further such an approach to NLO,
where $\eta$ is an input and to NNLO, where {\it both} $A_S$ and
$\eta$ should be fixed. We also show that in the cut-off theory the
NLO yields tiny corrections to deuteron properties whereas the NNLO
dominates. This proves that, at least perturbatively and in the
absence of the $\Delta$, the (integer) power counting to NNLO in
deuteron properties is obviously not convergent. To some extent this
result resembles qualitatively the findings of
Ref.~\cite{Fleming:1999ee} based on the idea that OPE and TPE can be
included perturbatively~\cite{Kaplan:1998we}. Of course, an
amelioration of the convergence in the cut-off theory when $\Delta$'s
are included is not precluded, and deserves further
investigation. However, the very need of a finite cut-off will still
hold, as our analytical study shows. 

\begin{figure}[]
\begin{center}
\epsfig{figure=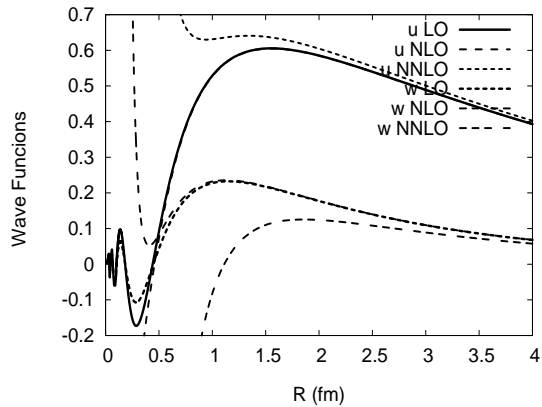,height=5.5cm,width=7.5cm} 
\end{center}
\caption{Deuteron wave functions at LO, NLO and NNLO in perturbation
theory when a short distance cut-off $r_c =0.5 {\rm fm} $ is
considered for the perturbative corrections. At LO we fix $\gamma$, at
NLO we fix $\gamma $ and $\eta$ and at NNLO $\gamma$, $\eta$ and $A_S$
are fixed. Solutions are perturbatively normalized. 
The LO wave functions are normalized by taking $A_S$ = 1}
\label{fig:u+w_pert}
\end{figure}

In Fig.~\ref{fig:u+w_pert} we show LO, NLO and NNLO order wave
functions when a finite short distance cut-off $r_c=0.5 {\rm fm} $ is
considered. The strong divergence of the wave function at the origin
can be clearly seen.

\section{Conclusions} 
\label{sec:conc} 

In the present work we have extended the coordinate space
renormalization of central waves in NN interaction discussed in our
previous work~\cite{PavonValderrama:2005gu} for the OPE potential to
the TPE potential. As we have stressed along the paper, the main
advantage of such a framework is that the (renormalized) potential is
finite everywhere except at the origin where a Van der Waals
attractive singularity takes place. This suggests using a radial
cut-off which provides a compact support for the short range part of
the potential thus making scheme dependent contact interactions
innocuous for the long range solution. Thus, there is no need to
device different regularization methods for the potential and the wave
functions both for bound state and scattering state solutions. As a
result model independent long range correlations between NN
observables can be deduced if the potential is iterated to all orders.

Important constraints can be deduced from the requirement of a small
wave function in the unknown short distance region. As a consequence
the boundary condition for the wave function at short distances
becomes energy independent if the long range contribution to the
potential is also energy independent. We stress here that such
requirements, although quite natural from a physical viewpoint, may
not appear obvious within the so far established EFT framework, and it
would be very interesting to provide further arguments within EFT
itself supporting our unconventional framework~\cite{Pavon2005}.
Actually, we find that the singularity structure of the potential at
short distances determines uniquely how many parameters must be
regarded as unknown, non predictable, information. This is done in
terms of short distance phases or equivalently via suitable mixed
boundary conditions at the origin. Moreover, for an energy independent
potential the orthogonality of wave functions precludes a possible
energy dependence of the boundary conditions. In the particular cases
studied in this paper, namely $^1S_0$ and $^3S_1 -^3 D_1 $ channels,
we have found that besides the NNLO TPE potential parameters, one can
use the S-wave scattering lengths in both channels as well as the
deuteron binding energy and the asymptotic $D/S$ ratio of the deuteron
wave functions as independent input information. The remaining
scattering or bound state properties in the triplet channel are then
predicted unambiguously. Based on the superposition principle of
boundary conditions, we have found analytical and simple universal
rational relations which clearly exhibit these features.  These
universal relations would be very difficult to deduce in momentum
space and, moreover, are free from uncertainties attributable to
finite cut-off effects. So, the cut-off has been effectively
eliminated. On a numerical level, the fact that our problem is an
initial value problem for the Schr\"odinger equation starting at
infinity, makes possible to obtain any solution by competitive
algorithms with adaptable integration steps with any prescribed
accuracy. This allows to faithfully describe the short distance
oscillations of the wave function. This is in contrast to the standard
Lippmann-Schwinger treatments, where matrix inversion methods may
eventually run into computer space limitations with a natural loss of
space resolution as a side-effect.  The non-trivial oscillating
structure of the wave functions with ever decreasing periods of the
wave functions close to the origin would actually be very difficult to
reproduce within a momentum space framework.

According to our analysis, there are finite cut-off effects in
previous works dealing also with TPE potentials both in coordinate as
well as in momentum space. The induced corrections are larger than the
experimental uncertainty of the computed observables, so that in some
cases agreement with data may be clearly attributed to the choice of a
finite cut-off. In our energy independent boundary condition treatment
we found short distance cut-offs of about $r_c=0.1-0.2 {\rm fm} $ to be
rather innocuous.  Within a Wilsonian viewpoint of renormalization,
changes in the cut-off should correspond to decimation, i.e. halving,
and not to linear changes in the scale. If one associates this
coordinate space cut-off to a momentum space ultraviolet cut-off of
$\Lambda = \pi / 2 r_c $~\cite{PavonValderrama:2004td} we are dealing
with an equivalent momentum scale of about $1.5-3 {\rm GeV}$, much
larger than the scales below $1{\rm GeV}$ usually employed in momentum
space calculations where only linear sensitivity to changes of the
cut-off was implemented. Nevertheless, it is fair to say that the
calculations based on Sets III and IV provide not too large
discrepancies.

As one naturally expects in a renormalized theory, errors are
dominated by uncertainties in the input data, and not by cut-off
uncertainties. Indeed, we seem to reach a limit in the accuracy of the
predictions, paralleling the findings in ChPT for mesons at the two
loop level. At the OPE level, one can predict bound state and
scattering properties in the singlet $^1S_0$ and triplet $^3S_1-^3D_1$
channels solely from the deuteron energy and the $^1S_0$ scattering
length.  At the TPE level, one needs not only the additional chiral
constants $c_1$, $c_3$ and $c_4$ but also the triplet S-wave
scattering length and the asymptotic $D/S$ deuteron ratio. Although
the TPE central values predictions improve, the induced TPE errors
turn out to be {\it larger} than the OPE uncertainties. In fact, this
large uncertainties make that, within errors, the TPE calculation
becomes compatible with experimental data at the $1\sigma$ level.
This suggests that in order to see in a statistically significant
sense other effects, such as electromagnetic, relativistic and three
pion effects one must first improve on the input data. Otherwise,
predictive power is lost. Nevertheless, given the finite cut-off
effects detected in previous works, the role of these corrections
beyond TPE should be reanalyzed within the present approach.

One of the important consequences of our treatment is that the chiral
constants $c_1$, $c_3$ and $c_4$ can be determined from {\it low
energy data} and {\it deuteron properties}. Specifically, we have used
the singlet and triplet effective ranges as well as the asymptotic
S-wave deuteron wave function to $c_1$, $c_3$ and $c_4$ {\it with
errors} varying all input data within their experimental
uncertainties. The decision on what set of data should be used to pin
down the chiral coefficients is not entirely trivial, because it
should become clear which hypothesis we want to verify or
to refute. The absence of cut-off effects makes this test cleaner; we
just check whether the TPE potential holds from zero to
infinity. Obviously, this cannot be literally true, but one expects
that at low energies other short range effects can be considered
negligible. Let us remind that error analysis within NN calculations
was only carried out in a large scale partial wave analysis to data in
Ref.~\cite{Rentmeester:1999vw}. The determinations of chiral constants
based on a fit to NN
databases~\cite{Stoks:1993tb,Stoks:1994wp,Arndt:1994br} for phase
shifts lack any error estimates because the databases themselves are
treated as errorless. The determination of chiral constants from
peripheral waves has similar drawbacks. From the chiral theory point
of view we see that it is possible to determine these parameters
precisely in the regime where we trust the theory most, namely in the
description of low energy NN data. A fit becomes possible, and the
values it yields only differ by $2 \sigma$ with the determination from
$\pi N$ data. We do not exclude that our values for the chiral
constants may eventually spoil the successful overall fit of phase
shifts in all channels presented in the past, after all
renormalization has been carried out. If so, the situation on the
effectiveness of effective field theory would be in a less optimistic
shape than assumed hitherto. A preliminary analysis of the problem
shows what Van der Waals coefficients in the TPE potential correspond
to attractive short range interactions, and hence what phase shifts
are completely determined in terms of coupled channel scattering
lengths. This issue is very relevant and would require a detailed
channel by channel analysis and renormalization, taking as input the
scattering lengths documented in our previous
work~\cite{PavonValderrama:2004se} and integrating in from large
distances along the lines of the present approach. Full details are
reported elsewhere~\cite{PavonValderrama:2005uj}.

Nevertheless, despite of the good convergence in the $^1S_0$ channel
for LO, NLO and NNLO calculations, we have noted a difficulty for the
triplet $^3S_1-^3D_1$ channel at NLO of the potential. In contrast to
dimensional power counting expectations one cannot use the scattering
length $\alpha_0$, the effective range $r_0$ and $\alpha_{02}$ as
arbitrary input parameters at NLO in the potential (one could equally
take $\gamma$, $\eta$ and $\alpha_0$) but they are entirely predicted
from the potential as required by finiteness of the
phase-shifts. Otherwise, the scattering amplitude diverges, as we have
shown. We have also seen that even if one assumes a finite value of
the cut-off the NLO is worse than the LO, suggesting that the problem
may indeed be related to the power counting on the long distance
potential. Remarkably, these parameters must be fixed at NNLO where,
according to the standard approach, no further low energy parameters
should be fixed. This mismatch in orders can be understood if one
considers the $N\Delta$ splitting to be a small parameter, making much
of the NNLO contributions to the potential become NLO ones, because
$c_3$ and $c_4 $ would be order minus one. In such a case, our
interpretation goes hand in hand with the standard approach; one needs
three independent low energy parameters at NLO in this counting. The
consequences of this $\Delta$-counting to higher orders within the
context of renormalization will be explored elsewhere. Of course, we
should point out that despite the rather tantalizing description
achieved at NNLO the existence of a consistent power counting
guaranteeing the success of the present approach to all orders remains
to be proved. A key ingredient of such a power counting would be the
correct incorporation of all long range physics. Apparently, within
Weinberg's power counting the NLO in the deuteron channel misses
important contributions. 

Finally, we have analyzed the consequences of a perturbative expansion
of TPE effects taking the OPE results as a zeroth order approximation
as suggested recently~\cite{Nogga:2005hy}. Our non-perturbative
calculations based on iterating a perturbative potential to all orders
exhibit unequivocal non-analytic dependence on the expansion
parameter, due to the singular character of the chiral potentials at
the origin. This is equivalent to a non-integer enhancement of the
power counting $\lambda^\alpha $ with $0 < \alpha < 1$ in the
potential, and it would be interesting to know the general rules of
such a counting {\it a priori}~\cite{Alvaro}. Thus, perturbation
theory based on standard power counting becomes divergent and can only
yield finite results at the expense of introducing more perturbative
counterterms than are needed in a non-perturbative treatment. This is
just a manifestation of the fact that singular potentials require
infinite counterterms in perturbation theory, while only a few ones
are needed non-perturbatively. Specifically, our analysis shows that
it would be necessary to include at least 4 counterterms for the
singlet $^1S_0$ and 6 counterterms for the triplet $^3S_1-^3D_1$
channel at NNLO. This proliferation of counterterms is expected to
occur also in other partial waves because the singularity of the
potential dominates over the centrifugal barrier at short
distances. In the $^1S_0$ channel, we have seen that adding more
counterterms in fact worsens the results for the effective range
expansion parameters.  In contrast, our non-perturbative calculations
are based on just 1 and 3 counterterms respectively.  The good quality
of our results suggests that our choice of less counterterms cannot be
refuted on the basis of phenomenology. In the deuteron case we have
made a calculation to NNLO in perturbation theory. Our analysis shows
that such a perturbative approach only makes sense if a finite cut-off
is introduced. In any case, the cut-off theory has less predictive
power, does not provide a better phenomenological description of the
deuteron than our non-perturbative renormalized results and is
non-convergent since NNLO corrections are numerically {\it much
larger} (two or three orders of magnitude) than NLO ones, despite
being parametrically small. In our view this is a perturbative
manifestation of the short distance dominance which has been unveiled
non-perturbatively. In addition, the difficulties faced by a
perturbative treatment are simply absent in the non-perturbative
approach.

One of the main goals of nuclear physics is the determination of the
nucleon-nucleon interaction. From a theoretical viewpoint the
disentanglement of such an interaction in terms of pion exchanges
based on chiral symmetry requires dealing with non-trivial and, to
some extent, unconventional non-perturbative renormalization issues in
the continuum, but it is crucial because it shows our quantitative
understanding of the underlying theory of quarks and gluons in the
chirally symmetric broken phase.  Our results also show that the
singular chiral Van der Waals forces are not necessarily spurious and
inconvenient features of the chiral potential. Instead, as we have
shown, the singularities alone in conjunction with renormalization
ideas explain much of the observed $S-$waves phase shifts with natural
values of the chiral constants, and provide an appealing physical
picture. In this regard, it is interesting to realize that based on
the analogy with molecular systems, which also exhibit a long range
Van der Waals force, the liquid drop model was formulated more than 60
years ago. Chiral dynamics may provide not only a closer analogy and
perhaps more quantitative insights into the hydrodynamical and
thermodynamical properties of nuclei but also a theoretical
justification from the underlying theory of strong interactions.

\begin{acknowledgments}

One of us (E.R.A.) thanks M. Rentmeester, R. Machleidt, E. Epelbaum,
N. Kaiser and G. Colangelo for useful correspondence. We thank them
and also R. Higa and A. Nogga for discussions and D. Phillips for
stressing the role of the $\Delta$. We thank J. Nieves for reading an
early version of the manuscript.  This work is supported in part by
funds provided by the Spanish DGI with grant no. FIS2005-00810, Junta
de Andaluc\'{\i}a grant no. FM-225 and EURIDICE grant number
HPRN-CT-2003-00311.

\end{acknowledgments}

\appendix

\section{The deuteron in OPE-distorted perturbation theory with a cut-off to NNLO}
\label{sec:second}

In this appendix we illustrate the situation discussed in
Sect.~\ref{sec:pert} by solving numerically the set of perturbative
equations (\ref{eq:pert_first}).  As we have mentioned such a
calculation makes only sense within a finite cut-off scheme.

In practice, we integrate from large distances ($\sim 25 {\rm fm}$)
with the conditions specified by Eq.~(\ref{eq:long_pert}) with some
prescribed values of $c_S$ and $c_D$~\footnote{This is numerically
more efficient and stable procedure than a direct use of the explicit
expressions Eq.~(\ref{eq:cS_cD_wronsk}) involving determinants.}. This
can be advantageously done using the superposition principle of
boundary conditions, Eq.~(\ref{eq:sup_bound}), yielding in
perturbation theory
\begin{eqnarray}
u_\gamma (r) &=& u_\gamma^{(0)}(r) + u_\gamma^{(2)}(r) +
u_\gamma^{(3)}(r) +\dots \nonumber \\ 
w_\gamma (r) &=& u_\gamma^{(0)}(r) + u_\gamma^{(2)}(r) +
u_\gamma^{(3)}(r) +\dots 
\end{eqnarray} 
At LO the wave function can be written as 
\begin{eqnarray}
u^{(0)}_\gamma (r) &=& u_S^{(0)} (r) + \eta^{(0)} u_D^{(0)} (r) \nonumber   \\ 
w^{(0)}_\gamma (r) &=& w_S^{(0)} (r) + \eta^{(0)} w_D^{(0)} (r) 
\end{eqnarray} 
and $\eta^{(0)}$ is determined from the regularity condition at the
origin~\cite{PavonValderrama:2005gu}. At LO the normalization
factor is
\begin{eqnarray}
\frac1{(A_S^{(0)})^2} = \int_0^\infty dr (u^{(0)}_\gamma (r)^2 +
w^{(0)}_\gamma (r)^2 )
\end{eqnarray}
The NLO and NNLO contributions are 
\begin{eqnarray}
u^{(2)}_\gamma (r) &=& c_S^{(2)} u_S^{(2)} (r) + \eta^{(0)} c_D^{(2)}
u_D^{(2)} (r) \nonumber \\ w^{(2)}_\gamma (r) &=& c_S^{(2)} w_S^{(2)}
(r) + \eta^{(0)} c_D^{(2)} w_D^{(2)} (r) \nonumber \\ 
u^{(3)}_\gamma (r) &=& c_S^{(3)} u_S^{(3)} (r) + \eta^{(0)} c_D^{(3)}
u_D^{(3)} (r) \nonumber \\ w^{(3)}_\gamma (r) &=& c_S^{(3)} w_S^{(3)}
(r) + \eta^{(0)} c_D^{(3)} w_D^{(3)} (r)
\end{eqnarray} 
The advantage is that the functions appearing here only depend on the
potential and the deuteron binding energy, whereas the coefficients
must be determined by some additional conditions. In the first place,
normalization to NLO and NNLO requires orthogonality of the wave
functions to the LO solution,
\begin{eqnarray}
0 &=& \int_0^\infty dr (u^{(0)} (r) u^{(2)} (r) + w^{(0)} (r) w^{(2)}
(r) ) \nonumber \\ 0 &=& \int_0^\infty dr (u^{(0)} (r) u^{(3)} (r) +
w^{(0)} (r) w^{(3)} (r) )
\end{eqnarray} 
This implies the couple of linear relations~\footnote{For instance, at
$r_c=0.5 {\rm fm} $ we get $ c_D^{(2)}= -5.865 c_S^{(2)}$ and $
c_D^{(3)}= -5.545 c_S^{(3)}$ }
\begin{eqnarray}
-\eta^{(0)} \frac{c_D^{(2)}}{c_S^{(2)}} &=& \frac{\int_0^\infty dr (u^{(0)} (r)
  u_S^{(2)} (r) + w^{(0)} (r) w_S^{(2)} (r) )}{\int_0^\infty dr
  (u^{(0)} (r) u_D^{(2)} (r) + w^{(0)} (r) w_D^{(2)} (r) )} \nonumber \\ 
-\eta^{(0)} \frac{c_D^{(3)}}{c_S^{(3)}} &=& \frac{\int_0^\infty dr (u^{(0)} (r)
  u_S^{(3)} (r) + w^{(0)} (r) w_S^{(3)} (r) )}{\int_0^\infty dr
  (u^{(0)} (r) u_D^{(3)} (r) + w^{(0)} (r) w_D^{(3)} (r) )} \nonumber \\ 
\label{eq:cS_cD_ort}
\end{eqnarray} 
Further relations can be obtained by imposing renormalization
conditions. Note that the required number of conditions increases with
the order. This is similar in spirit to the procedure of adding more
counterterms for the scattering problem discussed in
Sect.~\ref{sec:pert}. For instance, using the perturbative expansion for  
$A_S$ and $A_D$ 
\begin{eqnarray}
A_S &=& A_S^{(0)} \left( 1 + c_S^{(2)} + c_S^{(3)} + \dots \right)
\nonumber \\ A_D &=& A_D^{(0)} \eta^{(0)} \left( 1 + c_D^{(2)} +
c_D^{(3)} + \dots \right) = \eta A_S \nonumber \\
\end{eqnarray} 
In practice, we use a short distance cut-off $r_c$ for the NLO and
NNLO contributions only. Deuteron properties can be written to NNLO as follows 
\begin{eqnarray}
r_m &=& r_m^{(0)} + c_S^{(2)} r_m^{(2,S)} + \eta^{(0)} c_D^{(2)}
  r_m^{(2,D)} \nonumber \\ &+& c_S^{(3)} r_m^{(3,S)} + \eta^{(0)}
  c_D^{(3)} r_m^{(3,D)} + \dots \\ Q_d &=& Q_d^{(0)} + c_S^{(2)}
  Q_d^{(2,S)} + \eta^{(0)} c_D^{(2)} Q_d^{(2,D)} \nonumber \\ &+&
  c_S^{(3)} Q_d^{(3,S)} + \eta^{(0)} c_D^{(3)} Q_d^{(3,D)} + \dots
\label{eq:rm_Qd_pert}
\end{eqnarray}
where the potential contributions have explicitly been factored
out. The numerical solution requires some care, due to the short
distance instabilities and oscillations. This requires using an
adaptive grid to optimize the convergence.  Since solutions of
different orders must be mixed in the evaluation of the orthogonality
conditions, Eq.~(\ref{eq:cS_cD_ort}), and observables,
Eq.~(\ref{eq:rm_Qd_pert}), we solve all LO, NLO and NNLO equations
simultaneously to provide all functions on the same grid.
 
\begin{figure*}[]
\begin{center}
\epsfig{figure=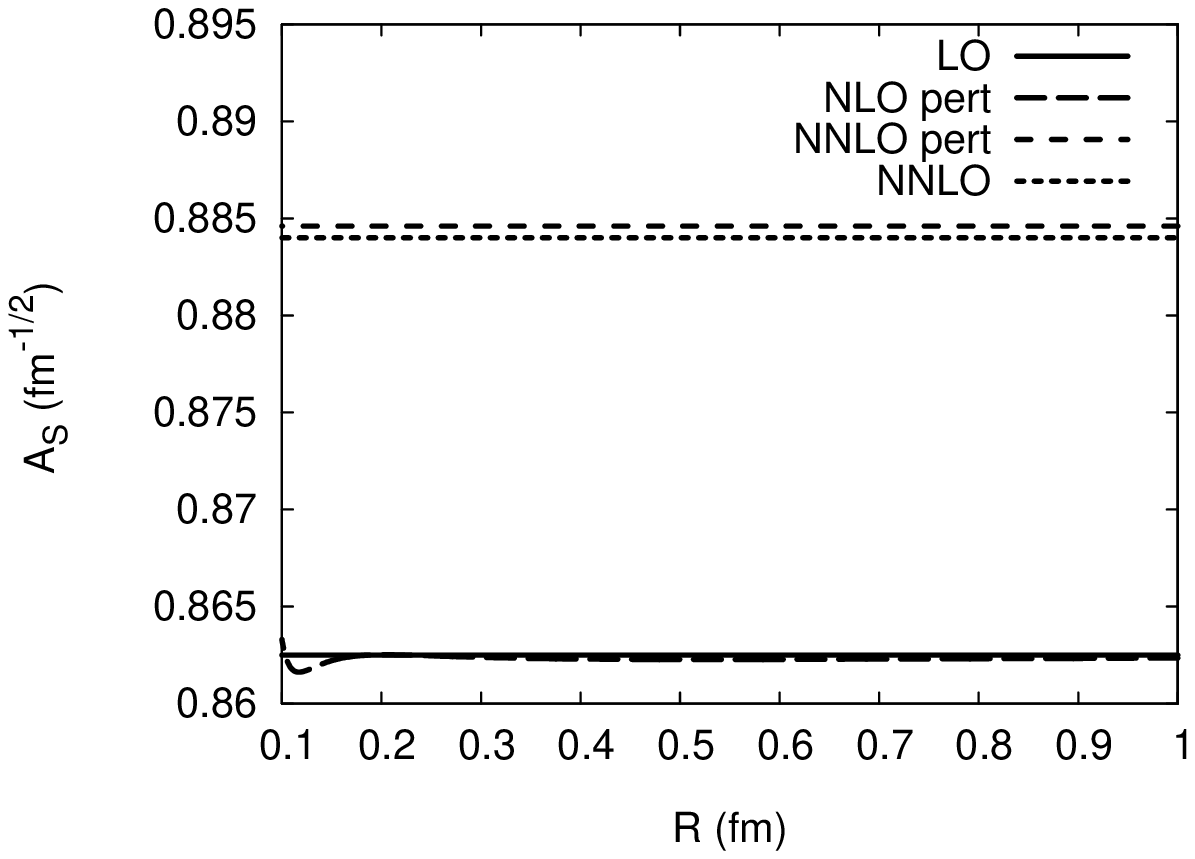,height=5.5cm,width=6.5cm} 
\epsfig{figure=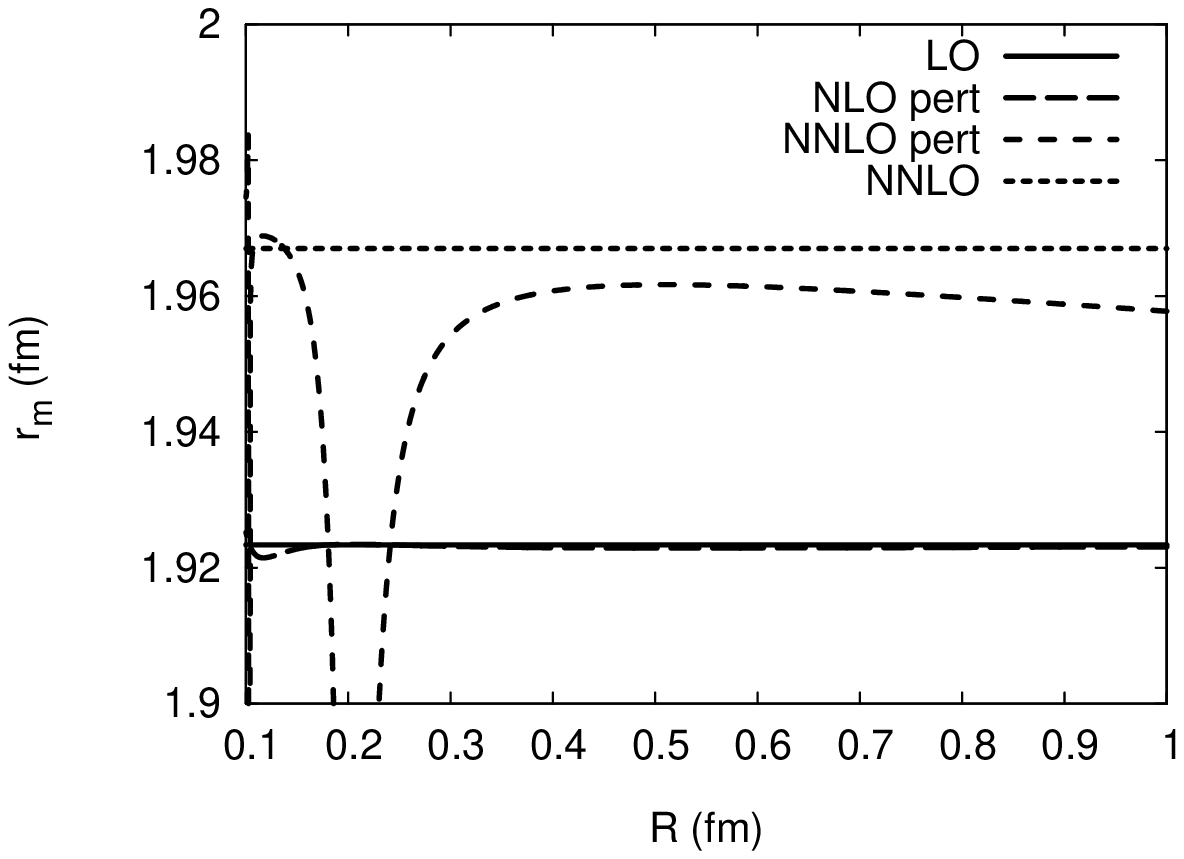,height=5.5cm,width=6.5cm} \\ 
\epsfig{figure=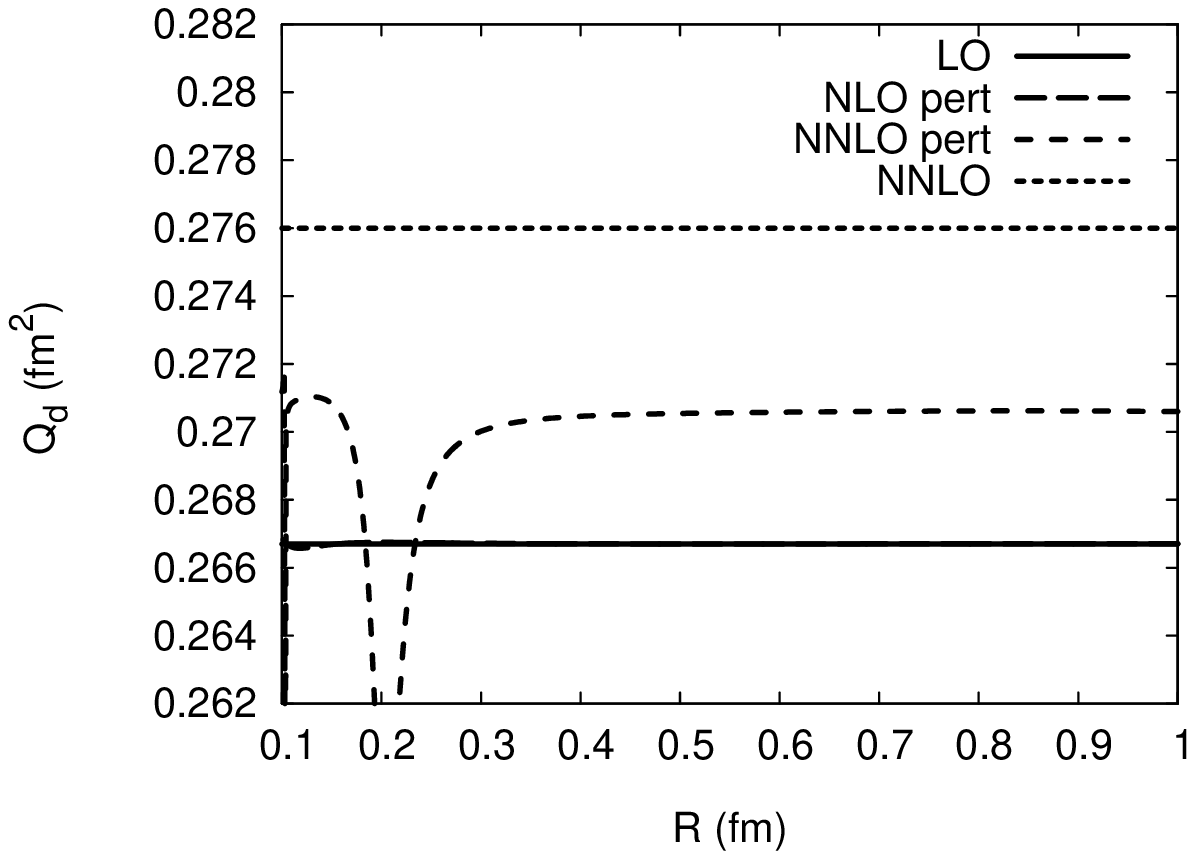,height=5.5cm,width=6.5cm} 
\epsfig{figure=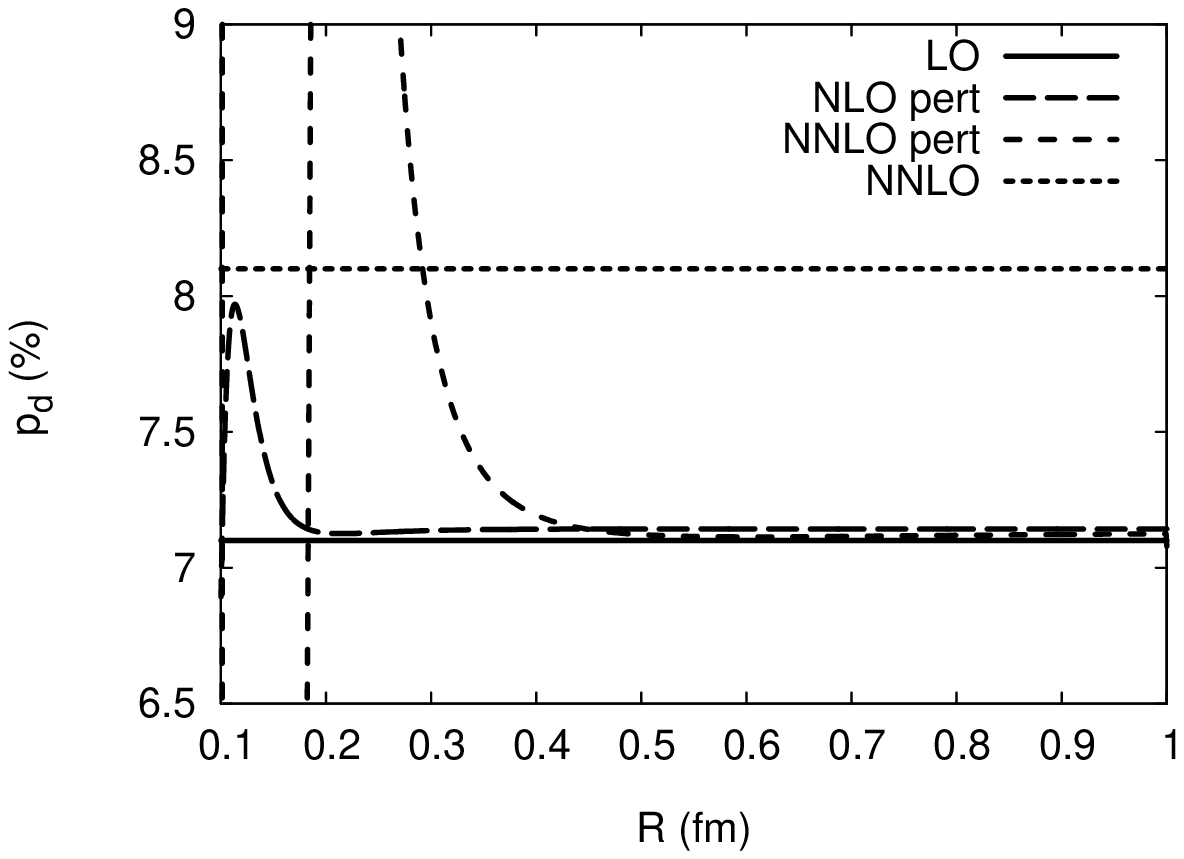,height=5.5cm,width=6.5cm} 
\end{center}
\caption{Dependence of the s-wave function normalization $ A_S (r_c)$,
the matter radius $r _m ( r_c) $, quadrupole moment $Q_d (r_c) $ and the 
D-wave probability $P_D(r_c) $ in the deuteron in perturbation theory
to NLO and NNLO on the short distance cut-off. We use Set IV for the
NNLO. In any case, the first order solution is fixed to the be
normalized and to reproduce the asymptotic D/S ratio, $\eta$. We
compare with the experimental value and the OPE result.}
\label{fig:pert_rc}
\end{figure*}

At NLO and fixing $r= r_c$ we demand the experimental value of $\eta$,
from Eq.~(\ref{eq:eta_pert}) and Eq.~(\ref{eq:cS_cD_ort}). This way, a
solution which we denote by $ ( c_S^{(2)}|_{\rm NLO}, c_D^{(2)}|_{\rm
NLO}) $ can be obtained. From there we can obtain deuteron properties
to NLO, as a function of the perturbative cut-off $r_c$ In
Fig.~\ref{fig:pert_rc} we show the dependence of $A_S$, $r_m$, $Q_d$
and $p_d$ on $r_c$.  As we see the NLO correction is tiny and stable
for $r_c > 0.2 {\rm fm}$. At NNLO we fix $\eta$ and $A_S$. The
solution is now $ ( c_S^{(2)}|_{\rm NNLO}, c_D^{(2)}|_{\rm NNLO}) $
and $ ( c_S^{(3)}|_{\rm NNLO}, c_D^{(3)}|_{\rm NNLO}) $. Note that in
general the NLO coefficients $c_S^{(2)}$ and $c_D^{(2)}$ must be
readjusted. In this case the correction is much larger than the NLO
case (see Fig.~\ref{fig:pert_rc}) and the cut-off dependence is
stronger due to the the $1/r^6$ singularity of the NNLO potential. As
a curiosity, we mention that at short distances worry-some negative
D-wave probabilities show up below $r_c =0.17 {\rm fm}$ at NNLO, a
spurious feature which can only take place in perturbation theory and
sets a unitarity bound on the short distance perturbative cut-off.
Numerical results are provided in Table~\ref{tab:table5} for Set
IV. Typically, we find that results do not depend dramatically on the
chosen chiral couplings . We take $r_c=0.5 {\rm fm}$ as a standard
choice. As we see, finite cut-off perturbation theory does not work
better than our non-perturbative results of Sect.~\ref{sec:triplet},
and in fact requires one more counterterm. Actually, this is a
perturbative indication that NNLO is more important than NLO, casting
doubts on the convergence of the approach.

Finally, we have checked that taking the LO to be the {\it full} OPE
potential, ${\bf U}_{1\pi}$ and the perturbation to be ${\bf
U}_{2\pi}^{(2)} + {\bf U}_{2\pi}^{(3)}$ as a whole and keeping the
cut-off $r_c > 0.1 {\rm fm}$ does not change the results
significantly. Actually, the perturbative result does not account for
the value obtained non-perturbatively, despite the apparent linear
behaviour observed when changing numerically the scaling parameter
$\lambda$ in the region $ \lambda \gg 0.1$ (see
Fig.~\ref{fig:Asr0-3S1.lambda.NNLO}). This supports our conclusion
that perturbation theory does not compute the slope of $A_S( \lambda)
$ at the origin. In addition, even if we disregard the divergence by
introducing a cut-off, the perturbative calculation does not account
for the non-perturbative renormalized result.

\begin{table*}
\caption{\label{tab:table5} Comparison of finite cut-off perturbation
theory, $r_c=0.5 {\rm fm}$, with renormalized non-perturbative results
for the Deuteron properties. We use the non-relativistic relation $
\gamma= \sqrt{ 2 \mu_{np} B} $ with $B=2.224575(9)$. 
The errors quoted in the perturbative calculations reflect the uncertainty
in the input parameters $\gamma$, $\eta$ and $A_S$.
Similarly, the errors quoted in the TPE reflect the uncertainty 
in the non-potential parameters
$\gamma$ and $\eta$. We use Set IV of low energy constants $c_1$,$c_3$
and $c_4$.}
\begin{ruledtabular}
\begin{tabular}{|c|c|c|c|c|c|c|c|c|}
\hline & $r_c $ & $\gamma ({\rm fm}^{-1})$ & $\eta$ & $A_S ( {\rm
fm}^{-1/2}) $ & $r_m ({\rm fm})$ & $Q_d ( {\rm fm}^2) $ & $P_D $ & $
\langle r^{-1} \rangle $ ( ${\rm fm}^{-1}$ ) \\ \hline \hline Non-perturbative \\ \hline 
$U_{1\pi}$  & 0 & Input & 0.02633 & 0.8681(1) & 1.9351(5) & 0.2762(1)
& 7.31(1)\% &  0.476(3) \\ \hline 

$U_{1\pi}+ U_{2\pi} $  & 0 &  Input & Input & 0.884(4) & 1.967(6) & 0.276(3)
& 8(1)\%  &  0.447(5) \\ \hline \hline Perturbative \\ \hline 

$U_{1\pi}^{(0)}$ &   0 & Input & 0.02555 & 0.8625(2) & 1.9233(5) & 0.2667(1) 
& 7.14(1)\%  & 0.484(3) \\ \hline 
$U_{1\pi}^{(2)}+U_{2\pi}^{(2)}$ & 0.5 fm  & Input & Input & 0.862(2) & 1.923(4) & 0.2667(3)
& 7.14(1)\% & 0.484(3) \\ \hline  
$U_{1\pi}^{(3)}+U_{2\pi}^{(3)}$ & 0.5 fm   & Input & Input & Input &
1.962(2) & 0.2705(2) & 7.12(1)\% & 0.484(3) \\ \hline \hline Potentials \\ \hline 
NijmII & --- & 0.231605 & 0.02521 & 0.8845(8) & 1.9675 & 0.2707 & 
5.635\% & 0.4502  \\
Reid93 & --- & 0.231605 & 0.02514 & 0.8845(8) & 1.9686 & 0.2703 & 
5.699\% & 0.4515 \\ \hline 
Exp. &  --- & 0.231605 &  0.0256(4)  & 0.8846(9)  & 1.971(6)  &
0.2859(3) & --- & ---   \\
\end{tabular}
\end{ruledtabular}
\end{table*}


\end{document}